\documentclass[11pt]{article} 
\usepackage{indentfirst,amsmath,latexsym,amssymb,amsbsy,amscd}

\newcommand{\be}{\begin{equation}}
\newcommand{\ee}{\end{equation}}
\newcommand{\ra}{\rightarrow}

\newcommand{\lra}{\leftrightarrow}
\newcommand{\llra}{\longleftrightarrow}

\newcommand{\qed}{\hfill $\bullet$}

\newcommand{\bllt}{\bullet}

\newtheorem{prop}{Proposition}
\newtheorem{thm}[prop]{Theorem}
\newtheorem{alg}[prop]{Algorithm}
\newtheorem{lem}[prop]{Lemma}
\newtheorem{cor}[prop]{Corollary}
\newtheorem{defn}[prop]{Definition}

\newenvironment{prf}{\trivlist \item[\hskip \labelsep{\bf Proof.}]}{\qed \endtrivlist}

\begin{document}

\newcommand{\je}[1]{j={#1},\ldots,\ell}
\newcommand{\ifof}{if and only if}
\newcommand{\bone}{\mathbf{1}}
\newcommand{\bzero}{\mathbf{0}}

\newcommand{\bmcpz}{{\mathbf{Z}}}
\newcommand{\bmd}{{\mathbf{d}}}
\newcommand{\bmb}{{\mathbf{b}}}
\newcommand{\bma}{{\mathbf{a}}}

\newcommand{\bmy}{{\mathbf{y}}}
\newcommand{\bmv}{{\mathbf{v}}}
\newcommand{\bmm}{{\mathbf{m}}}
\newcommand{\bmz}{{\mathbf{z}}}
\newcommand{\bms}{{\mathbf{s}}}
\newcommand{\bmcpw}{{\mathbf{W}}}

\newcommand{\bmc}{{\mathbf{c}}}
\newcommand{\bmu}{{\mathbf{u}}}
\newcommand{\bmr}{{\mathbf{r}}}
\newcommand{\bmg}{{\mathbf{g}}}

\newcommand{\bmbht}{{\hat{\bmb}}}
\newcommand{\bmcht}{{\hat{\bmc}}}
\newcommand{\bmuht}{{\hat{\bmu}}}
\newcommand{\bmrht}{{\hat{\bmr}}}
\newcommand{\bmght}{{\hat{\bmg}}}

\newcommand{\bmctld}{{\tilde{\bmc}}}

\newcommand{\bmcck}{{\check{\bmc}}}
\newcommand{\bmuck}{{\check{\bmu}}}
\newcommand{\bmrck}{{\check{\bmr}}}
\newcommand{\bmgck}{{\check{\bmg}}}

\newcommand{\bmcbr}{{\bar{\bmc}}}
\newcommand{\bmubr}{{\bar{\bmu}}}
\newcommand{\bmrbr}{{\bar{\bmr}}}
\newcommand{\bmgbr}{{\bar{\bmg}}}

\newcommand{\bmcddt}{{\ddot{\bmc}}}
\newcommand{\bmuddt}{{\ddot{\bmu}}}
\newcommand{\bmrddt}{{\ddot{\bmr}}}
\newcommand{\bmgddt}{{\ddot{\bmg}}}

\newcommand{\bmcdt}{{\dot{\bmc}}}
\newcommand{\bmudt}{{\dot{\bmu}}}
\newcommand{\bmrdt}{{\dot{\bmr}}}
\newcommand{\bmgdt}{{\dot{\bmg}}}

\newcommand{\bmcac}{{\acute{\bmc}}}
\newcommand{\bmuac}{{\acute{\bmu}}}
\newcommand{\bmrac}{{\acute{\bmr}}}
\newcommand{\bmgac}{{\acute{\bmg}}}

\newcommand{\bmcgr}{{\grave{\bmc}}}
\newcommand{\bmugr}{{\grave{\bmu}}}
\newcommand{\bmrgr}{{\grave{\bmr}}}
\newcommand{\bmggr}{{\grave{\bmg}}}

\newcommand{\bme}{{\mathbf{e}}}
\newcommand{\bmvht}{{\hat{\bmv}}}

\newcommand{\bmcpu}{{\mathbf{U}}}
\newcommand{\bmcpr}{{\mathbf{R}}}
\newcommand{\bmcpb}{{\mathbf{B}}}
\newcommand{\bmcpbht}{{\hat{\mathbf{B}}}}
\newcommand{\bmcps}{{\mathbf{S}}}
\newcommand{\bmcpv}{{\mathbf{V}}}
\newcommand{\bmw}{{\mathbf{w}}}
\newcommand{\bmx}{{\mathbf{x}}}

\newcommand{\cht}{{\hat{c}}}
\newcommand{\cac}{{\acute{c}}}
\newcommand{\cgr}{{\grave{c}}}
\newcommand{\cddt}{{\ddot{c}}}
\newcommand{\cdt}{{\dot{c}}}

\newcommand{\rht}{{\hat{r}}}
\newcommand{\rac}{{\acute{r}}}
\newcommand{\rgr}{{\grave{r}}}
\newcommand{\rddt}{{\ddot{r}}}
\newcommand{\rdt}{{\dot{r}}}

\newcommand{\uht}{{\hat{u}}}
\newcommand{\uac}{{\acute{u}}}
\newcommand{\ugr}{{\grave{u}}}
\newcommand{\uddt}{{\ddot{u}}}
\newcommand{\udt}{{\dot{u}}}

\newcommand{\qht}{{\hat{q}}}
\newcommand{\qac}{{\acute{q}}}
\newcommand{\qgr}{{\grave{q}}}
\newcommand{\qddt}{{\ddot{q}}}
\newcommand{\qdt}{{\dot{q}}}

\newcommand{\dottheta}{{\dot{\theta}}}
\newcommand{\bartheta}{{\bar{\theta}}}
\newcommand{\dotomega}{{\dot{\omega}}}
\newcommand{\hatomega}{{\hat{\omega}}}
\newcommand{\baromega}{{\bar{\omega}}}
\newcommand{\bhat}{{\hat{b}}}
\newcommand{\btld}{{\tilde{b}}}
\newcommand{\bdot}{{\dot{b}}}
\newcommand{\bbar}{{\bar{b}}}
\newcommand{\bddot}{{\ddot{b}}}
\newcommand{\bact}{{\acute{b}}}
\newcommand{\bgrv}{{\grave{b}}}
\newcommand{\gbr}{{\bar{g}}}
\newcommand{\hbr}{{\bar{h}}}
\newcommand{\cbr}{{\bar{c}}}
\newcommand{\rbr}{{\bar{r}}}
\newcommand{\ubr}{{\bar{u}}}
\newcommand{\qbr}{{\bar{q}}}
\newcommand{\cpqbr}{{\bar{Q}}}
\newcommand{\cpcbr}{{\bar{C}}}
\newcommand{\cpbbr}{{\bar{B}}}
\newcommand{\cpgbr}{{\bar{G}}}
\newcommand{\cphbr}{{\bar{H}}}
\newcommand{\cpnbr}{{\bar{N}}}
\newcommand{\phibr}{{\bar{\phi}}}
\newcommand{\sovrcpx}{{\overset{*}{X}}}

\newcommand{\bmghat}{{\hat{\mathbf{g}}}}
\newcommand{\bmgbar}{{\bar{\mathbf{g}}}}
\newcommand{\bmehat}{{\hat{\mathbf{e}}}}

\newcommand{\ssr}{Schreier series}
\newcommand{\fssr}{forward Schreier series}
\newcommand{\bssr}{backward Schreier series}
\newcommand{\gm}{generator matrix}
\newcommand{\gms}{generator matrices}
\newcommand{\stm}{static matrix}
\newcommand{\stms}{static matrices}
\newcommand{\sm}{shift matrix}
\newcommand{\sms}{shift matrices}
\newcommand{\gvec}{generator vector}
\newcommand{\gvecs}{generator vectors}
\newcommand{\svec}{shift vector}
\newcommand{\svecs}{shift vectors}
\newcommand{\mchn}{matrix chain}

\newcommand{\cdc}{coset decomposition chain}
\newcommand{\creps}{coset representatives}
\newcommand{\crep}{coset representative}
\newcommand{\crepc}{coset representative chain}
\newcommand{\compset}{complete set of coset representatives}
\newcommand{\fss}{full symmetry system}
\newcommand{\consub}{consistent subsystem}
\newcommand{\nss}{natural symmetry system}
\newcommand{\nsss}{natural symmetry systems}

\newcommand{\xj}{{\{X_j\}}}
\newcommand{\yi}{{\{Y_i\}}}
\newcommand{\xjt}{{\{X_j^t\}}}
\newcommand{\yit}{{\{Y_i^t\}}}
\newcommand{\yj}{{\{Y_j\}}}
\newcommand{\yk}{{\{Y_k\}}}
\newcommand{\bi}{{\{B^{(i)}\}}}

\newcommand{\msfc}{{\mathsf{C}}}
\newcommand{\calz}{\mathcal{Z}}
\newcommand{\barcaln}{{\bar{\caln}}}
\newcommand{\calp}{\mathcal{P}}
\newcommand{\calf}{{\mathcal{F}}}
\newcommand{\calt}{{\mathcal{T}}}
\newcommand{\calv}{{\mathcal{V}}}
\newcommand{\ellctl}{$\ell$-controllable}
\newcommand{\cald}{{\mathcal{D}}}
\newcommand{\calb}{{\mathcal{B}}}
\newcommand{\cals}{{\mathcal{S}}}
\newcommand{\calq}{{\mathcal{Q}}}
\newcommand{\calm}{{\mathcal{M}}}
\newcommand{\caln}{\mathcal{N}}
\newcommand{\calnht}{{\hat{\caln}}}

\newcommand{\calr}{{\mathcal{R}}}
\newcommand{\calrht}{{\hat{\calr}}}

\newcommand{\calu}{{\mathcal{U}}}
\newcommand{\calx}{{\mathcal{X}}}
\newcommand{\cale}{{\mathcal{E}}}
\newcommand{\call}{{\mathcal{L}}}

\newcommand{\cpuht}{{\hat{U}}}
\newcommand{\cprht}{{\hat{R}}}
\newcommand{\calbht}{{\hat{\calb}}}

\newcommand{\rmdef}{\stackrel{\rm def}{=}}

\newcommand{\bigtridn}{{\bigtriangledown}}
\newcommand{\tridn}[2]{{\bigtridn_{#1,#2}}}
\newcommand{\btridn}[2]{{-\!\!\!\!\!\bigtridn_{#1,#2}}}
\newcommand{\tridnut}[3]{{\tridn{#1}{#2}(\bmu^{#3})}}
\newcommand{\tridncut}[3]{{\tridn{#1}{#2}(\bmcpu^{#3})}}
\newcommand{\tridncrt}[3]{{\tridn{#1}{#2}(\bmcpr^{#3})}}
\newcommand{\btridnut}[3]{{\btridn{#1}{#2}(\bmu^{#3})}}
\newcommand{\btridncut}[3]{{\btridn{#1}{#2}(\bmcpu^{#3})}}

\newcommand{\tridnrt}[3]{{\tridn{#1}{#2}(\bmr^{#3})}}
\newcommand{\tridnrtddt}[3]{{\tridn{#1}{#2}(\bmrddt^{#3})}}
\newcommand{\tridnrtht}[3]{{\tridn{#1}{#2}(\bmrht^{#3})}}
\newcommand{\btridnrt}[3]{{\btridn{#1}{#2}(\bmr^{#3})}}
\newcommand{\btridnrtht}[3]{{\btridn{#1}{#2}(\bmrht^{#3})}}
\newcommand{\btridnrtdt}[3]{{\btridn{#1}{#2}(\bmrdt^{#3})}}
\newcommand{\btriuprtddt}[3]{{\btriup{#1}{#2}(\bmrddt^{#3})}}
\newcommand{\btridnrtddt}[3]{{\btridn{#1}{#2}(\bmrddt^{#3})}}

\newcommand{\tridnuty}[3]{{\tridn{#1}{#2}(\bmu_Y^{#3})}}
\newcommand{\tridncrty}[3]{{\tridn{#1}{#2}(\bmcpr_Y^{#3})}}
\newcommand{\tridnrty}[3]{{\tridn{#1}{#2}(\bmr_Y^{#3})}}
\newcommand{\tridnrthty}[3]{{\tridn{#1}{#2}(\bmrht_Y^{#3})}}

\newcommand{\tridnrtdt}[3]{{\tridn{#1}{#2}(\bmrdt^{#3})}}
\newcommand{\btriuprtdt}[3]{{\btriup{#1}{#2}(\bmrdt^{#3})}}
\newcommand{\btridnbonet}[3]{{\btridn{#1}{#2}(\bone^{#3})}}

\newcommand{\tridnutbs}[3]{{\tridn{#1}{#2}(\bsigma\bmu^{#3})}}
\newcommand{\btridnutbs}[3]{{\btridn{#1}{#2}(\bsigma\bmu^{#3})}}
\newcommand{\tridnwt}[3]{{\tridn{#1}{#2}(\bmw^{#3})}}

\newcommand{\bigtriup}{{\bigtriangleup}}
\newcommand{\triup}[2]{{\bigtriup_{#1,#2}}}
\newcommand{\btriup}[2]{{-\!\!\!\!\!\bigtriup_{#1,#2}}}
\newcommand{\triuput}[3]{{\triup{#1}{#2}(\bmu^{#3})}}
\newcommand{\triupcut}[3]{{\triup{#1}{#2}(\bmcpu^{#3})}}
\newcommand{\btriuput}[3]{{\btriup{#1}{#2}(\bmu^{#3})}}
\newcommand{\btriupbonet}[3]{{\btriup{#1}{#2}(\bone^{#3})}}
\newcommand{\btriupcut}[3]{{\btriup{#1}{#2}(\bmcpu^{#3})}}

\newcommand{\triupwt}[3]{{\triup{#1}{#2}(\bmw^{#3})}}
\newcommand{\btriupwt}[3]{{\btriup{#1}{#2}(\bmw^{#3})}}

\newcommand{\tridnutp}[3]{{\tridn{#1}{#2}(\bmu^{#3})'}}
\newcommand{\tridnutd}[3]{{\tridn{#1}{#2}({\bmudt}^{#3})}}
\newcommand{\tridnuth}[3]{{\tridn{#1}{#2}({\bmuht}^{#3})}}
\newcommand{\btridnutd}[3]{{\btridn{#1}{#2}(\bmudt^{#3})}}
\newcommand{\btridnuth}[3]{{\btridn{#1}{#2}(\bmuht^{#3})}}
\newcommand{\btridnsut}[3]{{\btridn{#1}{#2}(\bsigma\bmu^{#3})}}
\newcommand{\omegbtridnut}[3]
{{\omega_{#1,#2}^{#3}(u_{#1,#2}^{#3};\btridnut{#1}{#2}{#3})}}

\newcommand{\btriuptblt}[3]{{\btriup{#1}{#2}(\bullet^{#3})}}


\newcommand{\tridntht}[3]{{\tridn{#1}{#2}(\theta^{#3})}}
\newcommand{\tridnnt}[3]{{\tridn{#1}{#2}(\caln^{#3})}}

\newcommand{\bbmcu}{{-\!\!\!\!\!\bmcpu}}
\newcommand{\bcaln}{{-\!\!\!\!\!\caln}}

\newcommand{\fracfjk}[2]
{{
\frac{\calf^{#1}(\Delta_{#2}^t)}{\calf^{#1}(\Delta_{{#2}-1}^t)}
}}

\newcommand{\balpha}{{\boldsymbol{\alpha}}}
\newcommand{\bsigma}{{\boldsymbol{\sigma}}} 
\newcommand{\blambda}{{\boldsymbol{\lambda}}}
\newcommand{\bomga}{{\boldsymbol{\omega}}}
\newcommand{\bvarpi}{{\boldsymbol{\varpi}}}
\newcommand{\bomgaht}{{\hat{\bomga}}}
\newcommand{\bcomega}{{\boldsymbol{\Omega}}}

\newcommand{\rjkt}{{r_{j,k}^t}}
\newcommand{\rmnt}{{r_{m,n}^t}}
\newcommand{\sjkt}{{s_{j,k}^t}}
\newcommand{\smnt}{{s_{m,n}^t}}

\newcommand{\uj}[1]{{\bmu_{[#1,\ell]}^t}}
\newcommand{\ujt}[2]{{\bmu_{[#1,\ell]}^{#2}}}
\newcommand{\cuj}[1]{{\bmcpu_{[#1,\ell]}^t}}
\newcommand{\cujt}[2]{{\bmcpu_{[#1,\ell]}^{#2}}}

\newcommand{\duj}{{\cald_{\cuj{j}}}}
\newcommand{\dujinf}{{(\duj)^\infty}}
\newcommand{\dujpone}{{\cald_{\cuj{j+1}}}}

\newcommand{\uu}[1]{\bmu_{#1},\ldots,\bmu_\ell}
\newcommand{\uut}[2]{\bmu_{#1}^{#2},\ldots,\bmu_\ell^{#2}}
\newcommand{\cuu}[1]{\bmcpu_{#1} \times \cdots \times \bmcpu_\ell}
\newcommand{\cuut}[2]{\bmcpu_{#1}^{#2} \times \cdots \times \bmcpu_\ell^{#2}}

\newcommand{\rrt}[2]{\bmr_{#1}^{#2},\ldots,\bmr_\ell^{#2}}
\newcommand{\crrt}[2]{\bmcpr_{#1}^{#2} \times \cdots \times \bmcpr_\ell^{#2}}

\newcommand{\vphie}[1]{\varphi_{e,#1}}
\newcommand{\vphis}[1]{\varphi_{s,#1}}

\newcommand{\pv}[1]{\sigma\bmu_{#1}}
\newcommand{\pvt}[2]{\sigma\bmu_{#1}^{#2}}
\newcommand{\pvpv}[1]{\pv{#1},\ldots,\pv{\ell-1}}
\newcommand{\pvpvt}[2]{\pvt{#1}{#2},\ldots,\pvt{\ell-1}{#2}}
\newcommand{\pvphi}[1]{\sigma\varphi_{#1}}
\newcommand{\pvphit}[2]{\sigma\varphi_{#1}^{#2}}

\newcommand{\ssl}{{/ \!\! /}} 

\newcommand{\drbinf}{{\cald^\infty(\calr,\bmcpb)}}
\newcommand{\drbinfy}{{\cald^\infty(\calr_Y,\bmcpb_Y)}}
\newcommand{\drbinfone}{{\cald^\infty(\calr_1,\bmcpb_1)}}
\newcommand{\drbinftwo}{{\cald^\infty(\calr_2,\bmcpb_2)}}
\newcommand{\drbinfcone}{{\cald^\infty(\calr_{c,1},\bmcpb_{c,1})}}
\newcommand{\drbinfctwo}{{\cald^\infty(\calr_{c,2},\bmcpb_{c,2})}}
\newcommand{\drtbt}{{\cald(\bmcpr^t,\calb^t)}}
\newcommand{\drtbty}{{\cald(\bmcpr_Y^t;\calb_Y^t)}}

\newcommand{\edrbinf}{{E(\cald^\infty(\calr,\bmcpb))}}
\newcommand{\edrbinfs}{{E_s(\cald^\infty(\calr,\bmcpb))}}
\newcommand{\edrbinfy}{{E_Y(\cald^\infty(\calr_Y,\bmcpb_Y))}}
\newcommand{\edrbinfsy}{{E_{s,Y}(\cald^\infty(\calr_Y,\bmcpb_Y))}}
\newcommand{\edrbinfone}{{E(\cald^\infty(\calr_1,\bmcpb_1))}}
\newcommand{\edrbinftwo}{{E(\cald^\infty(\calr_2,\bmcpb_2))}}
\newcommand{\edrbinfcone}{{E(\cald^\infty(\calr_{c,1},\bmcpb_{c,1}))}}
\newcommand{\edrbinfctwo}{{E(\cald^\infty(\calr_{c,2},\bmcpb_{c,2}))}}
\newcommand{\edrbinftwos}{{E_s(\cald^\infty(\calr_2,\bmcpb_2))}}
\newcommand{\edrbinfctwos}{{E_s(\cald^\infty(\calr_{c,2},\bmcpb_{c,2}))}}
\newcommand{\edrbinfc}{{E(\cald^\infty(\calr_c,\bmcpb_c))}}
\newcommand{\edrbinfcy}{{E_Y(\cald^\infty(\calr_{Y,c},\bmcpb_{Y,c}))}}

\newcommand{\duinf}{{\cald^\infty(\calu)}}
\newcommand{\duinfy}{{\cald^\infty(\calu_Y)}}
\newcommand{\dut}{{\cald(\bmcpu^t)}}
\newcommand{\duty}{{\cald(\bmcpu_Y^t)}}

\title{Time and harmonic study of strongly controllable 
group systems, group shifts, and group codes}
 
\author{K. M. Mackenthun Jr. (email:  {\tt ken1212576@gmail.com})}
\maketitle

\vspace{3mm}
{\bf ABSTRACT} 
\vspace{3mm}

In this paper we give a complementary view of some of the results
on group systems by Forney and Trott.  We find an encoder of a group system
which has the form of a time convolution.  We consider this to be a
time domain encoder while the encoder of Forney and Trott is a spectral
domain encoder.  We study the outputs of time and spectral domain encoders
when the inputs are the same, and also study outputs when the same input
is used but time runs forward and backward.  In an abelian group system,
all four cases give the same output for the same input, but this may not be
true for a nonabelian system.  Moreover, time symmetry and harmonic symmetry
are broken for the same reason.  We use a canonic form, a set of tensors,
to show how the outputs are related.  These results show there is a time
and harmonic theory of group systems.

\newpage
\vspace{3mm}
{\bf 1.  INTRODUCTION}
\vspace{3mm}

The idea of group shifts and group codes is important in several areas
of mathematics and engineering such as symbolic dynamics,
linear systems theory, and coding theory.  
Research in this area started with the 
work of Kitchens \cite{KIT}, Willems \cite{JW},
Forney and Trott \cite{FT}, and Loeliger and Mittelholzer \cite{LM}.

Kitchens \cite{KIT} introduced the idea of a group shift \cite{LMr}
and showed that a group shift has finite memory, i.e., 
it is a shift of finite type \cite{LMr}.  Using
the work of Willems \cite{JW} on linear systems, Forney and Trott \cite{FT} 
describe the state group and state code of a set of sequences with 
a group property, which they term a group code $\msfc$.
A time invariant group code is essentially a group shift.
They show that any group code that is complete
(any global constraints can be determined locally, see \cite{FT}) can be wholely
specified by a sequence of connected labeled group trellis sections 
(which may vary in time) which form a group trellis $C$.  
They explained the important idea of ``shortest length code sequences''
or generators.  A generator is a code sequence which is not a combination
of shorter sequences.  In a strongly controllable group code, the 
nontrivial portions of all generators have a bounded length.  They
give an encoder whose inputs are generators and whose outputs are codewords 
in the group code.  At each time $t$, a finite set of generators is used to 
give a symbol in the codeword.

Loeliger and Mittelholzer \cite{LM} obtain an analog of the derivation
of Forney and Trott starting with a group trellis $C$ instead of the
group of sequences $\msfc$.  To derive their encoder, 
they use an intersection of paths 
which split and merge to the identity path in the trellis, an analog
of the quotient group of code sequences (granule) used in \cite{FT}.

Forney and Trott also suggest the term group system in place of group code.
Here we generally use the term group system rather than group code
because some results have analogues in classical systems theory and
harmonic analysis.  We only consider time
invariant group systems; therefore the results here also apply 
to group shifts.  In addition, we only consider strongly controllable 
group systems, in which there is
a fixed integer $\ell$ such that for any time $t$, for any sequence
on $(-\infty,t]$ there exists a valid path of length $\ell$ to any 
sequence on $[t+\ell,\infty)$.  Then the nontrivial lengths of
the generators are at most $\ell$. 

Forney and Trott have shown that any group system $\msfc$ can be reduced
to a group trellis $C$ whose vertices are the states of the group system.
The states are defined using a group theoretic construction as quotient
groups.  Each component of the trellis is a trellis section, a collection
of branches which forms a branch group $B^t$ at time $t$.
We call group trellis $C$ the {\it first canonic form} of the
group system.  The group system can be implemented with an encoder.
The encoder has a shift register structure and 
the outputs give a trellis which is graph isomorphic to $C$.

In this paper, we consider several problems that arise from their 
discussion.  First, their encoder is implemented going forward in time.
It is natural to ask what is the encoder if we go backwards in time, and 
if both forward and backward encoders are filled with the same sequence 
of generators, are their outputs the same.  We answer this question here.

The Forney and Trott encoder does not have the form of a time convolution.
Next, we find another encoder which has the form of a convolution.
For this reason, we call this encoder a time domain encoder, and the
Forney and Trott encoder a spectral domain encoder.  The time domain
encoder can be implemented for forward and backward time, and the same
question applies as for the spectral domain encoder.  The time domain
encoder uses the same input sequences of generators as the spectral domain
encoder.  So we may also compare the outputs of the time and spectral domain 
encoder if both use the same input.

In this paper, we show how the time and spectral domain encoders, and forward
and backward time encoders, are related.  In the abelian group system, we
show that all four encoders give the same outputs if the same input is used.
But in the nonabelian system, these symmetries can break, and we do not
necessarily get the same output for the same input.  Moreover,
time symmetry and harmonic symmetry break for the same reason. 
It is interesting to observe how these symmetries
break since a group system is possibly the most elementary
nonlinear system in mathematics with a time and spectral domain 
interpretation.

When time symmetry or harmonic symmetry breaks, we show how the two 
different outputs are related.  To do this we use a {\it second canonic
form} of the group system.  The second canonic form is a set of tensors $\calr$.
Each tensor is a sequence of generators.  At each time $t$, a component
of the tensor is a matrix, called a \stm.  Each \stm\ 
is formed by $\ell+1$ \sms\ at times $t-j$, for $j=0,1,\ldots,\ell$.
A row in a \sm\ is a \gvec, the nontrivial components of a
generator.    

The entries in a \stm\ are components of different \gvecs.  We show that
these elements are the representatives of a \cdc\ of the branch group
$B^t$ at time $t$.  And so each tensor is a sequence of branches
which is a path in the group trellis $C$.  Moreover, the \stm\ at time $t$
can be used to define group theoretic input and output states which are 
isomorphic to the quotient group states defined for $C$.  This means a
group trellis $C$ can be reduced to a set of tensors $\calr$.

We believe $\calr$ is more revealing of the structure of a group system
than group trellis $C$.  The group trellis $C$ emphasizes the branch
group $B^t$ of a trellis section.  But the set of tensors $\calr$ shows that
$B^t$ is a secondary object which is a snapshot at time $t$ of
$\ell+1$ \sms\ formed by \gvecs.  In addition, time reversal appears 
deceptively simple in $C$, but canonic form $\calr$ shows that it is not.

The canonic form $\calr$ has a natural shift structure which arises from quotient 
groups in the \cdc\ of $B^t$.  Then $\calr$ can also be written as a trellis, which
is graph isomorphic to $C$.  The labels of the branches in the trellis
are matrices.

The spectral domain encoder has a set theoretic description of its
states which is graph isomorphic to the group theoretic states of $C$,
but the isomorphism has not been described.
There is also a set theoretic construction of the states of $\calr$
which matches the set theoretic construction of the spectral domain encoder.
This explains the isomorphism between states of the Forney
and Trott encoder and states of $C$.  Therefore each tensor $\bmr\in\calr$ 
can be used as an input to any of the four encoders.  

The representatives in a tensor set $\calr$ can be replaced with integers.  
This gives a tensor set $\calu$.  There is a 1-1 correspondence $\bmu\lra\bmr$
between a tensor $\bmu\in\calu$ and a tensor $\bmr\in\calr$, and
between \svecs\ in $\bmu$ and \svecs\ in $\bmr$.
$\calu$ can also be realized as a trellis.
If the tensors in $\calr$ are used as inputs to each of the four encoders,
the outputs form $C$.  The outputs of one encoder are related to the outputs 
of another encoder by a graph automorphism of the trellis of $\calu$.

A selection of a set of generator vectors at each time $t$ 
that is necessary and sufficient to generate $C$ forms a basis $\bmcpb$.
Each basis $\bmcpb$ gives a tensor set $\calr$.  Two different bases give two
different tensor sets; this is called a change of basis.  The two different
tensor sets can be used as inputs to the same encoder.    
The tensor set $\calu$ is independent of basis, and when there is a change
of basis, the outputs of the same encoder are related by a graph automorphism
of the trellis of $\calu$.

The set of all graph automorphisms of $\calu$ forms a permutation group under 
composition.  This is termed the {\it full symmetry system} in \cite{SMT}.  
We calculate the full symmetry system of $\calu$.  
Any symmetry is specified by a finite set of separating permutations at
each time $t$.  Using the separating permutations, we give an algorithm
to construct any symmetry.

We show that any symmetry in the \fss\ takes each tensor $\bmu\in\calu$ to another
tensor $\bmuht\in\calu$, and takes each \svec\ in $\bmu$ to another 
\svec\ in $\bmuht$ of the same length $k$, for the same time $t$.
This induces a permutation of $\calr$ which 
takes each tensor $\bmr\in\calr$ to another
tensor $\bmrht\in\calr$, and takes each \gvec\ in $\bmr$ to another 
\gvec\ in $\bmrht$ of the same length $k$, for the same time $t$.
The permutation of a \gvec\ of length $k$ at time $t$ in $\bmr$
is only affected by \gvecs\ of length at least $k$ at time $t$ in $\bmr$.
The permutation of all tensors in tensor set $\calu$ or $\calr$ 
can be performed iteratively, starting with a permutation of the 
sequence of longest \gvecs, and working down.

The product $\bmc C$, where $\bmc$ is a path in $C$, permutes the paths 
of group trellis $C$, and therefore induces a symmetry of $\calu$.  The
set of symmetries induced by $\{\bmc C:  \bmc\in C\}$ forms a group which
we call the \nss\ $\caln$.  $\caln$ is a subgroup of the \fss, and $\caln$
is isomorphic to $C$.

Since the product $\bmc C$ induces a symmetry, we can study multiplication
in $C$ using the \nss\ $\caln$.  We show how two paths $\bmc_1$ and $\bmc_2$ 
multiply in terms of the two tensors $\bmr_1$ and $\bmr_2$ that encode
to $\bmc_1$ and $\bmc_2$, respectively.  We show that multiplication in $C$
implies that any group system has an underlying commutative property.

Since $C$ is time invariant, the \nss\ of $C$ is time invariant.
Therefore the \nss\ $\caln$ of $C$ can be specified by a finite
set of separating permutations which is constant for all time $t$.
This approach can be used to construct $C$.

This paper is organized as follows.  We start with a group system
$\msfc$, as in \cite{FT}.  Any group system $\msfc$ can be reduced 
to a group trellis $C$ with a group trellis section, or branch group
$B^t$ \cite{FT}; this is reviewed in Section 2.
We study an \ellctl\ group system and group trellis, in which each state 
can be reached from any other state in $\ell$ branches \cite{FT}.

In group trellis $C$, the sequence of branches that split from the 
identity path and merge to the identity path form two normal chains \cite{LM}.
The Schreier refinement theorem can be applied to these two normal
chains to obtain another normal chain, a refinement of the two
chains that we call a \ssr.  The \ssr\ is a normal chain of the
branch group at time $t$, $B^t$, of the group trellis.
The \ssr\ can be written in the form of a matrix, with rows and columns 
determined by branches of the splitting and merging trellis paths.
When the group trellis is strongly controllable, the matrix reduces to a 
triangular form, called the \stm.  The \stm\ is an echo of
matrix ideas used in classical linear systems analysis.

The \stm\ is defined over time interval $[t,t]$.
Since the group system is assumed to be time invariant,
we can replace the branches in column $j$ of the \stm\
with the same branches at time $t+j$.  The resulting matrix
is defined over the time interval $[t,t+\ell]$, and 
is called the \sm; it is also a triangular form.
We show the \sm\ has a natural shift property, and in
fact the \sm\ forms a part of the group trellis,
the truncation of the ray of paths splitting from the
identity path at time $t$.  This is discussed in Section 3.

We show that the rows of the \sm\ can be used to form quotient groups, and
the generator sequences of Forney and Trott are a transversal
of the quotient groups.  
The \creps\ of the generators in the transversal are also
a triangular form, a \sm\ which we call a \gm.  
The rows of the \gm\ are the nontrivial portion of a generator
sequence, called a \gvec.  At time $t$, the components of the generators
form a \compset\ for the \ssr\ decomposition of branch group $B^t$.
The same set of coset representatives
can be used for the \ssr\ decomposition of the branch group 
of the time reversed group trellis.  This is discussed in Section 4.

In Section 5, based on the \gm, we give a causal minimal encoder structure 
for a group trellis and group system.  We can think of the encoder as an estimator.
As in \cite{FT,LM}, the encoder uses shortest length generator sequences,
but here the components of the generator sequences give a time domain
convolution.  Therefore this appears to be a natural time domain encoder
for a group system, whereas the encoders in \cite{FT,LM} can be viewed
as spectral domain encoders.

In Section 6, we show the first canonic form, group trellis $C$,
can be reduced to the second canonic form $\calr$.  The tensor set $\calr$ 
depends on basis $\bmcpb$.  We find a tensor set and trellis $\calu$ 
which corresponds to $\calr$ but is independent of basis.  We show that
the four encoders are related by graph automorphisms of $\calu$; the
same holds for a change of basis.
In Section 7, we find the structure of graphs automorphisms of $\calu$,
a permutation group called the \fss.  In Section 8, we study the \nss\ of $C$
and multiplication in $C$ and $\calr$.

\newpage
\vspace{3mm}
{\bf 2.  GROUP SYSTEMS}
\vspace{3mm}

This section gives a very brief review of some fundamental concepts in 
\cite{FT}, and introduces some definitions used here.
We follow the notation of Forney and Trott as closely as possible.
One significant difference is that subscript $k$ in \cite{FT} denotes
time; we use $t$ (an integer) in place of $k$.  In any notation, a
superscript is used exclusively to indicate time; thus $t$ always 
appears as a superscript in any notation.

Forney and Trott study a collection of sequences with time axis defined on the
set of integers $\bmcpz$, whose components $a^t$ are taken from 
an {\it alphabet group} or {\it alphabet} $A^t$ at each time $t$, $t\in\bmcpz$.  
The set of sequences 
is a group under componentwise addition in $A^t$.  We call this a {\it group
system} or {\it group code} $\msfc$ \cite{FT}.  In this paper, we assume 
the group system is time invariant, so for each $t$, $A^t$ is
the same as a fixed common group $A$.
A sequence $\bma$ in $\msfc$ is given by
\be
\label{c9}
\bma=\ldots,a^{t-1},a^t,a^{t+1},\ldots,
\ee
where $a^t\in A^t$ is the component at time $t$.   

The group system $\msfc$ is assumed to be {\it complete} \cite{JW,FT}; 
an important consequence is that
local behavior is sufficient to describe global behavior.  Completeness
is the same as closure in symbolic dynamics \cite{LMr}.  Therefore
a time invariant complete group system $\msfc$ is the same thing as a
{\it group shift} in symbolic dynamics.  In this paper, we use the
language associated with group systems \cite{FT} rather than group shifts
\cite{LMr}.

Define $\msfc^{t^+}$ to be the set of all codewords in $\msfc$
for which $b^n=\bone^n$ for $n<t$, where $\bone^n$ is the identity
component at time $n$.    
Define $\msfc^{t^-}$ to be the set of all codewords in $\msfc$
for which $b^n=\bone^n$ for $n\ge t$.
The group system satisfies the {\it axiom of state}:  whenever
two sequences pass through the same state at a given time,
the concatenation of the past of either with the future of the other
is a valid sequence \cite{FT}.
The {\it canonic state space} $\Sigma^t$ at time $t$ is defined to be 
$$
\Sigma^t\rmdef\frac{\msfc}{\msfc^{t^-}\msfc^{t^+}}.
$$
The canonic state space is unique.  For a time invariant group system, 
for each time $t$, the state space $\Sigma^t$ is the same as a common
fixed group $\Sigma$.

\begin{figure}[h]
\centering

\begin{picture}(100,80)(0,-10)

\put(0,0){\line(1,0){100}}
\put(0,2){\line(1,0){50}}
\put(50,4){\line(1,0){50}}
\put(50,2){\line(1,1){50}}
\put(50,4){\line(-1,1){50}}

\put(30,-9){\makebox(0,0)[t]{$t-1$}}
\put(50,-9){\makebox(0,0)[t]{$t$}}
\put(70,-9){\makebox(0,0)[t]{$t+1$}}
\put(30,-5){\line(0,1){5}}
\put(50,-5){\line(0,1){5}}
\put(70,-5){\line(0,1){5}}
\put(10,50){\makebox(0,0)[l]{$\msfc^{t^-}$}}
\put(90,50){\makebox(0,0)[r]{$\msfc^{t^+}$}}

\end{picture}

\caption{Definition of $\msfc^{t^+}$ and $\msfc^{t^-}$.}
\label{statedefn}

\end{figure}

The state $\sigma^t(\bma)$ of a system sequence $\bma$ at time $t$
is determined by the natural map
$$
\sigma^t:  \msfc\ra \msfc/(\msfc^{t^-}\msfc^{t^+})=\Sigma^t,
$$
a homomorphism.  There is therefore a well defined state 
sequence $\bsigma(\bma)=\{\sigma^t(\bma): t\in\bmcpz\}$
associated with each $\bma\in \msfc$, and a well defined state code
$\bsigma(\msfc)=\{\bsigma(\bma): \bma\in \msfc\}$ associated with $\msfc$.
The {\it canonic realization} $C$ of a group system $\msfc$ is the 
set of all pairs of sequences $(\bma,\bsigma(\bma))$:
\be
\label{s90}
\{(\bma,\bsigma(\bma)):  \bma\in\msfc\},
\ee
where $\bsigma(\bma)$ is the state sequence of $\msfc$.
The state spaces of the canonic realization are $\Sigma^t$.
The canonic realization is a {\it minimal} realization of a group system.

An element of the canonic realization $C$ is denoted
\be
\label{eqn0}
\bmb=\ldots,b^{t-1},b^t,b^{t+1},\ldots,
\ee
where component $b^t$ is given by $b^t=(s^t,a^t,s^{t+1})$, where 
$s^t\in\Sigma^t$ is the canonic state at time $t$, and 
$s^{t+1}\in\Sigma^{t+1}$ is the canonic state at time $t+1$; 
we think of component $b^t$ stretching over the time interval $[t,t+1]$.
We say $s^t$ is the {\it left state} of $b^t$, and use notation
$(b^t)^-=s^t$.  In addition, we say $s^{t+1}$ is the {\it right state} 
of $b^t$, and use notation $(b^t)^+=s^{t+1}$.  For any path $\bmb$,
as given in (\ref{eqn0}), it is clear that for 
$b^t=(s^t,a^t,s^{t+1})$ and $b^{t+1}=({\hat{s}}^{t+1},a^{t+1},s^{t+2})$,
we must have $s^{t+1}={\hat{s}}^{t+1}$ or equivalently
$(b^t)^+=(b^{t+1})^-$.

Let $\bsigma(C)$ be the state code of $C$, the sequences of states
$\ldots,s^{t-1},s^t,s^{t+1},\ldots$ in each $\bmb\in C$.

\begin{thm}
There is a group isomorphism from $\msfc$ to $C$ given by the 1-1 
correspondence $\bma\lra\bmb$, where $\bma\in\msfc$ and $\bmb\in C$.
If 
$$
\bma=\ldots,a^{t-1},a^t,a^{t+1},\ldots,
$$
and 
$$
\bmb=\ldots,b^{t-1},b^t,b^{t+1},\ldots,
$$
then for each time $t$, $a^t\mapsto b^t=(s^t,a^t,s^{t+1})$
is the assignment of the group isomorphism.
\end{thm}

\begin{prf}
There is a well defined state sequence $\bsigma(\bma)$
associated with each $\bma\in \msfc$.  This means each $\bma\in \msfc$
is assigned to a well defined $\bmb\in C$ by the assignment
$a^t\mapsto b^t=(s^t,a^t,s^{t+1})$ for each time $t$.  This map
is a bijection since if $\bma\in \msfc$ and $\hat{\bma}\in \msfc$
are both assigned to the same $\bmb\in C$, then we must have
$a^t={\hat{a}}^t$ for each time $t$, so $\bma$ and $\hat{\bma}$
are the same.
\end{prf}
We will be interested in canonic realization $C$ rather than group system 
$\msfc$ in the remainder of the paper.  There is no loss in generality
in considering $C$ rather than $\msfc$ because of the above 1-1
correspondence and isomorphism.

The canonic realization can be described with a graph \cite{FT}.
Any other minimal realization is graph isomorphic to the canonic
realization \cite{FT}.
We think of component $b^t$ as a {\it branch} in a 
{\it trellis section} $T^t$ or an element in {\it branch group} $B^t$.
Trellis section $T^t$ is a bipartite graph
where the left vertices are states in $\Sigma^t$, the right vertices
are states in $\Sigma^{t+1}$, and the {\it label} of a 
branch $(s^t,a^t,s^{t+1})$
between state $s^t$ and state $s^{t+1}$ is $a^t\in A^t$.  $B^t$
is the group of branches $b^t$, which is a subdirect product,
a subgroup of the direct product group
$\Sigma^t\times A^t\times\Sigma^{t+1}$.  
Clearly there is a branch $(s^t,a^t,s^{t+1})$ in $T^t$, 
with label $a^t$ between two vertices $s^t$ and $s^{t+1}$,
\ifof\ $(s^t,a^t,s^{t+1})\in B^t$.  Then $C$ can be described by a
{\it group trellis}, a connected sequence of trellis sections, 
where $T^t$ and $T^{t+1}$ are joined
together using the common states in $\Sigma^{t+1}$ \cite{FT}.
We refer to this as {\it group trellis} $C$.  We regard
group trellis $C$ as the {\it first canonic form} of group system $\msfc$.

The states of $B^t$ are $\Sigma^t$ and $\Sigma^{t+1}$.
We now describe state groups of $B^t$ isomorphic to $\Sigma^t$ and $\Sigma^{t+1}$.
Consider the projection map $\pi_L:  B^t\ra\Sigma^t$
onto the left states of $B^t$,
given by the assignment $(s^t,a^t,s^{t+1})\mapsto s^t$.  
This is a homomorphism with 
kernel $X_0^t$, where $X_0^t$ is the subgroup of all elements of $B^t$ 
of the form $(\bone^t,a^t,s^{t+1})$, where $\bone^t$ is the identity 
of $\Sigma^t$.  Then by the first homomophism theorem 
$B^t/X_0^t\simeq\Sigma^t$.  Also consider the projection map 
$\pi_R:  B^t\ra\Sigma^{t+1}$ onto the right states of $B^t$,
given by the assignment
$(s^t,a^t,s^{t+1})\mapsto s^{t+1}$.  This is a homomorphism with 
kernel $Y_0^t$, where $Y_0^t$ is the subgroup of all elements of $B^t$ 
of the form $(s^t,a^t,\bone^{t+1})$, where $\bone^{t+1}$ is the identity 
of $\Sigma^{t+1}$.  Then by the first homomophism theorem 
$B^t/Y_0^t\simeq\Sigma^{t+1}$.  Thus any branch $b^t\in B^t$
is of the form $b^t=(s^t,a^t,s^{t+1})$ where 
$s^t\in\Sigma^t\simeq B^t/X_0^t$ and 
$s^{t+1}\in\Sigma^{t+1}\simeq B^t/Y_0^t$.
These results show there is a {\it state group isomorphism}
$B^t/Y_0^t\simeq\Sigma^{t+1}\simeq B^{t+1}/X_0^{t+1}$ at each time $t+1$.

Since $C$ is time invariant, we can regard $C$ as the sofic shift
\cite{LMr} of a graph $T$ which is graph isomorphic to $T^t$ for all $t$.
The branches of $T$ form a branch group $B$ which is isomorphic to $B^t$,
and the states of $T$ form a state group $\Sigma$ which is isomorphic to $\Sigma^t$,
for all $t$.  We can regard $C$ as the edge shift of $T$ and 
$\bsigma(C)$ as the vertex shift of $T$ \cite{SMT}.

Let $C$ be a group trellis, and let $\bmb$ be a trellis path in $C$.
Using (\ref{eqn0}), define the projection map at time $t$, 
$\chi^t:  C\ra B^t$, by the assignment $\bmb\mapsto b^t$.  
Define the projection map 
$\chi^{[t_1,t_2]}:  C\ra B^{t_1}\times\cdots\times B^{t_2}$ 
by the assignment $\bmb\mapsto (b^{t_1},\ldots,b^{t_2})$.
We say that $(b^{t_1},\ldots,b^{t_2})$ is a {\it trellis path segment}
of {\it length} $t_2-t_1+1$.
We say that codeword $\bmb$ has {\it span} $t_2-t_1+1$ if 
$b^{t_1}\ne\bone$, $b^{t_2}\ne\bone$, and $b^n=\bone$ for $n<t_1$
and $n>t_2$.

For any integer $l>0$, we say a group trellis $C$ is {\it $l$-controllable} 
if for any time epoch $t$, and any pair of states $s$ and $s'$, 
where $s\in \Sigma^t$ and $s'\in \Sigma^{t+l}$, there is a trellis 
path segment of length $l$ connecting the two states.  A group trellis
$C$ is {\it strongly controllable} if it is $l$-controllable for some integer $l$.
The least integer $l$ for which a group trellis is strongly controllable
is denoted as $\ell$.  In this paper, we only study the case $l=\ell$.

\newcommand{\calh}{{\mathcal{H}}}
\newcommand{\autd}{{\rm Aut}(\cald)}

\newcommand{\cvd}{\mathcal{V}(\cald)}
\newcommand{\ced}{\mathcal{E}(\cald)}

\newcommand{\argu}{{\,\cdot\,}}
\newcommand{\tldcb}{{\tilde{B}}}
\newcommand{\lamttpk}{{\Lambda^{[t,t+k]}}}

\newpage

\vspace{3mm}
{\bf 3.  THE STATIC MATRIX AND SHIFT MATRIX}
\vspace{3mm}

In this section, we write the \cdc\ of a group as a matrix, and call
this a {\it matrix chain}.  We study a \mchn\ called a \stm.
There is another matrix of group elements called a \sm.  We use both
matrices to construct a tensor.

In like manner to $\msfc$, define $C^{t^+}$ to be the set of all 
codewords in $C$ for which $b^n=\bone^n$ for $n<t$, where $\bone^n$ 
is the identity component of $B^n$ at time $n$.    
Define $C^{t^-}$ to be the set of all codewords in $C$
for which $b^n=\bone^n$ for $n\ge t$.  For all integers $j$, define
\be
\label{eqxt}
X_j^t\rmdef \{\chi^t(\bmb) : \bmb\in C^{(t-j)^+}\}.
\ee
Note that $X_0^t$ is consistent with the definition previously
given in Section 2.  We have $X_j^t=\bone^t$ for $j<0$.  
For all integers $i$, define
\be
\label{eqyt}
Y_i^t\rmdef \{\chi^t(\bmb) : \bmb\in C^{(t+i+1)^-}\}.
\ee
Note that $Y_0^t$ is consistent with the definition previously
given in Section 2.  We have $Y_i^t=\bone^t$ for $i>0$.  
It is clear that $X_j^t\lhd B^t$, and $Y_i^t\lhd B^t$ for any time $t$ 
and any integer $j$. 

The groups $X_j^t$ and $Y_i^t$ were first introduced in \cite{LM}.
The group intersections $X_j^t\cap Y_{\ell-j}^t$, for $0\le j\le\ell$,
are the groups used in \cite{LM} to give an abstract characterization of 
the branch group of an \ellctl\ group trellis.

For any set $H^t\in B^t$, define $(H^t)^+$ to be the set of
{\it right states} of $H^t$, or $\{s^{t+1}:  b^t=(s^t,a^t,s^{t+1})\in H^t\}$,
and define $(H^t)^-$ to be the set of {\it left states} of $H^t$, 
or $\{s^t:  b^t=(s^t,a^t,s^{t+1})\in H^t\}$.

For sets $H_1^t\subset B^t$ and $H_1^{t+1}\subset B^{t+1}$ such that 
$(H_1^t)^+=(H_2^t)^-$, 
define the {\it concatenation} of $H_1^t$ and $H_2^{t+1}$,
$H_1^t\wedge H_2^{t+1}$, to be all the (valid) trellis path segments 
of length two with first component in $H_1^t$ 
and second component in $H_2^{t+1}$.

Note that $(X_j^t)^+=(X_{j+1}^{t+1})^-$ and $(Y_i^t)^+=(Y_{i-1}^{t+1})^-$ 
for all integers $i,j$.  Then $X_j^t\wedge X_{j+1}^{t+1}$ and
$Y_i^t\wedge Y_{i-1}^{t+1}$ are sets of trellis path segments of length two.

The next result follows directly from Proposition 7.2 of \cite{LM},
using our notation.

\begin{prop}
\label{prop1}
The group trellis $C$ is \ellctl\ \ifof\ $X_\ell^t=B^t$, or equivalently, 
\ifof\ $Y_\ell^t=B^t$, for each time $t$.
\end{prop}

The group $B^t$ has two normal series (and chief series)
$$
\bone^t=X_{-1}^t\lhd X_0^t\lhd X_1^t\lhd\cdots\lhd X_\ell^t=B^t,
$$
and
$$
\bone^t=Y_{-1}^t\lhd Y_0^t\lhd Y_1^t\lhd\cdots\lhd Y_\ell^t=B^t.
$$
We denote these normal series by $\xjt$ and $\yit$.

The Schreier refinement theorem used to prove the Jordan-H\"{o}lder 
theorem \cite{ROT} shows
how to obtain a refinement of $\xjt$ by inserting $\yit$;
we call this the {\it forward Schreier series} of $\xjt$ and $\yit$.
Since $\xjt$ and $\yit$ are chief series, 
the \fssr\ of $\xjt$ and $\yit$ is a chief series.
In equation (\ref{sm}), we have written
the \fssr\ as a matrix of $\ell+1$ columns and $\ell+2$ rows.
Note that the terms in the bottom row form the
sequence $X_{-1}^t,X_0^t,X_1^t, \ldots X_{\ell-2}^t,X_{\ell-1}^t$,
and the terms in the top row form the sequence
$X_0^t,X_1^t,X_2^t, \ldots X_{\ell-1}^t,X_\ell^t$.
Thus (\ref{sm}) is indeed a refinement of the normal series $\xjt$.
We call (\ref{sm}) the \mchn\ of the \fssr\ of $\xjt$ and $\yit$.

\begin{prop}
\label{prop1x}
If the group trellis $C$ is \ellctl, then 
\be
\label{eq112c}
X_{j-1}^t(X_j^t\cap Y_{\ell-j}^t)=X_j^t,
\ee
for each $t$, for $j\ge 0$.
\end{prop}

\begin{prf}
If the group trellis $C$ is \ellctl, then from Proposition 7.2
of \cite{LM}, in our notation,
\be
\label{eq112b}
(X_0^t\cap Y_\ell^t)(X_1^t\cap Y_{\ell-1}^t)\cdots (X_j^t\cap Y_{\ell-j}^t)=X_j^t
\ee
for all $j\ge 0$.  This means we can rewrite (\ref{eq112b}) 
as (\ref{eq112c}).
\end{prf}

The {\it diagonal terms} of the \mchn\ (\ref{sm}) are 
$X_{j-1}^t(X_j^t\cap Y_{\ell-j}^t)$ for $j=0,\ldots\ell$.  Proposition 
\ref{prop1x} shows that the diagonal terms satisfy
$X_{j-1}^t(X_j^t\cap Y_{\ell-j}^t)=X_j^t$ for $j=0,\ldots\ell$,
if the group trellis is \ellctl.  For $j\in [1,\ell]$,
this means all column terms above the diagonal term are the same as
the diagonal term.  Then we can reduce the \mchn\ to a triangular form
as shown in (\ref{xttstm}).  A triangle can be formed in two ways, depending on
whether the columns in (\ref{sm}) are shifted up or not; we have shifted
the columns up since it is more useful here.  We call (\ref{xttstm})
the $X^{[t,t]}$ {\it static matrix}.  To make this notation 
clearer, the bracketed term $[t,t]$ only appears in the paper
as the superscript of a matrix defined over the time interval $[t,t]$
(except in this sentence).  A typical entry
in the matrix is $X_{j-1}^t(X_j^t\cap Y_{k-j}^t)$.

\begin{thm}
\label{thm6}
The $X^{[t,t]}$ \stm\ is a description (normal chain and chief series) 
of the branch group $B^t$ of an \ellctl\ group trellis.
\end{thm}

\begin{prf}
Both $\xjt$ and $\yit$ are normal chains of the branch group $B^t$.
Then by the Schreier refinement theorem, the \fssr\ is a
normal chain of $B^t$.
\end{prf}

For each $t$, we can replace $B^t$ in the group trellis $C$ with $X^{[t,t]}$.
We denote the resulting structure by $\bmx$; note that $\bmx$ is a tensor.
Since $X^{[t,t]}$ is a \cdc\ of $B^t$,
then $\bmx$ is a description of the coset structure of group trellis $C$.
Each path $\bmb\in C$ traverses some sequence of cosets in $\bmx$.

Note that the first column of (\ref{xttstm}) is a description of $X_0$,
which we can think of as an input.  The remaining columns are a
description of $B^t/X_0^t$, which is isomorphic to the state
$\Sigma^t$.  Thus columns of the \stm\ contain information about
the input and state.  Therefore an isomorphic copy of the state 
code $\bsigma(C)$ is embedded in $\bmx$.

\begin{figure}

\be
\label{sm}
\begin{array}{cccccc}
  \cup & \cup && \cup && \\

  X_{-1}^t(X_0^t\cap Y_\ell^t) & X_0^t(X_1^t\cap Y_\ell^t) & \cdots & X_{j-1}^t(X_j^t\cap Y_\ell^t) & \cdots & X_{\ell-1}^t(X_\ell^t\cap Y_\ell^t) \\

  \cup & \cup && \cup && \cup \\

  X_{-1}^t(X_0^t\cap Y_{\ell-1}^t) & X_0^t(X_1^t\cap Y_{\ell-1}^t) & \cdots & X_{j-1}^t(X_j^t\cap Y_{\ell-1}^t) & \cdots & X_{\ell-1}^t(X_\ell^t\cap Y_{\ell-1}^t) \\

  \cup & \cup && \cup && \cup \\

  X_{-1}^t(X_0^t\cap Y_{\ell-2}^t) & X_0^t(X_1^t\cap Y_{\ell-2}^t) & \cdots & X_{j-1}^t(X_j^t\cap Y_{\ell-2}^t) & \cdots & X_{\ell-1}^t(X_\ell^t\cap Y_{\ell-2}^t) \\

  \cup & \cup && \cup && \cup \\

  \cdots & \cdots & \cdots & \cdots & \cdots & \cdots \\

  \cup & \cup && \cup && \cup \\

  X_{-1}^t(X_0^t\cap Y_{\ell-j}^t) & X_0^t(X_1^t\cap Y_{\ell-j}^t) & \cdots & X_{j-1}^t(X_j^t\cap Y_{\ell-j}^t) & \cdots & X_{\ell-1}^t(X_\ell^t\cap Y_{\ell-j}^t) \\

  \cup & \cup && \cup && \cup \\

  X_{-1}^t(X_0^t\cap Y_{\ell-j-1}^t) & X_0^t(X_1^t\cap Y_{\ell-j-1}^t) & \cdots & X_{j-1}^t(X_j^t\cap Y_{\ell-j-1}^t) & \cdots & X_{\ell-1}^t(X_\ell^t\cap Y_{\ell-j-1}^t) \\

  \cup & \cup && \cup && \cup \\

  \cdots & \cdots & \cdots & \cdots & \cdots & \cdots \\

  \cup & \cup && \cup && \cup \\

  X_{-1}^t(X_0^t\cap Y_1^t) & X_0^t(X_1^t\cap Y_1^t) & \cdots & X_{j-1}^t(X_j^t\cap Y_1^t) & \cdots & X_{\ell-1}^t(X_\ell^t\cap Y_1^t) \\

  \cup & \cup && \cup && \cup \\

  X_{-1}^t(X_0^t\cap Y_0^t) & X_0^t(X_1^t\cap Y_0^t) & \cdots & X_{j-1}^t(X_j^t\cap Y_0^t) & \cdots & X_{\ell-1}^t(X_\ell^t\cap Y_0^t) \\

  \cup & \cup && \cup && \cup \\

  X_{-1}^t(X_0^t\cap Y_{-1}^t) & X_0^t(X_1^t\cap Y_{-1}^t) & \cdots & X_{j-1}^t(X_j^t\cap Y_{-1}^t) & \cdots & X_{\ell-1}^t(X_\ell^t\cap Y_{-1}^t)
\end{array}
\ee
\end{figure}

\begin{figure}

\be
\label{xttstm}
\begin{array}{ccccccc}
  \shortparallel & \shortparallel && \shortparallel && \shortparallel & \\

  X_{-1}^t(X_0^t\cap Y_\ell^t) & X_0^t(X_1^t\cap Y_{\ell-1}^t) & \cdots & X_{j-1}^t(X_j^t\cap Y_{\ell-j}^t) & \cdots & X_{\ell-1}^t(X_\ell^t\cap Y_0^t) & X_\ell^t(\bone^t) \\

  \cup & \cup && \cup && \cup & \\

  X_{-1}^t(X_0^t\cap Y_{\ell-1}^t) & X_0^t(X_1^t\cap Y_{\ell-2}^t) & \cdots & X_{j-1}^t(X_j^t\cap Y_{\ell-j-1}^t) & \cdots & X_{\ell-1}^t(\bone^t) & \\

  \cup & \cup && \cup &&& \\
  
  \cdots & \cdots & \cdots & \cdots & \cdots && \\

  \cup & \cup && \cup &&& \\

  X_{-1}^t(X_0^t\cap Y_k^t) & X_0^t(X_1^t\cap Y_{k-1}^t) & \cdots  & X_{j-1}^t(X_j^t\cap Y_{k-j}^t) & \cdots && \\
 
  \cup & \cup && \cup &&& \\

  \cdots & \cdots & \cdots & \cdots & \cdots && \\

  \cup & \cup && \cup &&& \\

  \cdots & \cdots & \cdots & X_{j-1}^t(X_j^t\cap Y_0^t) & X_j^t(\bone^t) && \\

  \cup & \cup && \cup &&& \\

  \cdots & \cdots & \cdots & X_{j-1}^t(\bone^t) &&& \\

  \cup & \cup &&&&& \\

  X_{-1}^t(X_0^t\cap Y_1^t) & X_0^t(X_1^t\cap Y_0^t) & X_1^t(\bone^t) &&&& \\

  \cup & \cup &&&&& \\
   
  X_{-1}^t(X_0^t\cap Y_0^t) & X_0^t(\bone^t) &&&&& \\

  \cup &&&&&& \\

  X_{-1}^t(\bone^t) &&&&&&
\end{array}
\ee

\end{figure}


\newpage

Since $C$ is time invariant, for any $t$, 
the elements in $X_{j-1}^t$ and $X_{j-1}^{t+j}$ are the same,
the elements in $X_j^t$ and $X_j^{t+j}$ are the same,
and the elements in $Y_{k-j}^t$ and $Y_{k-j}^{t+j}$ are the same.  Therefore
we replace the column containing $X_{j-1}^t(X_j^t\cap Y_{k-j}^t)$ in (\ref{xttstm})
with the column containing $X_{j-1}^{t+j}(X_j^{t+j}\cap Y_{k-j}^{t+j})$ 
in (\ref{xtshftm}).  Doing this for each column in (\ref{xttstm}) gives
the matrix shown in (\ref{xtshftm}).  Since the time index is changed
from one column to the next in (\ref{xtshftm}), we no longer have the
inclusion from one column to the next as in (\ref{xttstm}).  However the
coset decomposition within each column is preserved.  We call
(\ref{xtshftm}) the $X^{[t,t+\ell]}$ {\it shift matrix}.  
Notice the \sm\ extends over the time interval $[t,t+\ell]$.
A typical entry in the matrix is $X_{j-1}^{t+j}(X_j^{t+j}\cap Y_{k-j}^{t+j})$.

For $j=0,\ldots,\ell$, the $j$-th column of \stm\ $X^{[t,t]}$
is the $j$-th column of a \sm\ $X^{[t-j,t-j+\ell]}$ at time $t-j$.
Thus the \stm\ $X^{[t,t]}$ is a composite of columns of $\ell+1$ \sms.

\begin{figure}

\be
\label{xtshftm}
\begin{array}{ccccccc}
  &&&&&& \\

  X_{-1}^t(X_0^t\cap Y_\ell^t) & X_0^{t+1}(X_1^{t+1}\cap Y_{\ell-1}^{t+1}) & \cdots & X_{j-1}^{t+j}(X_j^{t+j}\cap Y_{\ell-j}^{t+j}) & \cdots & X_{\ell-1}^{t+\ell}(X_\ell^{t+\ell}\cap Y_0^{t+\ell}) & X_\ell^{t+\ell+1}(\bone^{t+\ell+1}) \\

  \cup & \cup && \cup && \cup & \\

  X_{-1}^t(X_0^t\cap Y_{\ell-1}^t) & X_0^{t+1}(X_1^{t+1}\cap Y_{\ell-2}^{t+1}) & \cdots & X_{j-1}^{t+j}(X_j^{t+j}\cap Y_{\ell-j-1}^{t+j}) & \cdots & X_{\ell-1}^{t+\ell}(\bone^{t+\ell}) & \\

  \cup & \cup && \cup &&& \\
  
  \cdots & \cdots & \cdots & \cdots & \cdots && \\

  \cup & \cup && \cup &&& \\

  X_{-1}^t(X_0^t\cap Y_k^t) & X_0^{t+1}(X_1^{t+1}\cap Y_{k-1}^{t+1}) & \cdots  & X_{j-1}^{t+j}(X_j^{t+j}\cap Y_{k-j}^{t+j}) & \cdots && \\
 
  \cup & \cup && \cup &&& \\

  \cdots & \cdots & \cdots & \cdots & \cdots && \\

  \cup & \cup && \cup &&& \\

  \cdots & \cdots & \cdots & X_{j-1}^{t+j}(X_j^{t+j}\cap Y_0^{t+j}) & \cdots && \\

  \cup & \cup && \cup &&& \\

  \cdots & \cdots & \cdots & X_{j-1}^{t+j}(\bone^{t+j}) &&& \\

  \cup & \cup &&&&& \\

  X_{-1}^t(X_0^t\cap Y_1^t) & X_0^{t+1}(X_1^{t+1}\cap Y_0^{t+1}) & \cdots &&&& \\

  \cup & \cup &&&&& \\
   
  X_{-1}^t(X_0^t\cap Y_0^t) & X_0^{t+1}(\bone^{t+1}) &&&&& \\

  \cup &&&&&& \\

  X_{-1}^t(\bone^t) &&&&&&
\end{array}
\ee

\end{figure}


The \fssr\ evolves forward in time.  There is a dual of the \fssr\ that
evolves backward in time.  The {\it backward Schreier series}
of $\xjt$ and $\yit$ is a refinement of $\yit$ obtained by inserting $\xjt$.
The \stm\ of the \bssr\ is $Y^{[t,t]}$, the dual of $X^{[t,t]}$,
and the \sm\ is $Y^{[t-\ell,t]}$, the dual of $X^{[t,t+\ell]}$.
As an example, the \stm\ $Y^{[t,t]}$ is shown in (\ref{yttstm}).  
$Y^{[t,t]}$ is a reflection of $X^{[t,t]}$ about the vertical axis.
In (\ref{xttstm}), index $j$ increases from left to right, while
in (\ref{yttstm}), index $i$ increases from right to left.
This reflects the symmetry in the definitions of $\xjt$ and $\yit$.

\begin{figure}

\be
\label{yttstm}
\begin{array}{ccccccc}
  & \shortparallel && \shortparallel && \shortparallel & \shortparallel \\

  Y_\ell^t(\bone^t) & Y_{\ell-1}^t(Y_\ell^t\cap X_0^t) & \cdots & Y_{i-1}^t(Y_i^t\cap X_{\ell-i}^t) & \cdots & Y_0^t(Y_1^t\cap X_{\ell-1}^t) & Y_{-1}^t(Y_0^t\cap X_\ell^t) \\

  & \cup && \cup && \cup & \cup \\

  & Y_{\ell-1}^t(\bone^t) & \cdots & Y_{i-1}^t(Y_i^t\cap X_{\ell-i-1}^t) & \cdots & Y_0^t(Y_1^t\cap X_{\ell-2}^t) & Y_{-1}^t(Y_0^t\cap X_{\ell-1}^t) \\

  &&& \cup && \cup & \cup \\
  
  && \cdots & \cdots & \cdots & \cdots & \cdots \\

  &&& \cup && \cup & \cup \\

  && \cdots  & Y_{i-1}^t(Y_i^t\cap X_{k-i}^t) & \cdots & Y_0^t(Y_1^t\cap X_{k-1}^t) & Y_{-1}^t(Y_0^t\cap X_k^t) \\
 
  &&& \cup && \cup & \cup \\

  && \cdots & \cdots & \cdots & \cdots & \cdots \\

  &&& \cup && \cup & \cup \\

  && Y_i^t(\bone^t) & Y_{i-1}^t(Y_i^t\cap X_0^t) & \cdots & \cdots & \cdots \\

  &&& \cup && \cup & \cup \\

  &&& Y_{i-1}^t(\bone^t) & \cdots & \cdots & \cdots \\

  &&&&& \cup & \cup \\

  &&&& Y_1^t(\bone^t) & Y_0^t(Y_1^t\cap X_0^t) & Y_{-1}^t(Y_0^t\cap X_1^t)\\

  &&&&& \cup & \cup \\
   
  &&&&& Y_0^t(\bone^t) & Y_{-1}^t(Y_0^t\cap X_0^t) \\

  &&&&&& \cup \\

  &&&&&& Y_{-1}^t(\bone^t) 
\end{array}
\ee

\end{figure}


We now show that the $X^{[t,t+\ell]}$ \sm\ (\ref{xtshftm}) has a kind of 
shift property, after some preliminary results.  The discussion will show 
that the \sm\ has a physical interpretation as the quotient group of 
certain paths that split from the identity path.

For any time $t$, for each branch $b^t\in B^t$, we define the 
{\it following branch set} $\calf(b^t)$ to be the set of branches that can 
follow $b^t$ at the next time epoch $t+1$ in valid trellis paths.  
In other words, branch $b^{t+1}\in\calf(b^t)$ \ifof\ $(b^t)^+=(b^{t+1})^-$.
Then the following branch set $\calf(b^t)$ represents the contraction, 
correspondence, and expansion given by
$$
b^t\mapsto b^tY_0^t\stackrel{\eta}{\llra}b^{t+1}X_0^{t+1},
$$
where $\eta$ is the 1-1 correspondence 
$B^t/Y_0^t\stackrel{\eta}{\llra} B^{t+1}/X_0^{t+1}$
given by the state group isomorphism $B^t/Y_0^t\simeq B^{t+1}/X_0^{t+1}$.

It is clear that $b^t\in B^t$ and $\calf(b^t)\subset B^{t+1}$.
However note that $\calf$ is {\bf not} a function with domain
$B^t$ and range $B^{t+1}$.  But we can think of $\calf$ as a relation
on $B^t\times B^{t+1}$.  In this relation, we can think of $\calf$
as an assignment of set $\calf(b^t)$ to branch $b^t$, or 
$\calf:  b^t\mapsto\calf(b^t)$.

\begin{prop}
If $(b^t)^+=(b^{t+1})^-$, the following branch set $\calf(b^t)$ of 
a branch $b^t$ in $B^t$ is the coset $b^{t+1}X_0^{t+1}$ in $B^{t+1}$,
or the assignment $\calf:  b^t\mapsto b^{t+1}X_0^{t+1}$.
\end{prop}

Define the following branch set $\calf:  B^t\ra B^{t+1}$ such that for any set
$H^t\subset B^t$, the set $\calf(H^t)$ is
the union $\cup_{b^t\in H^t} \calf(b^t)$.  The set $\calf(H^t)$ always
consists of cosets of $X_0^{t+1}$.  In particular, 
$\calf(X_j^t)=X_{j+1}^{t+1}$ for all integers $j\ge -1$.

For a set $H^t\subset B^t$ and integer $j>0$, define $\calf^j(H^t)$
to be the $j$-fold composition 
$\calf^j(H^t)=\calf\circ\calf\circ\cdots\circ\calf(H^t)$.
For $j=0$, define $\calf^j(H^t)=\calf^0(H^t)$ to be just $H^t$.
If $H^t$ is a set of trellis branches at time epoch $t$,
then $\calf^j(H^t)$ is the set of trellis branches at time
epoch $t+j$, such that for each $b^{t+j}\in\calf^j(H^t)$ there is
a $b^t\in H^t$ and a path in the trellis from $b^t$ to $b^{t+j}$.
Note that $X_j^{t+j}=\calf^j(X_0^t)$.

For a set $H^t\subset B^t$ and integer $k\ge 0$, define
$\calf^{[0,k]}(H^t)$ to be the set of all trellis path segments
$(b^t,\ldots,b^{t+k})$ on time interval $[t,t+k]$ that start with a
branch $b^t\in H^t$.

\begin{prop} 
\label{prop5plus}
For any subsets $G^t,H^t$ of $B^t$, we have $(G^tH^t)^+=(G^t)^+(H^t)^+$, 
$(G^tH^t)^-=(G^t)^-(H^t)^-$,  and $\calf(G^tH^t)=\calf(G^t)\calf(H^t)$.
\end{prop}

\begin{prf}
It is clear that $(G^tH^t)^+=(G^t)^+(H^t)^+$.  Then it follows that
$\calf(G^tH^t)=\calf(G^t)\calf(H^t)$.
\end{prf}

\begin{prop}
\label{prop5}
For any subsets $G^t,H^t$ of $B^t$, we have $(G^t\cap H^t)^+=(G^t)^+\cap (H^t)^+$, 
$(G^t\cap H^t)^-=(G^t)^-\cap (H^t)^-$, and $\calf(G^t\cap H^t)=\calf(G^t)\cap\calf(H^t)$.
\end{prop}

We index the rows and columns of (\ref{xtshftm}), and denote terms, in a definite
way.  We index the columns with $j$, for $0\le j\le\ell$, and rows with
$k$, for $0\le k\le\ell$, starting with $(j,k)=(0,0)$ in the bottom left corner.
In general, we indicate a term in the \sm\ by
$X_{j-1}^{t+j}(X_j^{t+j}\cap Y_{k-j}^{t+j})$, 
where the subscripts mean definite things.  
The subscript $\alpha$ of $X$ in the
factor term $(X_\alpha\cap Y_\beta)$ always indicates the column, 
and the sum of the subscripts $\alpha+\beta$ of $X$ and $Y$ in the
factor term always indicates the row.  So the term 
$X_{j-1}^{t+j}(X_j^{t+j}\cap Y_{k-j}^{t+j})$ is
in column $j$ and row $k$.  We do not include terms of the 
form $X_{j-1}^{t+j}(\bone^{t+j})$.
For example, $X_{-1}^t(X_0^t\cap Y_0^t)$ is 
the bottom left corner term, in column $j=0$
and row $k=0$.  As other examples, the factor term
$(X_j^{t+j}\cap Y_{k-j-1}^{t+j})$ is in column $j$ and row $k-1$,   
and the factor term $(X_{j-1}^{t+j-1}\cap Y_{k-j}^{t+j-1})$ is in
column $j-1$ and row $k-1$.  Note that row $k$ of the \sm\ has (length) $k+1$
terms, ignoring the last term $X_k^{t+k+1}(\bone^{t+k+1})$.

We now show the $X^{[t,t+\ell]}$ \sm\ preserves shifts, that is, it has a
{\it shift property}.

\begin{prop}
\label{pro10}
Fix $k$, $0\le k\le\ell$, and fix $j$, $0\le j\le k$.
The \sm\ has a shift property:  the term 
$X_j^{t+j+1}(X_{j+1}^{t+j+1}\cap Y_{k-j-1}^{t+j+1})$
in column $j+1$ and row $k$ is a shift of the term 
$X_{j-1}^{t+j}(X_j^{t+j}\cap Y_{k-j}^{t+j})$
in column $j$ and row $k$, that is
\be
\label{pro10b}
\calf(X_{j-1}^{t+j}(X_j^{t+j}\cap Y_{k-j}^{t+j}))
=X_j^{t+j+1}(X_{j+1}^{t+j+1}\cap Y_{k-j-1}^{t+j+1}).
\ee
\end{prop}

\begin{prf}
Fix $k$, $0\le k\le\ell$, and fix $j$, $0\le j\le k$.  We have
\begin{align*}
\calf(X_{j-1}^{t+j}(X_j^{t+j}\cap Y_{k-j}^{t+j}))
&=\calf(X_{j-1}^{t+j})\calf(X_j^{t+j}\cap Y_{k-j}^{t+j}) \\
&=X_j^{t+j+1}(\calf(X_j^{t+j})\cap\calf(Y_{k-j}^{t+j})) \\
&=X_j^{t+j+1}(X_{j+1}^{t+j+1}\cap X_0^{t+j+1}Y_{k-j-1}^{t+j+1}) \\
&=X_j^{t+j+1}X_0^{t+j+1}(X_{j+1}^{t+j+1}\cap Y_{k-j-1}^{t+j+1}) \\
&=X_j^{t+j+1}(X_{j+1}^{t+j+1}\cap Y_{k-j-1}^{t+j+1}),
\end{align*}
where the fourth equality follows from the Dedekind Law 
(if $H$, $K$, and $L$ are subgroups of group $G$ with
$H\subset L$, then $HK\cap L=H(K\cap L)$).
\end{prf}

The first column of the \sm\ (\ref{xtshftm}) will be important to us
so we define $\Delta_k^t=X_0^t\cap Y_k^t$, for $-1\le k\le\ell$.  
We now show that row $k$ of the \sm\ is just $\calf^{[0,k]}(\Delta_k^t)$,
or just the trellis path segments in $C$ on time interval $[t,t+k]$
that start with a branch $b^t\in\Delta_k^t$.

\begin{thm}
\label{pro11}
Fix $k$, $0\le k\le\ell$.  We have
\be
\label{pro11a}
\calf^j(\Delta_k^t)=X_{j-1}^{t+j}(X_j^{t+j}\cap Y_{k-j}^{t+j}),
\ee
for $0\le j\le k$.  And $\calf^j(\Delta_k^t)=X_{j-1}^{t+j}$ for $k<j\le\ell$,
$\calf^j(\Delta_k^t)=X_\ell^{t+j}$ for $j>\ell$.
The $k$-th row of the \sm\ is just
the terms in $\calf^{[0,k]}(\Delta_k^t)$.
\end{thm}

\begin{prf}
We prove (\ref{pro11a}) by induction.  Assume it is true for $j=n$.
Then use (\ref{pro10b}) to show it is true for $j=n+1$.  Then (\ref{pro11a})
shows the $k$-th row of the \sm\ is just
the terms in $\calf^{[0,k]}(\Delta_k^t)$.
\end{prf}

Note that $\chi^{[t,t+\ell]}(C^{t^+})$ are the trellis path segments in a 
truncated {\it ray}, paths in the trellis which split from the identity state at
time epoch $t$.  Further we have $\chi^{[t,t+\ell]}(C^{t^+})=\calf^{[0,\ell]}(X_0^t)$.

\begin{thm}
The $X^{[t,t+\ell]}$ \sm\ describes the coset structure of the truncated ray
$\chi^{[t,t+\ell]}(C^{t^+})=\calf^{[0,\ell]}(X_0^t)$ of an \ellctl\ group trellis.
\end{thm}

Using Theorem \ref{pro11}, we can represent a quotient group
of adjacent terms in the same column of \sm\ (\ref{xtshftm})
in two equivalent ways:
\be
\label{eq888}
\frac{\calf^j(\Delta_k^t)}{\calf^j(\Delta_{k-1}^t)}=
\frac{X_{j-1}^{t+j}(X_j^{t+j}\cap Y_{k-j}^{t+j})}
{X_{j-1}^{t+j}(X_j^{t+j}\cap Y_{k-j-1}^{t+j})},
\ee
for $0\le j\le k$.

\begin{prop}
$\calf^{[0,k]}(\Delta_{k-1}^t)$ and $\calf^{[0,k]}(\Delta_k^t)$ are groups.
\end{prop}

\begin{prf}
$\Delta_k^t=X_0^t\cap Y_k^t$ is a group so the trellis path
segments in $\calf^{[0,k]}(\Delta_k^t)$ are a group.
\end{prf}

\begin{prop}
$\calf^{[0,k]}(\Delta_{k-1}^t)\lhd\calf^{[0,k]}(\Delta_k^t)$ \ifof\
$\Delta_{k-1}^t\lhd\Delta_k^t$.  Then as a result
$\calf^{[0,k]}(\Delta_{k-1}^t)\lhd\calf^{[0,k]}(\Delta_k^t)$. 
\end{prop}

\begin{thm}
\label{thm17}
We have
$$  
\frac{\calf^{[0,k]}(\Delta_k^t)}{\calf^{[0,k]}(\Delta_{k-1}^t)}\simeq\frac{\Delta_k^t}{\Delta_{k-1}^t}.
$$
\end{thm}

\begin{prf}
The projection $\chi^t:  \calf^{[0,k]}(\Delta_k^t)\ra\Delta_k^t$
is onto.  It is a homomorphism with kernel $\calf^{[0,k]}(\bone^t)$.
The projection $\chi^t:  \calf^{[0,k]}(\Delta_{k-1}^t)\ra\Delta_{k-1}^t$
is onto.  It is a homomorphism with kernel $\calf^{[0,k]}(\bone^t)$.
Therefore, by the first homomorphism theorem,
\begin{align*}
\frac{\calf^{[0,k]}(\Delta_k^t)}{\calf^{[0,k]}(\bone^t)} &\simeq\Delta_k^t, \\
\frac{\calf^{[0,k]}(\Delta_{k-1}^t)}{\calf^{[0,k]}(\bone^t)} &\simeq\Delta_{k-1}^t.
\end{align*}
Now use the correspondence theorem and third 
isomorphism theorem to complete the proof.
\end{prf}

\begin{prop}
$\calf^j(\Delta_{k-1}^t)$ and $\calf^j(\Delta_k^t)$ are groups.
\end{prop}

\begin{prf}
See (\ref{pro11a}) or note that $\calf^j(\Delta_k^t)$ is the projection
of group $\calf^{[0,k]}(\Delta_k^t)$ on the time interval $[t+j,t+j]$.
\end{prf}

\begin{prop}
$\calf^j(\Delta_{k-1}^t)\lhd\calf^j(\Delta_k^t)$.
\end{prop}

\begin{prf}
See (\ref{eq888}).
\end{prf}

\begin{thm}
\label{thm40}
For $0\le j\le k$, we have
$$  
\frac{\calf^{[0,k]}(\Delta_k^t)}{\calf^{[0,k]}(\Delta_{k-1}^t)}\simeq
\frac{\calf^j(\Delta_k^t)}{\calf^j(\Delta_{k-1}^t)}.
$$
\end{thm}

\begin{prf}
The projection $\chi^{t+j}:  \calf^{[0,k]}(\Delta_k^t)\ra\calf^j(\Delta_k^t)$
is onto.  It is a homomorphism with kernel $K_k$, the path segments
in $\calf^{[0,k]}(\Delta_k^t)$ that are the identity at time $t+j$.
The projection $\chi^{t+j}:  \calf^{[0,k]}(\Delta_{k-1}^t)\ra\calf^j(\Delta_{k-1}^t)$
is onto.  It is a homomorphism with kernel $K_{k-1}$, the path segments
in $\calf^{[0,k]}(\Delta_{k-1}^t)$ that are the identity at time $t+j$.  
Therefore, by the first homomorphism theorem,
\begin{align*}
\frac{\calf^{[0,k]}(\Delta_k^t)}{K_k} &\simeq\calf^j(\Delta_k^t), \\
\frac{\calf^{[0,k]}(\Delta_{k-1}^t)}{K_{k-1}} &\simeq\calf^j(\Delta_{k-1}^t).
\end{align*}

We now show $K_k=K_{k-1}$; we first show $K_k\subset K_{k-1}$.
Let $(b^t,\ldots,b^{t+j},\ldots,b^{t+k})$ be a path segment in
$\calf^{[0,k]}(\Delta_k^t)$ that is the identity at time $t+j$,
$0\le j\le k$.  But then $b^t$ must be in $(X_0^t\cap Y_{j-1}^t)=\Delta_{j-1}^t$.
Since $j\le k$, then $\Delta_{j-1}^t\subset\Delta_{k-1}^t$ and
$b^t\in\Delta_{k-1}^t$.  Then
$(b^t,\ldots,b^{t+j},\ldots,b^{t+k})\in\calf^{[0,k]}(\Delta_{k-1}^t)$
and $(b^t,\ldots,b^{t+j},\ldots,b^{t+k})\in K_{k-1}$.
Therefore $K_k\subset K_{k-1}$.

We now show $K_{k-1}\subset K_k$.
Let $(b^t,\ldots,b^{t+j},\ldots,b^{t+k})$ be a path segment in
$\calf^{[0,k]}(\Delta_{k-1}^t)$ that is the identity at time $t+j$,
$0\le j\le k$.  But then $b^t$ must be in $(X_0^t\cap Y_{j-1}^t)=\Delta_{j-1}^t$.
Since $j\le k$, then $\Delta_{j-1}^t\subset\Delta_k^t$ and
$b^t\in\Delta_k^t$.  Then
$(b^t,\ldots,b^{t+j},\ldots,b^{t+k})\in\calf^{[0,k]}(\Delta_k^t)$
and $(b^t,\ldots,b^{t+j},\ldots,b^{t+k})\in K_k$.
Therefore $K_{k-1}\subset K_k$.

We have just shown $K_k=K_{k-1}$.  Now use the correspondence theorem 
and third isomorphism theorem to complete the proof.
\end{prf}

Note that the proof breaks down if we try to go further.  In other words,
we cannot show that for $0\le j\le k$, we have
$$  
\frac{\calf^{[0,k]}(\Delta_k^t)}{\calf^{[0,k]}(\Delta_{k-2}^t)}\simeq
\frac{\calf^j(\Delta_k^t)}{\calf^j(\Delta_{k-2}^t)}.
$$

Define
$$  
\lamttpk\rmdef \frac{\calf^{[0,k]}(\Delta_k^t)}{\calf^{[0,k]}(\Delta_{k-1}^t)}.
$$

\begin{cor}
\label{cor42}
For $0\le j\le k$,
the $t+j$-th components of a transversal of $\lamttpk$ are a transversal of 
\be
\label{star11}  
\frac{\calf^j(\Delta_k^t)}{\calf^j(\Delta_{k-1}^t)}.
\ee
\end{cor}

\begin{prf}
Theorem \ref{thm40} shows that the projection $\chi^{t+j}(\lamttpk)$ 
gives a 1-1 correspondence between cosets of $\lamttpk$ and cosets of 
(\ref{star11}).  Therefore the projection $\chi^{t+j}$ 
of a transversal of $\lamttpk$ is a transversal of (\ref{star11}).
\end{prf}

\begin{cor}
\label{cor51}
For $0\le k\le\ell$, and $0\le j\le k$, we have
$$  
\frac{\Delta_k^t}{\Delta_{k-1}^t}\simeq
\frac{\calf^{[0,k]}(\Delta_k^t)}{\calf^{[0,k]}(\Delta_{k-1}^t)}\simeq
\frac{\calf^j(\Delta_k^t)}{\calf^j(\Delta_{k-1}^t)}.
$$
\end{cor}

{\it Remark:}  This result can be regarded as a rectangle criterion 
for a \sm, with $\Delta_k^t$, $\Delta_{k-1}^t$, $\calf^j(\Delta_k^t)$, 
and $\calf^j(\Delta_{k-1}^t)$ as the corners of a rectangle in 
(\ref{xtshftm}).  It is similar in spirit to a quadrangle criterion for
a Latin square \cite{DK} or a configuration theorem for a net \cite{DJ}.
In fact, the rectangle condition can be generalized
further by starting with groups $\Delta_k^t$ and $\Delta_{k-m}^t$, for $m>1$.
These more general results are not needed.

We can use (\ref{eq888}) and Corollary \ref{cor51} to create a
tensor.  Fix $j$ such that $0\le j\le\ell$, and define
$X_j^{t+j}\ssl X_{j-1}^{t+j}$ to be the column vector of 
quotient groups
\be
\label{qg1}
\frac{X_{j-1}^{t+j}(X_j^{t+j}\cap Y_{k-j}^{t+j})}
{X_{j-1}^{t+j}(X_j^{t+j}\cap Y_{k-j-1}^{t+j})},
\ee
for $k$ such that $j\le k\le\ell$.  This is the vector of
quotient groups formed from groups in the normal chain in
the center column of (\ref{xtshftm}).  Then using (\ref{eq888}),
$$
X_j^{t+j}\ssl X_{j-1}^{t+j}\rmdef\left(
\begin{array}{lllll}
\fracfjk{j}{\ell} & \!\!\cdots\!\! & \fracfjk{j}{k} & \!\!\cdots\!\! & \fracfjk{j}{j} 
\end{array}
\right)^T.
$$
For $j=0,\ldots,\ell$, we obtain the column vectors $X_j^{t+j}\ssl X_{j-1}^{t+j}$,
which can be used to form the {\it shift matrix} $X_\ssl^{[t,t+\ell]}$,
\be
\label{smtx1}
X_\ssl^{[t,t+\ell]}\rmdef\left(
\begin{array}{llllll}
X_0^t\ssl X_{-1}^t & X_1^{t+1}\ssl X_0^{t+1} & \!\!\cdots\!\! & X_j^{t+j}\ssl X_{j-1}^{t+j} & \!\!\cdots\!\! & X_\ell^{t+\ell}\ssl X_{\ell-1}^{t+\ell} 
\end{array}
\right).
\ee
This is a second example of a shift matrix.
The $k$-th row of \sm\ $X_\ssl^{[t,t+\ell]}$, $0\le k\le\ell$,
is a {\it shift vector}
$$
\left(
\begin{array}{llllll}
\fracfjk{0}{k} & \fracfjk{1}{k} & \!\!\cdots\!\! & \fracfjk{j}{k} & \!\!\cdots\!\! & \fracfjk{k}{k} 
\end{array}
\right).
$$
The \svec\ is just all the components of
$$
\lamttpk=\frac{\calf^{[0,k]}(\Delta_k^t)}{\calf^{[0,k]}(\Delta_{k-1}^t)}.
$$

Corollary \ref{cor51} shows the \sm\ $X_\ssl^{[t,t+\ell]}$ 
preserves isomorphism of quotient groups, and each shift of a 
quotient group in a row gives the next quotient group in the row.  
Therefore we can regard a \sm\ $X_\ssl^{[t,t+\ell]}$ as the natural 
shift structure of a strongly controllable group system.

Fix $j$ such that $0\le j\le\ell$, and define
$X_j^t\ssl X_{j-1}^t$ to be the column vector of 
quotient groups
\be
\label{qg1x}
\frac{X_{j-1}^t(X_j^t\cap Y_{k-j}^t)}
{X_{j-1}^t(X_j^t\cap Y_{k-j-1}^t)},
\ee
for $k$ such that $j\le k\le\ell$.  This is the vector of
quotient groups formed from groups in the normal chain in
the center column of (\ref{xttstm}).
For $j=0,\ldots,\ell$, we obtain the column vectors $X_j^t\ssl X_{j-1}^t$,
which can be used to form the {\it static matrix} $X_\ssl^{[t,t]}$,
\be
\label{cmtx1}
X_\ssl^{[t,t]}\rmdef\left(
\begin{array}{llllll}
X_0^t\ssl X_{-1}^t & X_1^t\ssl X_0^t & \!\!\cdots\!\! & X_j^t\ssl X_{j-1}^t & \!\!\cdots\!\! & X_\ell^t\ssl X_{\ell-1}^t
\end{array}
\right).
\ee
Note that the definition of $X_j^{t+j}\ssl X_{j-1}^{t+j}$ and
$X_j^t\ssl X_{j-1}^t$ is consistent since $X_j^{t+j}\ssl X_{j-1}^{t+j}$
is defined using (\ref{qg1}) and $X_j^t\ssl X_{j-1}^t$ is defined
using (\ref{qg1x}), and (\ref{qg1}) and (\ref{qg1x}) are consistent.
Note that
$$ X_j^t\ssl X_{j-1}^t=X_j^{(t-j)+j}\ssl X_{j-1}^{(t-j)+j}, $$
and we can think of $X_j^{(t-j)+j}\ssl X_{j-1}^{(t-j)+j}$ as
the definition $X_j^{t'+j}\ssl X_{j-1}^{t'+j}$ with time $t'$
defined by the parentheses term $(t-j)$.  Then we can
also think of \stm\ (\ref{cmtx1}) as
\be
\label{cmtx2}
X_\ssl^{[t,t]}\rmdef\left(
\begin{array}{llllll}
X_0^{(t)}\ssl X_{-1}^{(t)} & X_1^{(t-1)+1}\ssl X_0^{(t-1)+1} & \!\!\cdots\!\! & X_j^{(t-j)+j}\ssl X_{j-1}^{(t-j)+j} & \!\!\cdots\!\! & X_\ell^{(t-\ell)+\ell}\ssl X_{\ell-1}^{(t-\ell)+\ell}
\end{array}
\right).
\ee
Now it is clear that each term in (\ref{cmtx2}) is from one of $\ell+1$ 
different \sms.

We can relate a \stm\ $X_\ssl^{[t,t]}$ to a \sm\
$X_\ssl^{[t,t+\ell]}$ using the tensor description 
shown in (\ref{xtnsr}).  Time increases as we move up the page.
The vectors in the \sm\ (\ref{smtx1}) are the vectors along the
diagonal in (\ref{xtnsr}), and the vectors in the \stm\ (\ref{cmtx2})
are the vectors in a row of (\ref{xtnsr}).  
The superscript parentheses terms in (\ref{xtnsr}), like $(t-j)$,
indicate terms that all belong to the same \sm.  For example,
the diagonal terms 
$$
X_0^{(t-j)}\ssl X_{-1}^{(t-j)},
X_1^{(t-j)+1}\ssl X_0^{(t-j)+1},\ldots,
X_j^{(t-j)+j}\ssl X_{j-1}^{(t-j)+j},\ldots,
X_\ell^{(t-j)+\ell}\ssl X_{\ell-1}^{(t-j)+\ell},
$$
all belong to the \sm\ starting at time $t-j$, $X_\ssl^{[(t-j),(t-j)+\ell]}$.  
The center row in (\ref{xtnsr}) is (\ref{cmtx2}), which reduces to (\ref{cmtx1}),
which is just the \stm\ $X_\ssl^{[t,t]}$.  
\be
\label{xtnsr}
\setcounter{MaxMatrixCols}{7}
\begin{pmatrix}   
               &              &             & \vdots          & &          &             \\
               &              &             &                 & &          & X_\ell^{(t)+\ell}\ssl X_{\ell-1}^{(t)+\ell} \\
               &              &             &                 & &          & \vdots      \\
               &              & \cdots      & X_j^{(t)+j}\ssl X_{j-1}^{(t)+j} & \cdots &   &             \\
               &              &             & \vdots          & &          & X_\ell^{(t-j)+\ell}\ssl X_{\ell-1}^{(t-j)+\ell} \\
\cdots         & X_1^{(t)+1}\ssl X_0^{(t)+1} & \cdots    &    & &          &  \vdots     \\
X_0^{(t)}\ssl X_{-1}^{(t)}   & X_1^{(t-1)+1}\ssl X_0^{(t-1)+1} & \cdots  & X_j^{(t-j)+j}\ssl X_{j-1}^{(t-j)+j} & \cdots & & X_\ell^{(t-\ell)+\ell}\ssl X_{\ell-1}^{(t-\ell)+\ell} \\
\vdots         & \vdots       &             & \vdots          & &          & \vdots      \\
\cdots         & X_1^{(t-j)+1}\ssl X_0^{(t-j)+1} & \cdots  &  & &          &             \\
X_0^{(t-j)}\ssl X_{-1}^{(t-j)} & \cdots &   &                 & &          &             \\
               &              &             & \vdots          & &          &
\end{pmatrix}
\ee
We let $\bmx_\ssl$ denote the tensor in (\ref{xtnsr}), 
and say $\bmx_\ssl$ is a {\it chain tensor}.  For a given group trellis
$C$, there is only one chain tensor $\bmx_\ssl$. The tensor $\bmx_\ssl$ 
is a description of the coset structure of group trellis $C$.  
The tensor $\bmx_\ssl$ has a dual nature of having both \sms\ and \stms.
The most natural and important way to unnderstand $\bmx_\ssl$ is
to look at (\ref{xtnsr}) along the diagonals, in terms of \sms.

\begin{thm}
For each time $t$, the diagonals of (\ref{xtnsr}) are a description of the
quotient groups $\lamttpk$ for $k$ such that $0\le k\le\ell$.
\end{thm}

In the next two sections, we will show how to recover
paths $\bmb\in C$ from generators, which are representatives of the coset
structure described by $\bmx_\ssl$.

\newcommand{\cttpk}{{C^{[t,t+k]}}}
\newcommand{\gamttpk}{{\Gamma^{[t,t+k]}}}

\newpage
\vspace{3mm}
{\bf 4.  GENERATORS AND THE GENERATOR MATRIX}
\vspace{3mm}

We now show the Forney-Trott generators are a transversal
of $\lamttpk$, and components of the generators are
a transversal of (\ref{star11}), for $0\le j\le k$.
Forney and Trott \cite{FT} define a
generator for a group code $\msfc$ using the quotient group
$$
\calt^{[t,t+k]}\rmdef\frac{\msfc^{[t,t+k]}}{\msfc^{[t,t+k)}\msfc^{(t,t+k]}},
$$
for $0\le k\le\ell$, where $\calt^{[t,t+k]}$ is called a {\it granule}.  
A \crep\ of $\calt^{[t,t+k]}$ is called a {\it generator}.
The \crep\ of $\msfc^{[t,t+k)}\msfc^{(t,t+k]}$ is always taken
to be the identity sequence.  In case
$\calt^{[t,t+k]}$ is isomorphic to the identity group, the identity
sequence is the only coset representative.  
A nonidentity generator is an element of
$\msfc^{[t,t+k]}$ but not of $\msfc^{[t,t+k)}$ or of $\msfc^{(t,t+k]}$, so its span
is exactly $k+1$.  Thus every nonidentity generator
is a codeword that cannot be expressed as a combination of shorter
codewords \cite{FT}.
A {\it basis} of $\msfc$ is a minimal set of shortest length generators 
that is sufficient to generate the group system $\msfc$ \cite{FY1}.  
It is a set of \creps\ of $\calt^{[t,t+k]}$, for $0\le k\le\ell$.

Since a group trellis is a group system, we can transcribe the
generator approach of \cite{FT} to the group trellis $C$, used here, as
$$  
\gamttpk\rmdef \frac{\cttpk}{C^{[t,t+k)}C^{(t,t+k]}},
$$
where quotient group $\gamttpk$ is a granule.  
If $Q$ is any quotient group, let $[Q]$ denote a transversal of $Q$.
Let $[\gamttpk]$ be a transversal of $\gamttpk$.  A \crep\ of
$\gamttpk$, or an element of $[\gamttpk]$, is a
generator $\bmg^{[t,t+k]}$, or a generator at time $t$.
Then transversal $[\gamttpk]$ is a set of representatives $\bmg^{[t,t+k]}$
of $\gamttpk$ at time $t$.  For each time $t$, let 
{\it vector basis} $\calb^t$ be the set of generators
$\{\bmg^{[t,t+k]}\in [\gamttpk]:  0\le k\le\ell\}$ 
in all transversals at time $t$.  We allow $\calb^t$ to vary with time, 
e.g., $\calb^{t+1}$ need not be just a time shift of $\calb^t$.
The sequence of vector bases, $\ldots,\calb^t,\calb^{t+1},\ldots$, 
gives a {\it basis} $\bmcpb=\{\calb^t:  t\in\bmcpz\}$.
We also consider a {\it constant basis} $\bmcpb_c=\{\ldots,\calb,\calb,\ldots\}$
where $\calb^t$ is the same vector basis $\calb$ for all $t\in\bmcpz$.

We now show that the projection $\chi^{[t,t+k]}$ of generators in $[\gamttpk]$
is also a transversal of $\lamttpk$.  Therefore a basis $\bmcpb$ of $C$ 
can be found using representatives of either $\gamttpk$ or $\lamttpk$.

\begin{lem}
\label{lem23}
The set of paths formed by the concatenation of groups 
\be
\label{paths2}
\ldots,\bone^{t-2},\bone^{t-1}\wedge (X_0^t\cap Y_k^t)\wedge 
\cdots\wedge (X_j^{t+j}\cap Y_{k-j}^{t+j})\wedge (X_{j+1}^{t+j+1}\cap Y_{k-j-1}^{t+j+1})\wedge
\cdots\wedge (X_k^{t+k}\cap Y_0^{t+k})\wedge\bone^{t+k+1},\bone^{t+k+2},\ldots
\ee
is $\cttpk$.  
\end{lem}

\begin{prf}
From the proof of Proposition \ref{pro10}, we have
$$
\calf(X_j^{t+j}\cap Y_{k-j}^{t+j})
=X_0^{t+j+1}(X_{j+1}^{t+j+1}\cap Y_{k-j-1}^{t+j+1}).
$$
This means the set of paths formed by the concatenation of groups 
in (\ref{paths2}) is well defined:  for any branch 
$b^{t+j}\in X_j^{t+j}\cap Y_{k-j}^{t+j}$,
there is a branch $b^{t+j+1}\in X_{j+1}^{t+j+1}\cap Y_{k-j-1}^{t+j+1}$ 
such that $(b^{t+j})^+=(b^{t+j+1})^-$, and $(b^{t+j},b^{t+j+1})$ is 
a trellis path segment of length two.  The paths in (\ref{paths2})
consist of sequences which split from the identity state 
at time $t$ and merge to the identity state at time $t+k+1$.  
Therefore, any path in (\ref{paths2}) must be in $\cttpk$.

Fix integer $k$ such that $0\le k\le\ell$.  Let $\bmb$ be a 
sequence in $\cttpk$.  We now show $\bmb$ is in (\ref{paths2}).
If $\bmb\in\cttpk$, then 
for each $j$, $0\le j\le k$, $b^{t+j}$ must be in $X_j^{t+j}$, but
cannot be in $X_m^{t+j}$, $m>j$.  Similarly, $b^{t+j}$ must be 
in $Y_{k-j}^{t+j}$.  Then $b^{t+j}\in X_j^{t+j}\cap Y_{k-j}^{t+j}$ 
for all $j\in [0,k]$.  Since (\ref{paths2}) contains all code sequences 
whose component $b^{t+j}\in X_j^{t+j}\cap Y_{k-j}^{t+j}$ for all $j\in [0,k]$, 
then $\bmb$ is in (\ref{paths2}). 
\end{prf}

\begin{lem}
\label{lem24}
For $j$, $0\le j\le k$, we have $\chi^{t+j}(\cttpk)=X_j^{t+j}\cap Y_{k-j}^{t+j}$.  
For example, this means $X_0^t\cap Y_k^t=\chi^t(\cttpk)$ 
and $Y_0^{t+k}\cap X_k^{t+k}=\chi^{t+k}(\cttpk)$.
\end{lem}

\begin{prf}
From (\ref{paths2}), we know 
$\chi^{t+j}(\cttpk)\subset X_j^{t+j}\cap Y_{k-j}^{t+j}$.   

We now show $X_j^{t+j}\cap Y_{k-j}^{t+j}\subset\chi^{t+j}(\cttpk)$.  
The proof of Lemma \ref{lem23} shows that 
for any branch $b^{t+j}\in X_j^{t+j}\cap Y_{k-j}^{t+j}$,
there is a branch $b^{t+j+1}\in X_{j+1}^{t+j+1}\cap Y_{k-j-1}^{t+j+1}$ 
such that $(b^{t+j},b^{t+j+1})$ is a trellis path segment of length two.
We can continue this argument:
for any branch $b^{t+j+1}\in X_{j+1}^{t+j+1}\cap Y_{k-j-1}^{t+j+1}$, 
there is a branch $b^{t+j+2}\in X_{j+2}^{t+j+2}\cap Y_{k-j-2}^{t+j+2}$ 
such that $(b^{t+j+1},b^{t+j+2})$ is a trellis path segment of length two.
Continuing the argument further shows that  
for any branch $b^{t+j}\in X_j^{t+j}\cap Y_{k-j}^{t+j}$,
there is a trellis path segment of length $k-j+1$,
$(b^{t+j},b^{t+j+1},\ldots,b^{t+k})$, which merges to the
identity state at time $t+k+1$.  This argument works in
reverse time as well:  
for any branch $b^{t+j}\in X_j^{t+j}\cap Y_{k-j}^{t+j}$,
there is a branch $b^{t+j-1}\in X_{j-1}^{t+j-1}\cap Y_{k-j+1}^{t+j-1}$ 
such that $(b^{t+j-1},b^{t+j})$ is a trellis path segment of length two,
and so on.  Thus we see that 
for any $b^{t+j}\in X_j^{t+j}\cap Y_{k-j}^{t+j}$, there
is a sequence $\bmb\in\cttpk$ such that $\chi^{t+j}(\bmb)=b^{t+j}$.  
Thus we have shown $X_j^{t+j}\cap Y_{k-j}^{t+j}\subset\chi^{t+j}(\cttpk)$.  
\end{prf}  

\begin{lem}
\label{lem34}
We have 
\be
\label{eq33a}
\chi^{[t,t+k]}(\cttpk)\subset\calf^{[0,k]}(\Delta_k^t),
\ee
and 
\be
\label{eq33b}
\chi^{[t,t+k]}(C^{[t,t+k)}C^{(t,t+k]})\subset\calf^{[0,k]}(\Delta_{k-1}^t).
\ee
\end{lem}

\begin{prf}
We have (\ref{eq33a}) holds \ifof\ $\chi^t(C^{[t,t+k]})\subset\Delta_k^t$.
But this follows from Lemma \ref{lem24}.  
We have (\ref{eq33b}) holds \ifof\ 
$\chi^t(C^{[t,t+k)}C^{(t,t+k]})\subset\Delta_{k-1}^t$.
But $\chi^t(C^{[t,t+k)}C^{(t,t+k]})=\chi^t(C^{[t,t+k)})=\Delta_{k-1}^t$
from Lemma \ref{lem24}.
\end{prf}

\begin{thm}
\label{thm35}
There is an isomorphism
$$
\gamttpk\stackrel{\mu}{\simeq}\lamttpk,
$$
where the 1-1 correspondence $\mu$ between cosets of $\gamttpk$ and 
$\lamttpk$ is given by
\be
\label{eq34}
\mu:  C^{[t,t+k)}C^{(t,t+k]}\bmb
\mapsto\calf^{[0,k]}(\chi^t(C^{[t,t+k)}C^{(t,t+k]}\bmb)).
\ee
\end{thm}

\begin{prf}
Using Lemma \ref{lem34}, we have
\begin{align*}
\chi^t(C^{[t,t+k)}C^{(t,t+k]}\bmb) 
  &=\chi^t(C^{[t,t+k)}\bmb) \\
  &=\chi^t(C^{[t,t+k)})\chi^t(\bmb) \\
  &=\Delta_{k-1}^tb^t.
\end{align*}
Since $\lamttpk=\calf^{[0,k]}(\Delta_k^t)/\calf^{[0,k]}(\Delta_{k-1}^t)$,
this shows we can properly define the
1-1 correspondence $\mu$ between cosets of $\gamttpk$
and $\lamttpk$ as given in (\ref{eq34}).

Forney and Trott \cite{FT} define an {\it input chain} 
$F_0^t\subset F_1^t\subset\cdots\subset F_\ell^t$ by the projection
$F_k^t\rmdef \chi^t(\cttpk)$ for $k=0,1,\ldots,\ell$.
Using Lemma \ref{lem24}, this gives $F_k^t=\Delta_k^t$.
In their Input Granule Theorem \cite{FT}, Forney and Trott
show that $\gamttpk\simeq F_k^t/F_{k-1}^t$ for $k$ such that $0<k\le\ell$.
Then we have 
$$
\gamttpk\simeq F_k^t/F_{k-1}^t=\Delta_k^t/\Delta_{k-1}^t.
$$
Combining this with Theorem \ref{thm17} gives
$$
\gamttpk\simeq \Delta_k^t/\Delta_{k-1}^t\simeq\lamttpk.
$$
Then following the correspondences given in the Input Granule Theorem
of \cite{FT} and Theorem \ref{thm17} shows that the isomorphism
$\gamttpk\simeq\lamttpk$ is given by $\mu$.
\end{prf}

\begin{cor}
\label{cor36}
Let $[\gamttpk]$ be a set of generators which is a transversal of
$\gamttpk$.  Then 
$\{\chi^{[t,t+k]}(\bmg^{[t,t+k]}) : \bmg^{[t,t+k]}\in [\gamttpk]\}$
is a transversal of $\lamttpk$.
\end{cor}
The above corollary shows that any set of Forney-Trott generators
can equally well be found from a tranversal of $\lamttpk$.

If $Q$ is any quotient group, there is another way we denote a transversal
of $Q$ besides $[Q]$.  If $\{q\}$ is a set of \creps\ of $Q$ which is a
transversal of $Q$, we let $[\{q\}]$ denote a transversal of $Q$.

Fix $k$ such that $0\le k\le\ell$.  Let generator 
$\bmg^{[t,t+k]}$ be a representative in $\gamttpk$,
\be
\label{tps0}
\bmg^{[t,t+k]}=
\ldots,\bone^{t-2},\bone^{t-1},r_{0,k}^t,r_{1,k}^{t+1},\ldots,r_{j,k}^{t+j},\ldots,r_{k,k}^{t+k},\bone^{t+k+1},\bone^{t+k+2},\ldots,
\ee
From (\ref{paths2}) we know component $r_{j,k}^{t+j}$ 
is an element of $X_j^{t+j}\cap Y_{k-j}^{t+j}$, and from Corollaries 
\ref{cor36} and \ref{cor42} we know $r_{j,k}^{t+j}$ is a representative of
\be
\label{qg2}
\frac{\calf^j(\Delta_k^t)}{\calf^j(\Delta_{k-1}^t)}=
\frac{X_{j-1}^{t+j}(X_j^{t+j}\cap Y_{k-j}^{t+j})}{X_{j-1}^{t+j}(X_j^{t+j}\cap Y_{k-j-1}^{t+j})},
\ee
for $j=0,1,\ldots,k$.  If we pick a set of generators $\bmg^{[t,t+k]}$ which is 
a transversal of $\gamttpk$, $[\gamttpk]$, then $[\gamttpk]$ induces a 
transversal $[\{r_{j,k}^{t+j}\}]$ of (\ref{qg2}), for $j=0,1,\ldots,k$.

Pick a generator $\bmg^{[t,t+k]}$ in $\gamttpk$ for each $k$, $0\le k\le\ell$. 
We can arrange the nontrivial components of these generators in a matrix
as shown in (\ref{gmr}), which is called a \sm, or also a {\it generator matrix},
at time $t$, and denoted $R^{[t,t+\ell]}$.  
The $k$-th row of matrix $R^{[t,t+\ell]}$, $0\le k\le\ell$,
is a {\it shift vector}, also called a {\it generator vector}, 
denoted $\bmr^{[t,t+k]}$, where
\be
\label{gmrm}
\bmr^{[t,t+k]}
\rmdef (r_{0,k}^t,r_{1,k}^{t+1},\ldots,r_{j,k}^{t+j},\ldots,r_{k,k}^{t+k}).
\ee
A generator vector $\bmr^{[t,t+k]}$ is the nontrivial 
components of the generator $\bmg^{[t,t+k]}$.
\be
\label{gmr}
\begin{array}{llllllllll}
  r_{0,\ell}^t   & r_{1,\ell}^{t+1}   & \cdots & \cdots & r_{j,\ell}^{t+j}   & \cdots & \cdots & \cdots & r_{\ell-1,\ell}^{t+\ell-1}   & r_{\ell,\ell}^{t+\ell} \\
  r_{0,\ell-1}^t & r_{1,\ell-1}^{t+1} & \cdots & \cdots & r_{j,\ell-1}^{t+j} & \cdots & \cdots & \cdots & r_{\ell-1,\ell-1}^{t+\ell-1} & \\
  \vdots & \vdots & \vdots & \vdots & \vdots & \vdots & \vdots & \vdots && \\
  r_{0,k}^t & r_{1,k}^{t+1} & \cdots & \cdots & r_{j,k}^{t+j} & \cdots & r_{k,k}^{t+k} &&& \\
  \vdots & \vdots & \vdots & \vdots & \vdots & \vdots &&&& \\
  \cdots & \cdots & \cdots & \cdots & r_{j,j}^{t+j} &&&&& \\
  \vdots & \vdots & \vdots &&&&&&& \\
  r_{0,2}^t   & r_{1,2}^{t+1} & r_{2,2}^{t+2} &&&&&&& \\
  r_{0,1}^t   & r_{1,1}^{t+1} &&&&&&&& \\
  r_{0,0}^t   &&&&&&&&&
\end{array}
\ee
We define $\bmr_j^{t+j}$ to be a column vector in (\ref{gmr}), 
for $0\le j\le\ell$, where
$$
\bmr_j^{t+j}\rmdef\left(
\begin{array}{lllll}
r_{j,\ell}^{t+j} & \!\!\cdots\!\! & r_{j,k}^{t+j} & \!\!\cdots\!\! & r_{j,j}^{t+j} 
\end{array}
\right)^T.
$$
Then we can rewrite (\ref{gmr}) as
\be
\label{gmr1}
R^{[t,t+\ell]}=(\bmr_0^t,\bmr_1^{t+1},\ldots,\bmr_j^{t+j},\ldots,\bmr_\ell^{t+\ell}).
\ee

There is another related form, shown in (\ref{rttf}), called the
{\it static matrix} $R^{[t,t]}$, where component $r_{j,k}^t$ 
is just an element in $X_j^t\cap Y_{k-j}^t$.  As can be seen, all
components of the \stm\ occur at time $t$.  For a \gm, the first column
specifies the matrix completely. For a \stm, the first column does not 
determine the \stm\ uniquely.
\be
\label{rttf}
\begin{array}{llllllllll}
  r_{0,\ell}^t   & r_{1,\ell}^t   & \cdots & \cdots & r_{j,\ell}^t   & \cdots & \cdots & \cdots & r_{\ell-1,\ell}^t   & r_{\ell,\ell}^t \\
  r_{0,\ell-1}^t & r_{1,\ell-1}^t & \cdots & \cdots & r_{j,\ell-1}^t & \cdots & \cdots & \cdots & r_{\ell-1,\ell-1}^t & \\
  \vdots & \vdots & \vdots & \vdots & \vdots & \vdots & \vdots & \vdots && \\
  r_{0,k}^t & r_{1,k}^t & \cdots & \cdots & r_{j,k}^t & \cdots & r_{k,k}^t &&& \\
  \vdots & \vdots & \vdots & \vdots & \vdots & \vdots &&&& \\
  \cdots & \cdots & \cdots & \cdots & r_{j,j}^t &&&&& \\
  \vdots & \vdots & \vdots &&&&&&& \\
  r_{0,2}^t   & r_{1,2}^t & r_{2,2}^t &&&&&&& \\
  r_{0,1}^t   & r_{1,1}^t &&&&&&&& \\
  r_{0,0}^t   &&&&&&&&&
\end{array}
\ee
We can rewrite (\ref{rttf}) as
\be
\label{bmr1}
R^{[t,t]}=(\bmr_0^t,\bmr_1^t,\ldots,\bmr_j^t,\ldots,\bmr_\ell^t).
\ee

We can relate a \stm\ $R^{[t,t]}$ to a \gm\
$R^{[t,t+\ell]}$ using the tensor description shown in (\ref{rtnsr}).
Time increases as we move up the page.
The vectors in the \gm\ (\ref{gmr1}) are the vectors along the
diagonal in (\ref{rtnsr}), and the vectors in the \stm\ (\ref{bmr1})
are the vectors in a row of (\ref{rtnsr}).  
The superscript parentheses terms in (\ref{rtnsr}), like $(t-j)$,
indicate terms that all belong to the same \gm.  For example,
the diagonal terms 
$\bmr_0^{(t-j)},\bmr_1^{(t-j)+1},\ldots,\bmr_j^{(t-j)+j},\ldots,\bmr_\ell^{(t-j)+\ell}$
all belong to the \gm\ starting at time $t-j$, $R^{[(t-j),(t-j)+\ell]}$.  
The center row in (\ref{rtnsr}) is
\be
\label{ctrrow1}
(\bmr_0^{(t)},\bmr_1^{(t-1)+1},\ldots,\bmr_j^{(t-j)+j},\ldots,\bmr_\ell^{(t-\ell)+\ell}),
\ee
where each entry is itself a column; this reduces to 
\be
\label{ctrrow2}
(\bmr_0^t,\bmr_1^t,\ldots,\bmr_j^t,\ldots,\bmr_\ell^t),
\ee
which is just the \stm\ $R^{[t,t]}$.  Notice that each term in (\ref{ctrrow1})
and (\ref{ctrrow2}) is from one of $\ell+1$ different \sms.
\be
\label{rtnsr}
\setcounter{MaxMatrixCols}{7}
\begin{pmatrix}   
               &              &             & \vdots          & &          &             \\
               &              &             &                 & &          & \bmr_\ell^{(t)+\ell}  \\
               &              &             &                 & &          & \vdots      \\
               &              & \cdots      & \bmr_j^{(t)+j}  & \cdots &   &             \\
               &              &             & \vdots          & &          & \bmr_\ell^{(t-j)+\ell}  \\
\cdots         & \bmr_1^{(t)+1} & \cdots    &                 & &          &  \vdots     \\
\bmr_0^{(t)}   & \bmr_1^{(t-1)+1} & \cdots  & \bmr_j^{(t-j)+j}  & \cdots & & \bmr_\ell^{(t-\ell)+\ell} \\
\vdots         & \vdots       &             & \vdots          & &          & \vdots      \\
\cdots         & \bmr_1^{(t-j)+1} & \cdots  &                 & &          &             \\
\bmr_0^{(t-j)} & \cdots       &             &                 & &          &             \\
               &              &             & \vdots          & &          &
\end{pmatrix}
\ee

\begin{thm}
\label{thm31}
Fix time $t$.  A finite sequence of $\ell+1$ \gms\ $R^{[(t-j),(t-j)+\ell]}$ at
times $t-j$, for $j=0,\ldots,\ell$, uniquely determines a \stm\ 
$R^{[t,t]}$, where column $j$ of \gm\
$R^{[(t-j),(t-j)+\ell]}$, denoted $\bmr_j^{(t-j)+j}$, is column $j$ of 
\stm\  $R^{[t,t]}$, denoted $\bmr_j^t$.
\end{thm}

\begin{prf}
The center row in (\ref{rtnsr}) is (\ref{ctrrow1}), which reduces to 
(\ref{ctrrow2}), which is just \stm\ $R^{[t,t]}$.  But entry 
$\bmr_j^{(t-j)+j}$ in (\ref{ctrrow1}) is just the $(j+1)$-th column
of the \gm\ $R^{[(t-j),(t-j)+\ell]}$ at time $t-j$.
\end{prf}

We let $\bmr$ denote the tensor in (\ref{rtnsr}), 
and say $\bmr$ is a {\it representative tensor}.
We can regard $\bmr$ in two different ways, as a sequence of \stms\
or as a sequence of \sms.  In the first way we can write $\bmr$ as
\be
\label{interp}
\bmr=\ldots,\bmr^t,\bmr^{t+1},\ldots,
\ee
where each $\bmr^t$ is a \stm\ $R^{[t,t]}$ in the set of all
\stms, denoted $\bmcpr^t$.  Therefore (\ref{interp}) is
equivalent to
$$
\bmr=\ldots,R^{[t,t]},R^{[t+1,t+1]},\ldots.
$$

We have just seen from Theorem \ref{thm31} that each $\bmr^t$ is
determined by $\ell+1$ \sms.  Then tensor $\bmr$ 
in (\ref{interp}) is also determined by a sequence of \sms.  
We denote this interpretation of $\bmr$ using notation 
$$
\bmr\sim\ldots,R^{[t,t+\ell]},R^{[t+1,t+1+\ell]},\ldots,
$$
where each \sm\ $R^{[t,t+\ell]}$ is in the set of all possible \sms,
denoted $\bmcpr^{[t,t+\ell]}$.

We define tensor set $\calr$ to be the set of representative tensors 
$\bmr$ determined by the Cartesian product of all possible \sms,
$$
\calr\sim\prod_{t=-\infty}^\infty\bmcpr^{[t,t+\ell]}.
$$
Note that $\calr$ depends on choice of basis $\bmcpb$.
Because $\calr$ is the product of all possible \sms, we say
$\calr$ is {\it full}.

For a given group trellis $C$, there is only one coset tensor $\bmx_\ssl$,
and at each time $t$, there is only one \sm\ $X_\ssl^{[t,t+\ell]}$
and one \stm\ $X_\ssl^{[t,t]}$.  A group trellis $C$ can have many bases
$\bmcpb$.  Each basis $\bmcpb$ is a selection of one \crep\ (\gvec) 
from each of the cosets in each of the quotient groups $\{\lamttpk:  0\le k\le\ell\}$
in $X_\ssl^{[t,t+\ell]}$, at each time $t$.  Now fix basis $\bmcpb$ and
fix the corresponding tensor set $\calr$.  
Each tensor $\bmr\in\calr$ is a selection of one 
\crep\ from a single coset of each of the quotient groups $\{\lamttpk:  0\le k\le\ell\}$
in $X_\ssl^{[t,t+\ell]}$, at each time $t$.  Thus for each basis $\bmcpb$, there are
many possible $\bmr\in\calr$.

Each tensor $\bmr\in\calr$ gives 
one \sm\ $R^{[t,t+\ell]}$ and one \stm\ $R^{[t,t]}$
at each time $t$.  A different tensor $\bmrht\in\calr$ 
may have a different \sm\ $\cprht^{[t,t+\ell]}$ and different \stm\ $\cprht^{[t,t]}$
at each time $t$.  $R^{[t,t+\ell]}$ is a selection of one 
\crep\ from a single coset of each quotient group in $X_\ssl^{[t,t+\ell]}$.
Thus $R^{[t,t+\ell]}$ has the same
form and time indices as the $X_\ssl^{[t,t+\ell]}$ \sm.  Similarly 
$R^{[t,t]}$ is a selection of one \crep\ from a single coset of each quotient group
in $X_\ssl^{[t,t]}$.  Thus $R^{[t,t]}$ has the same
form and time indices as the $X_\ssl^{[t,t]}$ \stm.
This explains why tensor $\bmr$ in (\ref{rtnsr}) has the
same form as tensor $\bmx_\ssl$ in (\ref{xtnsr}).

A given $\bmr\in\calr$ produces a sequence of \sms\ $R^{[t,t+\ell]}$ and
a sequence of \stms\ $R^{[t,t]}$.  Any sequence of \sms\ corresponds to
some $\bmr\in\calr$ and uniquely determines a sequence of \stms.
But an arbitrary sequence of \stms\ may not correspond to a valid 
sequence of \gvecs\ and therefore an $\bmr\in\calr$.
In this paper we regard \sms\ and \svecs\ as the primary objects;
these have intrinsic meaning since they are related to generators.
The \stm\ is formed by an interleaving of columns of different \sms\
and is regarded as a secondary object.

\begin{lem}
\label{lem90f}
Fix $j$ such that $0\le j\le\ell$.  Fix $k$ such that $j\le k\le\ell$.
Let $[\Gamma^{[t-j,t-j+k]}]$
be a set of generators $\{\bmg^{[t-j,t-j+k]}\}$ which is a transversal of 
$\Gamma^{[t-j,t-j+k]}$.  The $(t-j)+j$-th components of generators 
$\bmg^{[t-j,t-j+k]}\in[\Gamma^{[t-j,t-j+k]}]$ form a transversal
\be
\label{keyeqf}
[\{\chi^t(\bmg^{[(t-j),(t-j)+k]})\}]=[\{r_{j,k}^{(t-j)+j}\}]=[\{r_{j,k}^t\}]
\ee
of
\be
\label{qg3}
\frac{X_{j-1}^t(X_j^t\cap Y_{k-j}^t)}{X_{j-1}^t(X_j^t\cap Y_{k-j-1}^t)}.
\ee
\end{lem}

\begin{prf}
Fix $j$, where $0\le j\le\ell$, and examine time $t-j$.  Fix $k$ such
that $j\le k\le\ell$.  Pick a set of generators $\bmg^{[(t-j),(t-j)+k]}$ which
is a transversal of $\Gamma^{[(t-j),(t-j)+k]}$, denoted 
$[\Gamma^{[(t-j),(t-j)+k]}]$.  Then $[\Gamma^{[(t-j),(t-j)+k]}]$ induces
a transversal $[\{r_{j,k}^{(t-j)+m}\}]$ of
\be
\label{qg2b}
\frac{X_{j-1}^{(t-j)+m}(X_j^{(t-j)+m}\cap Y_{k-j}^{(t-j)+m})}{X_{j-1}^{(t-j)+m}(X_j^{(t-j)+m}\cap Y_{k-j-1}^{(t-j)+m})},
\ee
for $m=0,1,\ldots,k$.  Choose $m=j$.  Then $[\{r_{j,k}^{(t-j)+m}\}]$ is a transversal 
$[\{r_{j,k}^{(t-j)+j}\}]=[\{r_{j,k}^t\}]$ of (\ref{qg2b}) for $m=j$,
which is the same as (\ref{qg3}).
\end{prf}

Note that the set of transversals $[\{r_{j,k}^{(t-j)+j}\}]$
for $k$ such that $j\le k\le\ell$ are the \creps\ of all cosets
in quotient groups in column $j$ of \sm\ $X_\ssl^{[(t-j),(t-j)+\ell]}$, 
which is column $X_j^{(t-j)+j}\ssl X_{j-1}^{(t-j)+j}$.
And the set of transversals $[\{r_{j,k}^t\}]$
for $k$ such that $j\le k\le\ell$ are the \creps\ of all cosets
in quotient groups in column $j$ of \stm\ $X_\ssl^{[t,t]}$, 
which is column $X_j^t\ssl X_{j-1}^t$.  
By selecting one \crep\ from each quotient group of $X_\ssl^{[t,t]}$,
we obtain a {\it complete set of coset representatives}
for the normal chain of $B^t$ given by the $X^{[t,t]}$ \stm.
This gives the following result.

\begin{thm}
\label{thm90f}
For $0\le j\le\ell$, for $k$ such that $j\le k\le\ell$, let $[\Gamma^{[t-j,t-j+k]}]$
be a set of generators $\{\bmg^{[t-j,t-j+k]}\}$ which is a transversal of 
$\Gamma^{[t-j,t-j+k]}$.  The $(t-j)+j$-th components of generators 
$\bmg^{[t-j,t-j+k]}\in[\Gamma^{[t-j,t-j+k]}]$ form a transversal (\ref{keyeqf})
of (\ref{qg3}) for $0\le j\le\ell$, for $j\le k\le\ell$.
The set of transversals, $[\{r_{j,k}^t\}]$, for $0\le j\le\ell$, 
for $j\le k\le\ell$, forms a \compset\ 
for the normal chain of $B^t$ given by the $X^{[t,t]}$ \stm.
\end{thm}

Any branch $b^t\in B^t$ can be written using elements of this \compset\ as
\be
\label{enctd}
b^t=\prod_{j=0}^\ell \left(\prod_{k=j}^\ell r_{j,k}^t\right).
\ee
By the convention used here, equation (\ref{enctd})
is evaluated as
\be
\label{enctd1}
b^t=r_{\ell,\ell}^t r_{\ell-1,\ell}^t r_{\ell-1,\ell-1}^t\cdots 
r_{j,\ell}^t\cdots r_{j,k}^t\cdots r_{j,j}^t\cdots
r_{2,2}^tr_{1,\ell}^t\cdots r_{1,1}^tr_{0,\ell}^t\cdots r_{0,2}^tr_{0,1}^tr_{0,0}^t.  
\ee
Note that $b^t$ is the product of terms in some \stm\ 
$R^{[t,t]}$, where the inner product in parentheses in (\ref{enctd})  
is just the product of terms in the $j$-th column of $R^{[t,t]}$.
Using (\ref{keyeqf}), (\ref{enctd}) can be written in equivalent forms as
\begin{align}
\label{enctda}
b^t
&=\prod_{j=0}^\ell \left(\prod_{k=j}^\ell r_{j,k}^t\right) \\
\label{enctdb}
&=\prod_{j=0}^\ell \left(\prod_{k=j}^\ell r_{j,k}^{(t-j)+j}\right) \\
\label{enctdc}
&=\prod_{j=0}^\ell \left(\prod_{k=j}^\ell \chi^t(\bmg^{[t-j,t-j+k]})\right).
\end{align}

We have just shown that for any time $t$, we can find any branch $b^t\in B^t$
using a selected set of generators at times $t-j$, for $j=0,\ldots,\ell$.
However we have not shown we can construct any path in $C$ this way.
We do this in the next section.

We now give a development dual to the \fssr\ using the \bssr.
We show that components 
of the same generators form a \compset\ for two normal chains.  
Define $\Delta_{Y,k}^t=Y_0^t\cap X_k^t$, 
for $-1\le k\le\ell$.  Define
the {\it previous branch set} $\calp(b)$ to be the time reversal
of $\calf(b)$.  The time reversal of quotient group $\lamttpk$ is 
$\Lambda_Y^{[t-k,t]}$,
$$  
\Lambda_Y^{[t-k,t]}\rmdef 
\frac{\calp^{[-k,0]}(\Delta_{Y,k}^t)}
{\calp^{[-k,0]}(\Delta_{Y,k-1}^t)},
$$
where $\calp^{[-k,0]}$ is the time reversal of $\calf^{[0,k]}$.
The time reversal of quotient group $\gamttpk$ is 
$\Gamma_Y^{[t-k,t]}$,
$$  
\Gamma_Y^{[t-k,t]}\rmdef\Gamma^{[t-k,t]}.
$$
The representatives of quotient group
$\Gamma_Y^{[t-k,t]}$ are generators $\bmg^{[t-k,t]}$.
Previously we defined a vector basis $\calb^t$ using generators
$\bmg^{[t,t+k]}$ which begin at time $t$, for $0\le k\le\ell$.
Now we define a vector basis $\calb_Y^t$ using generators
$\bmg^{[t-k,t]}$ which end at time $t$, for $0\le k\le\ell$.
This defines a basis $\bmcpb_Y$ and constant basis $\bmcpb_{c,Y}$.
The vector bases $\calb^t$ and $\calb_Y^t$ have an inherent asymmetry
with respect to time.  The asymmetry of $\calb^t$ and $\calb_Y^t$
is reflected in $\bmcpb$ and $\bmcpb_Y$ also.

Using these definitions, the arguments in Lemma \ref{lem34}
and Theorem \ref{thm35} can be reversed in time.  In place of
the input chain \cite{FT} in the proof of Theorem \ref{thm35}, 
the {\it last output chain} \cite{FT} is used.  This gives the 
following time reversed version of Theorem \ref{thm35} and 
Corollary \ref{cor36}.

\begin{thm}
\label{thm28}
There is an isomorphism
$$
\Gamma_Y^{[t-k,t]}\stackrel{\mu'}{\simeq}\Lambda_Y^{[t-k,t]},
$$
where the 1-1 correspondence $\mu'$ between cosets of $\Gamma_Y^{[t-k,t]}$ and 
$\Lambda_Y^{[t-k,t]}$ is given by
$$
\mu':  C^{[t-k,t)}C^{(t-k,t]}\bmb
\mapsto\calp^{[-k,0]}(\chi^t(C^{[t-k,t)}C^{(t-k,t]}\bmb)).
$$
\end{thm}

\begin{cor}
\label{cor37}
Let $[\Gamma_Y^{[t-k,t]}]$ be a set of generators which is a transversal of
$\Gamma_Y^{[t-k,t]}$.  Then 
$\{\chi^{[t-k,t]}(\bmg^{[t-k,t]}) : \bmg^{[t-k,t]}\in [\Gamma_Y^{[t-k,t]}]\}$
is a transversal of $\Lambda_Y^{[t-k,t]}$.
\end{cor}

The \gm\ of the \bssr\ is $R_Y^{[t-\ell,t]}$ and the \stm\ is
$R_Y^{[t,t]}$, shown in (\ref{rttb}).
To distinguish representatives in the forward and \bssr, 
we have added an additional subscript $Y$ to representatives in the \bssr.  

\be
\label{rttb}
\begin{array}{llllllllll}
  r_{Y,\ell,\ell}^t   & r_{Y,\ell-1,\ell}^t   & \cdots & \cdots & \cdots   & r_{Y,i,\ell}^t & \cdots & \cdots & r_{Y,1,\ell}^t   & r_{Y,0,\ell}^t \\
  & r_{Y,\ell-1,\ell-1}^t & \cdots & \cdots & \cdots & r_{Y,i,\ell-1}^t & \cdots & \cdots & r_{Y,1,\ell-1}^t & r_{Y,0,\ell-1}^t \\
  && \vdots & \vdots & \vdots & \vdots & \vdots & \vdots & \vdots & \vdots \\
  &&& r_{Y,k,k}^t & \cdots & r_{Y,i,k}^t & \cdots & \cdots & r_{Y,1,k}^t & r_{Y,0,k}^t \\
  &&&& \vdots & \vdots & \vdots & \vdots & \vdots & \vdots \\
  &&&&& r_{Y,i,i}^t & \cdots & \cdots & \cdots & \cdots \\
  &&&&&&& \vdots & \vdots & \vdots \\
  &&&&&&& r_{Y,2,2}^t   & r_{Y,1,2}^t & r_{Y,0,2}^t \\
  &&&&&&&& r_{Y,1,1}^t   & r_{Y,0,1}^t \\
  &&&&&&&&& r_{Y,0,0}^t   
\end{array}
\ee

The \gm\ $R_Y^{[t-\ell,t]}$ consists of representatives $r_{Y,i,k}^{t-i}$
from generators $\bmg^{[t-k,t]}$ for $0\le k\le\ell$.  From Theorem 
\ref{thm28} and Corollary \ref{cor37} we may use the same generators
for the forward and \bssr.  Then in the \fssr, $\bmg^{[t-k,t]}$ is a generator
which begins at time $t-k$ and ends at time $t$.  In the \bssr, 
we consider $\bmg^{[t-k,t]}$ to be a generator
which begins at time $t$ and ends at time $t-k$.
In the \fssr, the generator $\bmg^{[t-k,t]}$ is written as
\be
\label{repcmp}
\bmg^{[t-k,t]}=
\ldots,\bone^{t-k-2},\bone^{t-k-1},r_{0,k}^{t-k},r_{1,k}^{t-k+1},\ldots,r_{j,k}^{t-k+j},\ldots,r_{k,k}^t,\bone^{t+1},\bone^{t+2},\ldots,
\ee
while in the \bssr, the generator $\bmg^{[t-k,t]}$ is written as
\be
\label{repcmp1}
\bmg^{[t-k,t]}=
\ldots,\bone^{t-k-2},\bone^{t-k-1},r_{Y,k,k}^{t-k},r_{Y,k-1,k}^{t-k+1},\ldots,r_{Y,i,k}^{t-i},\ldots,r_{Y,0,k}^t,\bone^{t+1},\bone^{t+2},\ldots.
\ee
Note that $r_{j,k}^{t-k+j}=r_{Y,i,k}^{t-i}$ when $j=k-i$.

The first column $\bmr_0^t$ in $R^{[t,t+\ell]}$ and $R^{[t,t]}$ is composed
of representatives from generators $\bmg^{[t,t+k]}$ that begin at time $t$,
for $0\le k\le\ell$.  
The first column $\bmr_{Y,0}^t$ in $R_Y^{[t-\ell,t]}$ and $R_Y^{[t,t]}$ is composed
of representatives from generators $\bmg^{[t-k,t]}$ that end at time $t$
going forward in time, or begin at time $t$ going backward in time,
for $0\le k\le\ell$.  If the same generators are used for the \fssr\ and
\bssr, the first column $\bmr_0^t$ in $R^{[t,t+\ell]}$ and $R^{[t,t]}$
are the representatives in the diagonal terms $r_{Y,i,i}^t$ of $R_Y^{[t,t]}$
for $0\le i\le\ell$.  And the first column 
$\bmr_{Y,0}^t$ in $R_Y^{[t-\ell,t]}$ and $R_Y^{[t,t]}$
are the representatives in the diagonal terms $r_{j,j}^t$ of $R^{[t,t]}$
for $0\le j\le\ell$.

A representative tensor $\bmr_Y\in\calr_Y$ in the \bssr\ is constructed
using $R_Y^{[t-\ell,t]}$ and $R_Y^{[t,t]}$ in a dual manner to constructing 
$\bmr\in\calr$.  

Assume that basis $\bmcpb=\{\calb^t:  t\in\bmcpz\}$ is chosen.
Then tensor set $\calr$ can be found.  Fix time $t$.  For $k=0,\ldots,\ell$,
a generator $\bmg^{[t-k,t]}$ in vector basis $\calb^{t-k}$ of basis
$\bmcpb$ ends at time $t$.  We can use these generators to form a
vector basis $\calb_Y^t$.  The vector bases $\calb_Y^t$, for each
$t\in\bmcpz$, form a basis $\bmcpb_Y$, and we say $\bmcpb$ and
$\bmcpb_Y$ formed in this way have a {\it natural correspondence},
denoted $\bmcpb\equiv\bmcpb_Y$.  We can use $\bmcpb_Y$ to find a tensor set
$\calr_Y$ and we say $\calr\equiv\calr_Y$.  If $\bmcpb\equiv\bmcpb_Y$
and $\calr\equiv\calr_Y$, then there is a 1-1 correspondence 
$\calr\lra\calr_Y$ such that for each $\bmr\in\calr$, there is an
$\bmr_Y\in\calr_Y$ which uses the same sequence of generators.
In other words, $\bmr$ and $\bmr_Y$ are the same tensor, and we say
there is a natural correspondence $\bmr\equiv\bmr_Y$.

\begin{thm}
Fix basis $\bmcpb$ and tensor set $\calr$.  
We can find a basis $\bmcpb_Y$ and tensor set $\calr_Y$ such
that there is a natural correspondence $\bmcpb\equiv\bmcpb_Y$, 
$\calr\equiv\calr_Y$, and $\bmr\equiv\bmr_Y$ for each $\bmr\in\calr$.
\end{thm}
If $\bmr\equiv\bmr_Y$
then at each time $t$, the representatives in the \stms\ 
$R^{[t,t]}$ and $R_Y^{[t,t]}$ are the same aside from a change in index as
shown in (\ref{repcmp})-(\ref{repcmp1}).  In other words, a representative
$r_{j,k}^t$ in (\ref{rttf}) is the same as representative
$r_{Y,i,k}^t$ in (\ref{rttb}) when $j=k-i$, for $0\le j\le\ell$ 
and $j\le k\le\ell$.  We write this as $R^{[t,t]}\equiv R_Y^{[t,t]}$,
so if $\bmr\equiv\bmr_Y$, then $R^{[t,t]}\equiv R_Y^{[t,t]}$ at each time $t$.

Using a development dual to Theorem \ref{thm31} and Lemma \ref{lem90f}, 
we obtain the following theorem dual to Theorem \ref{thm90f}.

\begin{thm}
\label{thm90b}
For $0\le i\le\ell$, for $k$ such that $i\le k\le\ell$, let $[\Gamma_Y^{[t+i-k,t+i]}]$
be a set of generators $\{\bmg^{[t+i-k,t+i]}\}$ which is a transversal of 
$\Gamma_Y^{[t+i-k,t+i]}$.  The $(t+i)-i$-th components of generators 
$\bmg^{[t+i-k,t+i]}\in[\Gamma_Y^{[t+i-k,t+i]}]$ form a transversal
\be
\label{keyeqb}
[\{\chi^t(\bmg^{[(t+i)-k,(t+i)]})\}]=[\{r_{Y,i,k}^{(t+i)-i}\}]=[\{r_{Y,i,k}^t\}]
\ee
of
\be
\label{qg3a}
\frac{Y_{i-1}^t(Y_i^t\cap X_{k-i}^t)}{Y_{i-1}^t(Y_i^t\cap X_{k-i-1}^t)}
\ee
for $0\le i\le\ell$, for $i\le k\le\ell$.
The set of transversals, $[\{r_{Y,i,k}^t\}]$, for $0\le i\le\ell$, 
for $i\le k\le\ell$, forms a \compset\ 
for the normal chain of $B^t$ given by the $Y^{[t,t]}$ \stm.
\end{thm}

From (\ref{keyeqf}) we have
$$
\chi^t(\bmg^{[(t-j),(t-j)+k]})=r_{j,k}^{(t-j)+j},
$$
and from (\ref{keyeqb}) we have
$$
\chi^t(\bmg^{[(t+i)-k,(t+i)]})=r_{Y,i,k}^{(t+i)-i}.
$$
The generators $\bmg^{[(t-j),(t-j)+k]}$ and $\bmg^{[(t+i)-k,(t+i)]}$
have the same endpoints when $j=k-i$.  If $\bmr\equiv\bmr_Y$,
the generators are the same, and then $r_{j,k}^t=r_{Y,i,k}^t$ for $j=k-i$.
Then $R^{[t,t]}\equiv R_Y^{[t,t]}$.  Fix $i$ such that $0\le i\le\ell$.
Let $j=k-i$.  Then there is a 1-1 correspondence between
the set of transversals $[\{r_{j,k}^t\}]$ for $j\le k\le\ell$, and 
the set of transversals $[\{r_{Y,i,k}^t\}]$ for $i\le k\le\ell$, 
such that transversals with the same index $k$ are the same.

\begin{cor}
There is one set of transversals, either $[\{r_{j,k}^t\}]$ for $0\le j\le\ell$ 
and $j\le k\le\ell$, or $[\{r_{Y,i,k}^t\}]$ for $0\le i\le\ell$
and $i\le k\le\ell$, that forms a \compset\ for two normal chains, 
the normal chain of $B^t$ given by the $X^{[t,t]}$ \stm\ and
the normal chain of $B^t$ given by the $Y^{[t,t]}$ \stm.
\end{cor}

Note that for the \fssr, a generator $\bmg^{[t-j,t-j+k]}$ is selected
at time $t-j$, while for the \bssr, the same generator
$\bmg^{[t-j,t-j+k]}=\bmg^{[t+i-k,t+i]}$ where $j=k-i$, is selected at time
$t-j+k$.  Thus in both cases there is a causal collection of 
generators at time $t$.

We previously calculated a branch $b^t\in B^t$ using representatives in $R^{[t,t]}$
in (\ref{enctd}) and (\ref{enctd1}).  We now calculate a branch $b_Y^t\in B^t$ 
using representatives in $R_Y^{[t,t]}$.  Then
\be
\label{enctdrv}
b_Y^t=\prod_{i=0}^\ell \left(\prod_{k=i}^\ell r_{Y,i,k}^t\right).
\ee
By the convention used here, equation (\ref{enctdrv}) is evaluated as
\be
\label{enctdrv1}
b_Y^t=r_{Y,\ell,\ell}^t r_{Y,\ell-1,\ell}^t r_{Y,\ell-1,\ell-1}^t\cdots 
r_{Y,i,\ell}^t\cdots r_{Y,i,k}^t\cdots r_{Y,i,i}^t\cdots
r_{Y,2,2}^tr_{Y,1,\ell}^t\cdots r_{Y,1,1}^tr_{Y,0,\ell}^t\cdots r_{Y,0,2}^tr_{Y,0,1}^tr_{Y,0,0}^t.  
\ee
If $R^{[t,t]}\equiv R_Y^{[t,t]}$, then $r_{j,k}^t$ in (\ref{rttf}) 
is the same as $r_{Y,i,k}^t$ in (\ref{rttb}) when $j=k-i$, and 
we can rewrite $b_Y^t$ in terms of representatives $r_{j,k}^t$ in the
\fssr\ as
\be
\label{enctdrv2}
b_Y^t=r_{0,\ell}^t r_{1,\ell}^t r_{0,\ell-1}^t\cdots 
r_{\ell-i,\ell}^t\cdots r_{k-i,k}^t\cdots r_{0,i}^t\cdots
r_{0,2}^tr_{\ell-1,\ell}^t\cdots r_{0,1}^tr_{\ell,\ell}^t\cdots r_{2,2}^tr_{1,1}^tr_{0,0}^t.  
\ee

If $R^{[t,t]}\equiv R_Y^{[t,t]}$, 
product (\ref{enctdrv2}) is a rearrangement of product (\ref{enctd1}).
If $B^t$ is abelian, then rearrangements of the same terms give 
the same result, and then $b^t=b_Y^t$.  If $B^t$ is not abelian, this may
not be true.

\newpage
\vspace{3mm}
{\bf 5.  THE TIME DOMAIN ENCODER}
\vspace{3mm}

For $0\le j<\ell$, we know that in \sm\ $R^{[t,t+\ell]}$ there is
a column vector
\be
\bmr_j^{t+j}=\left(
\begin{array}{lllll}
r_{j,\ell}^{t+j} & \!\!\cdots\!\! & r_{j,k}^{t+j} & \!\!\cdots\!\! & r_{j,j}^{t+j} 
\end{array}
\right)^T,
\ee
and a column vector
\be
\bmr_{j+1}^{t+j+1}=\left(
\begin{array}{lllll}
r_{j+1,\ell}^{t+j+1} & \!\!\cdots\!\! & r_{j+1,k}^{t+j+1} & \!\!\cdots\!\! & r_{j+1,j+1}^{t+j+1} 
\end{array}
\right)^T.
\ee
Note that column $\bmr_{j+1}^{t+j+1}$ is completely determined by column 
$\bmr_j^{t+j}$.
Then we can think of $\bmr_{j+1}^{t+j+1}$ as a {\it shift} of $\bmr_j^{t+j}$.
For $0\le j\le\ell$, 
let $\bmcpr_j^{t+j},\bmcpr_{j+1}^{t+j+1}$ be the set of all columns
$\bmr_j^{t+j},\bmr_{j+1}^{t+j+1}$ in all possible \sms\ $\bmcpr^{[t,t+\ell]}$.
For $0\le j<\ell$, define a column shift map
$\bsigma: \bmcpr_j^{t+j} \ra \bmcpr_{j+1}^{t+j+1}$ by the assignment
$\bsigma: \bmr_j^{t+j}\ra\bmr_{j+1}^{t+j+1}$, where this 
assignment is given by 
$\sigma:  r_{j,k}^{t+j}\mapsto r_{j+1,k}^{t+j+1}$ for $j<k\le\ell$.
Note that $\sigma r_{j,j}^{t+j}$ is not defined since $r_{j,j}^{t+j}$ ``shifts out''.
We abbreviate $\bsigma(\bmr_j^{t+j})$ as $\bsigma \bmr_j^{t+j}$
and $\sigma(r_{j,k}^{t+j})$ as $\sigma r_{j,k}^{t+j}$.  
(The notation $\sigma r_{j,k}^{t+j}$ and $\bsigma\bmr_j^{t+j}$ is 
slightly inconsistent, but any ambiguity in $\sigma$ or $\bsigma$
is resolved by looking at its argument.
In addition $\sigma$ and $\bsigma$ should have a time index, but again this
ambiguity is resolved by looking at its argument.  Although somewhat
inconsistent and incomplete, this notation is simple and helps to clarify
the basic argument.)

Define
$$
\bsigma\bmr^t\rmdef 
(\bsigma\bmr_0^t,\bsigma\bmr_1^t,\ldots,\bsigma\bmr_j^t,\ldots,\bsigma\bmr_{\ell-1}^t,\bsigma\bmr_\ell^t).
$$

\begin{thm}
\label{thm29a}
Let $\bmw=\ldots,\bmr^t,\bmr^{t+1},\ldots$ be an arbitrary sequence, not
necessarily a tensor in $\calr$, where $\bmr^t\in\bmcpr^t$
for each time $t\in\bmcpz$.  Then $\bmw$ is a tensor in $\calr$
\ifof\ for each time $t$, $\bmr^{t+1}=(\bmr_0^{t+1},\bsigma\bmr^t)$ 
where input $\bmr_0^{t+1}$ is any element of $\bmcpr_0^{t+1}$.
\end{thm}

\begin{prf}
First assume $\bmw\in\calr$.  Then we know $\bmw$ is formed from a
sequence of \sms.  Consider $(\bmr^t,\bmr^{t+1})$ where $\bmr^t\in\bmcpr^t$
and $\bmr^{t+1}\in\bmcpr^{t+1}$.  Fix $0\le j<\ell$.  We know 
column $\bmr_j^t$ of $\bmr^t$ is a column $\bmr_j^{(t-j)+j}$ in \sm\
$R^{[(t-j),(t-j)+\ell]}$.  From the preceding discussion of shifts,
we know
\begin{align*}
\bsigma\bmr_j^t &=\bsigma\bmr_j^{(t-j)+j} \\
&=\bmr_{j+1}^{(t-j)+j+1} \\
&=\bmr_{j+1}^{t+1},
\end{align*}
where $\bmr_{j+1}^{(t-j)+j+1}$ is a column in \sm\ $R^{[(t-j),(t-j)+\ell]}$
and $\bmr_{j+1}^{t+1}$ is a column in $\bmr^{t+1}$.  Then
\begin{align*}
(\bmr_1^{t+1},\bmr_2^{t+1},\ldots,\bmr_{j+1}^{t+1},\ldots,\bmr_\ell^{t+1})
&=(\bsigma\bmr_0^t,\bsigma\bmr_1^t,\ldots,\bsigma\bmr_j^t,\ldots,\bsigma\bmr_{\ell-1}^t,\bsigma\bmr_\ell^t) \\
&=\bsigma\bmr^t
\end{align*}
and $\bmr^{t+1}=(\bmr_0^{t+1},\bsigma\bmr^t)$ where $\bmr_0^{t+1}\in\bmcpr^{t+1}$.

Conversely, if $\bmr^{t+1}=(\bmr_0^{t+1},\bsigma\bmr^t)$ for each $t\in\bmcpz$, 
then it can be shown $\bmw$ is a sequence of \sms, and therefore a tensor in 
$\calr$.
\end{prf}

Theorem \ref{thm29a} shows the tensor set $\calr$ has a natural shift structure.  
In the remainder of this section, we show that any path $\bmb\in C$ 
is the encoding of some $\bmr\in\calr$.  
Then the group trellis $C$ can be considered to
have a natural shift structure.  The fact that a group code $C$
has an encoder with a shift structure was first proven by Forney and Trott
\cite{FT} using a spectral domain encoder.  We prove $C$ has a
natural shift structure using a time domain approach.

An {\it encoder} of the group trellis is a finite state machine that,
given a sequence of inputs, can produce any path (any sequence of states
and branches) in the group trellis.  An encoder can help to explain the 
structure of a group trellis.  We give an encoder here which has a sliding
block structure and uses the same generators as in \cite{FT}, but the encoder
is different.  The encoder is given in (\ref{enctd}) and
(\ref{enctdb})-(\ref{enctdc}).  It is useful to think of (\ref{enctd}) and
(\ref{enctdb})-(\ref{enctdc}) as equivalent forms of the 
same encoder; each version is useful in the following discussion.

Assume we have found a basis $\bmcpb$.  Then we have found generators
$\bmg^{[t,t+k]}\in[\Gamma^{[t,t+k]}]$ for each $t\in\bmcpz$,
for $0\le k\le\ell$.  Fix time $t$.  The nontrivial components of the
selected generators in encoder (\ref{enctdc}) form a \gm\ $R^{[(t-j),(t-j)+\ell]}$, for
$j=0,\ldots,\ell$.  From Theorem \ref{thm31},  these \gms\
uniquely determine a \stm\  $R^{[t,t]}$, 
where column $j$ of \gm\
$R^{[(t-j),(t-j)+\ell]}$, $\bmr_j^{(t-j)+j}$, is column $j$ of 
\stm\  $R^{[t,t]}$, $\bmr_j^t$.  Then we can see that (\ref{rtnsr})
has the form of a sliding block encoder.
At each time $t$, we select a new \gm\ $R^{[(t),(t)+\ell]}$
whose column vectors are shown along the diagonals in (\ref{rtnsr}).
The column vectors $\bmr_j^{(t-j)+j}$ of the \gm\ at time $t-j$,
$$
R^{[(t-j),(t-j)+\ell]}=(\bmr_0^{(t-j)},\bmr_1^{(t-j)+1},\ldots,\bmr_j^{(t-j)+j},\ldots,\bmr_\ell^{(t-j)+\ell}),
$$
and column vectors $\bmr_j^{t+j}$ of the \gm\ at time $t$,
$$
R^{[(t),(t)+\ell]}=(\bmr_0^{(t)},\bmr_1^{(t)+1},\ldots,\bmr_j^{(t)+j},\ldots,\bmr_\ell^{(t)+\ell}),
$$
are shown along the diagonals of (\ref{rtnsr}).  As time increases, we slide
along the infinite matrix in (\ref{rtnsr}) from left to right.  
At time $t$, the output branch $b^t$ of the sliding block encoder 
is calculated from the \stm\
\begin{align*}
R^{[t,t]} &=(\bmr_0^{(t)},\bmr_1^{(t-1)+1},\ldots,\bmr_j^{(t-j)+j},\ldots,\bmr_\ell^{(t-\ell)+\ell}) \\
&=(\bmr_0^t,\bmr_1^t,\ldots,\bmr_j^t,\ldots,\bmr_\ell^t),
\end{align*}
whose terms are shown in the center row in (\ref{rtnsr}).  
The first term in the center row is the new {\it input} $\bmr_0^{(t)}$, 
the first column vector of the new \gm\ $R^{[(t),(t)+\ell]}$ selected at time $t$,
and the remaining terms 
$\bmr_1^{(t-1)+1},\ldots,\bmr_j^{(t-j)+j},\ldots,\bmr_\ell^{(t-\ell)+\ell}$
are from previous \gms\ selected at times
$t-1,\ldots,t-j,\ldots,t-\ell$, respectively.  To calculate
branch $b^t$ at time $t$, the sliding block encoder uses 
time window $[t-\ell,t]$, and therefore the encoder is causal.
We now show that we can use (\ref{rtnsr}) to implement (\ref{enctd})
as a sliding block encoder.

\begin{lem}
\label{lemx}
Fix $\bmr^t\in\bmcpr^t$.  Consider all 
$(\bmr^t,\bmr^{t+1})\in\bmcpr^t\times\bmcpr^{t+1}$ that appear in any
tensor $\bmr\in\calr$.  Then encoder (\ref{enctd}) encodes $(\bmr^t,\bmr^{t+1})$
into a trellis path segment $(b^t,b^{t+1})$ of length 2 in group 
trellis $C$.  In other words, $b^t\in B^t$, $b^{t+1}\in B^{t+1}$, and
$b^{t+1}\in\calf(b^t)$.
\end{lem}

\begin{prf}
Using (\ref{enctd}), the encoding of $(\bmr^t,\bmr^{t+1})$ is
\begin{multline}
\label{encdr0}
\left(\prod_{j=0}^\ell \left(\prod_{k=j}^\ell r_{j,k}^t\right),
\prod_{j=0}^\ell \left(\prod_{k=j}^\ell r_{j,k}^{t+1}\right)\right) \\
=\left(\prod_{j=0}^\ell \left(\prod_{k=j}^\ell r_{j,k}^t\right),
\left(\prod_{k=0}^\ell r_{0,k}^{t+1}\right)
\left(\prod_{j=0}^{\ell-1} \left(\prod_{k=j+1}^\ell r_{j+1,k}^{t+1}\right)\right)\right).
\end{multline}
We know that $\bmr^{t+1}$ is of the form $(\bmr_0^{t+1},\bsigma\bmr^t)$.
Then for $0\le j\le\ell-1$, $r_{j+1,k}^{(t+1)}$ is a shift of 
$r_{j,k}^{(t)}$.  Since 
$$
r_{j,k}^{(t)}=\chi^t(\bmg^{[t-j,t-j+k]}),
$$
then
$$
r_{j+1,k}^{(t+1)}=\chi^{t+1}(\bmg^{[t-j,t-j+k]}).
$$
This means that we can rewrite (\ref{encdr0}) in terms of generators
(see (\ref{enctdc})) as
\be
\label{encdr1}
\left(\prod_{j=0}^\ell \left(\prod_{k=j}^\ell \chi^t(\bmg^{[t-j,t-j+k]}) \right),
\left(\prod_{k=0}^\ell \chi^{t+1}(\bmg^{[t+1,t+1+k]}) \right)
\left(\prod_{j=0}^{\ell-1} \left(\prod_{k=j+1}^\ell \chi^{t+1}(\bmg^{[t-j,t-j+k]})\right)\right)\right).
\ee
Since $\chi^{t+1}(\bmg^{[t-j,t-j+k]})=\bone^{t+1}$ for $0\le j\le\ell$, we can
change the limits of the last double product in (\ref{encdr1}) as
\be
\label{encdr1a}
\left(\prod_{j=0}^\ell \left(\prod_{k=j}^\ell \chi^t(\bmg^{[t-j,t-j+k]}) \right),
\left(\prod_{k=0}^\ell \chi^{t+1}(\bmg^{[t+1,t+1+k]}) \right)
\left(\prod_{j=0}^\ell \left(\prod_{k=j}^\ell \chi^{t+1}(\bmg^{[t-j,t-j+k]}) \right)\right)\right).
\ee
Note that the term
$$
\left(\prod_{k=0}^\ell \chi^{t+1}(\bmg^{[t+1,t+1+k]})\right)
$$
involves generators from vector basis $\calb^{t+1}$, and the other
terms involve generators from
vector bases $\calb^{t-j}$ for $j=0,\ldots,\ell$.

First consider the case where $\bmr^{t+1}$ is 
$\bmrht^{t+1}=(\bone_0^{t+1},\bsigma\bmr^t)$.  Let $\bhat^{t+1}$ be
the encoding of $\bmrht^{t+1}$.  Since $\bmr_0^{t+1}=\bone_0^{t+1}$,
then components $r_{0,k}^{t+1}$ are the identity for $0\le k\le\ell$.  Then
we can rewrite (\ref{encdr1a}) as
\be
\label{encdr2}
(b^t,\bhat^{t+1})=\left(\prod_{j=0}^\ell \left(\prod_{k=j}^\ell \chi^t(\bmg^{[t-j,t-j+k]})\right),
\prod_{j=0}^\ell \left(\prod_{k=j}^\ell \chi^{t+1}(\bmg^{[t-j,t-j+k]})\right)\right).
\ee
Note that (\ref{encdr2}) just involve generators from
vector bases $\calb^{t-j}$ for $j=0,\ldots,\ell$.
We can pair terms in (\ref{encdr2}) as 
\be
\label{encdr3}
(b^t,\bhat^{t+1})=\prod_{j=0}^\ell \left(\prod_{k=j}^\ell [\chi^t(\bmg^{[t-j,t-j+k]}),
\chi^{t+1}(\bmg^{[t-j,t-j+k]})]\right),
\ee
where the product multiplication in the inner square bracket is component
by component, i.e., $[a,b]*[c,d]=[a*c,b*d]$.  But note that 
$$
[\chi^t(\bmg^{[t-j,t-j+k]}),\chi^{t+1}(\bmg^{[t-j,t-j+k]})]
$$
is a valid trellis path segment of length 2, for $0\le j\le\ell$, for 
$j\le k\le\ell$.  Then (\ref{encdr3}) is a product of trellis path segments
of length 2, and hence by properties of the group trellis,
$(b^t,\bhat^{t+1})$ is a trellis path segment of length 2.  This
means $\bhat^{t+1}\in\calf(b^t)$.

Now consider the case where $\bmr^{t+1}=(\bmr_0^{t+1},\bsigma\bmr^t)$.  
Let $b^{t+1}$ be the encoding of $\bmr^{t+1}$.  Then
using (\ref{encdr1a}) and (\ref{encdr2}), we have
\be
\label{encdr3a}
(b^t,b^{t+1})=
\left(b^t,\left(\prod_{k=0}^\ell \chi^{t+1}(\bmg^{[t+1,t+1+k]}) \right)
\bhat^{t+1}\right).
\ee
But
$$
\left(\prod_{k=0}^\ell \chi^{t+1}(\bmg^{[t+1,t+1+k]}) \right)=
\prod_{k=0}^\ell r_{0,k}^{t+1},
$$
and this is some branch $\btld^{t+1}\in X_0^{t+1}$.  Then 
$$
(b^t,b^{t+1})=
(b^t,\btld^{t+1}\bhat^{t+1})
$$
where $b^{t+1}=\btld^{t+1}\bhat^{t+1}$ and $b^{t+1}\in\calf(b^t)$.
\end{prf}

Notice that we can think of the encoder as an estimator.  The encoding
of $\bmrht^{t+1}=(\bone_0^{t+1},\bsigma\bmr^t)$ gives an initial estimate
$\bhat^{t+1}$ where $\bhat^{t+1}\in\calf(b^t)$.  Then at 
time $t+1$, we use new input $\bmr_0^{(t+1)}$ to find $\btld^{t+1}\in X_0^{t+1}$
to correct the initial estimate $\bhat^{t+1}$ so that 
$b^{t+1}=\btld^{t+1}\bhat^{t+1}$ and $b^{t+1}\in\calf(b^t)$.

\begin{cor}
\label{corb}
Fix $\bmr^t\in\bmcpr^t$.  Let $A_\calr^t$ be the set of components
$(\bmr^t,\bmr^{t+1})$ that appear in any $\bmr\in\calr$.
Let $\bmr^t$ encode to $b^t$ using (\ref{enctd}).  Let $A_C^t$ 
be the set of trellis path segments $(b^t,b^{t+1})$ of length 2 in $C$.
Encoder (\ref{enctd}) encodes $(\bmr^t,\bmr^{t+1})\in A_\calr^t$
into $(b^t,b^{t+1})\in A_C^t$.  This map is
1-1 and onto.
\end{cor}

\begin{prf}
We know from Lemma \ref{lemx} that $(\bmr^t,\bmr^{t+1})$ encodes to 
$(b^t,b^{t+1})$ using (\ref{enctd}).  Therefore (\ref{enctd}) maps
$A_\calr^t$ into $A_C^t$.  But $b^{t+1}$ is uniquely determined by 
$\bmr^{t+1}$, and specifically $\bmr_0^{t+1}$.  There are $|X_0^{t+1}|$
possible $\bmr_0^{t+1}$, and $|X_0^{t+1}|$ possible $b^{t+1}\in\calf(b^t)$.
Therefore the map from $A_\calr^t$ into $A_C^t$ is 1-1 and onto.
\end{prf}

\begin{cor}
\label{core}
Fix any $\bmb\in C$.  For any time $t$, consider a trellis path 
segment $(b^t,b^{t+1})$ of length 2 in group trellis $C$.
In other words, $b^t\in B^t$, $b^{t+1}\in B^{t+1}$, and
$b^{t+1}\in\calf(b^t)$.  Then there is some 
$(\bmr^t,\bmr^{t+1})\in\bmcpr^t\times\bmcpr^{t+1}$ such that 
$b^t$ decodes to $\bmr^t$ using (\ref{enctd}),
$b^{t+1}$ decodes to $\bmr^{t+1}$ using (\ref{enctd}), and 
$\bmr^{t+1}=(\bmr_0^{t+1},\bsigma\bmr^t)$.
\end{cor}

\begin{prf}
Use Corollary \ref{corb}.
\end{prf}

\begin{thm}
\label{thm29}
Each tensor $\bmr\in\calr$ can be encoded into a path $\bmb\in C$ 
using (\ref{enctd}).
\end{thm}

\begin{prf}
Lemma \ref{lemx} shows that for each time $t$, if $(\bmr^t,\bmr^{t+1})\in\bmr$,
then $(b^t,b^{t+1})$ is a trellis path segment of length 2 in $C$.
To show that we obtain a path $\bmb\in C$ using (\ref{enctd}), we have
to show the trellis path segments of length 2 can be connected.
Fix $\bmr$ and fix time $t$.  Then $(\bmr^t,\bmr^{t+1})$
gives a trellis path segment $(b^t,b^{t+1})$ of length 2.  Now use 
$(\bmr^{t+1},\bmr^{t+2})$ to obtain a trellis path segment 
$(\bhat^{t+1},b^{t+2})$ of length 2.  But the encoding of 
$\bmr^{t+1}$ using (\ref{enctd}) is unique so $b^{t+1}=\bhat^{t+1}$.
Therefore we have obtained a trellis path segment 
$(b^t,b^{t+1},b^{t+2})$ of length 3 in $C$.  

Continuing forward in this way, we can find a trellis path segment
$\bmb^{[t,\infty)}$ on $[t,\infty)$ in $C$.  Given $\bmr$,
the trellis path segment $\bmb^{[t,\infty)}$ is unique since for each time
$t$, (\ref{enctd}) is a unique function of $\bmr^t$.  But since we know how to
find a unique trellis path segment $\bmb^{[t,\infty)}$ on $[t,\infty)$ in $C$,
we can apply the same argument again starting with $\bmr^{t-1}$  
to find a unique trellis path segment $\bmbht^{[t-1,\infty)}$ on $[t-1,\infty)$.
Given $\bmr$, the trellis path segments $\bmb^{[t,\infty)}$ and $\bmbht^{[t-1,\infty)}$
must agree on $[t,\infty)$ since again (\ref{enctd}) is a unique function of 
$\bmr^t$.  Then we have found a unique trellis path segment on $[t-1,\infty)$.
Therefore, continuing in this way, we can encode 
$\bmr$ into a unique trellis path $\bmb\in C$ on $(-\infty,\infty)$.
\end{prf}

\begin{lem}
\label{lem30}
If tensor $\bmr\in\calr$ is encoded into $\bmb\in C$ 
using (\ref{enctd}), then $\bmr$ is the only tensor in $\calr$ that
encodes to $\bmb$ using (\ref{enctd}).
\end{lem}

\begin{prf}
Fix time $t$.  If $\bmr^t$ encodes to $b^t$ using (\ref{enctd}), 
$\bmr^t$ is unique because $b^t$ is a unique function of the \creps\ in $\bmr^t$
(see (\ref{enctd1})).  Since this holds for each $t$, $\bmr$ must be unique.
\end{prf}

\begin{thm}
\label{thm31a}
Each path $\bmb\in C$ can be decoded into a unique tensor $\bmr\in\calr$.
In other words, for each path $\bmb\in C$, there is a unique $\bmr\in\calr$ 
that can be encoded to $\bmb$ using (\ref{enctd}).
\end{thm}

\begin{prf}
The proof is analogous to the proof of Theorem \ref{thm29} but with
Corollary \ref{core} in place of Lemma \ref{lemx}.
\end{prf}

\begin{cor}
\label{cor36a}
There is a 1-1 correspondence $\calr\lra C$ given by $\bmr\lra\bmb$,
where $\bmb$ is an encoding of $\bmr$ using (\ref{enctd}).
\end{cor}

\begin{prf}
Combine Theorem \ref{thm29} and Theorem \ref{thm31a}.
\end{prf}

Consider the triple $(\calr,C;\bmcpb)$.  $\calr$ is a tensor set that
depends on choice of basis $\bmcpb$.  If $\bmcpb$ is fixed, then $\calr$
is fixed, and there is a 1-1 correspondence $\calr\lra C$.  Each 
$\bmr\in\calr$ can be encoded into a $\bmb\in C$, and each $\bmb\in C$
can be decoded into an $\bmr\in\calr$.  Note that the restriction
that $C$ be \ellctl\ is transparent from the structure of $\calr$.

We can reverse time in the argument just given for encoder $b^t$ in
(\ref{enctd}) and obain analogous results for encoder $b_Y^t$
in (\ref{enctdrv}).  In particular the analog of Corollary \ref{cor36a}
is the following.

\begin{cor}
\label{cor36arv}
There is a 1-1 correspondence $\calr_Y\lra C$ given by $\bmr_Y\lra\bmb$,
where $\bmb$ is an encoding of $\bmr_Y$ using (\ref{enctdrv}).
\end{cor}

We now review the encoder construction in \cite{FT}.
Forney and Trott \cite{FT} define the $k$-controllable subcode $\msfc_k$ 
of a group code $\msfc$.  We can transcribe their approach to the group
trellis $C$ used here.  The $k$-controllable subcode $C_k$ of a group trellis
$C$ is defined as the set of combinations of code sequences of 
span $k+1$ or less:
$$
C_k=\prod_t C^{[t,t+k]}.
$$
They show
$$
C_0\subset C_1\subset\ldots C_{k-1}\subset C_k\subset\ldots C_\ell=C
$$
is a normal series.  Then in their {\it Code Granule Theorem}, they 
show $C_k/C_{k-1}$ is isomorphic to a direct product,
$$
C_k/C_{k-1}\simeq\prod_t \Gamma^{[t,t+k]},
$$
where $\Gamma^{[t,t+k]}$ is a granule.
Let $\left[\Gamma^{[t,t+k}\right]=\{\bmg^{[t,t+k]}\}$ 
be a set of coset representatives for the granule $\Gamma^{[t,t+k]}$.
Then it follows (p.\ 1509) that the set 
$\prod_t \left[\Gamma^{[t,t+k]}\right]$
is a set of coset representatives for the cosets of $C_{k-1}$ in $C_k$.
This means ({\it Generator Theorem}) that every code sequence $\bmb$ 
can be uniquely expressed as a product
\be
\label{encft}
\bmb=\prod_{0\le k\le\ell} \prod_t \bmg^{[t,t+k]}
\ee
of generators $\bmg^{[t,t+k]}$.  Thus every code sequence $\bmb$
is a product of some sequence of generators, and conversely, every
sequence of generators corresponds to some code sequence $\bmb$.
It is clear that for any particular time $t$, only the generators 
$\bmg^{[t-j,t-j+k]}$ are relevant in calculating an output, for
$k$ such that $0\le k\le\ell$, for $0\le j\le k$.
Therefore the equation (\ref{encft}) can be realized as a minimal encoder
with a shift register structure, as discussed and diagrammed in \cite{FT}.
Using our notation, the output at time $t$, denoted as branch $b_s^t$ for the 
spectral domain encoder, is given by
\begin{align}
\label{encftc}
b_s^t
&=\prod_{k=0}^\ell \left(\prod_{j=k}^0 \chi^t(\bmg^{[t-j,t-j+k]})\right) \\
\label{encftb}
&=\prod_{k=0}^\ell \left(\prod_{j=k}^0 r_{j,k}^{(t-j)+j}\right) \\
\label{encfta}
&=\prod_{k=0}^\ell \left(\prod_{j=k}^0 r_{j,k}^t\right).
\end{align} 
The output at time $t$ for the time reversed spectral domain encoder,
denoted $b_{s,Y}^t$, is given by 
\be
\label{encftrv}
b_{s,Y}^t=\prod_{k=0}^\ell \left(\prod_{i=k}^0 r_{Y,i,k}^t\right).
\ee

We now compare the two forward time encoders (\ref{enctda}) and (\ref{encfta}).
In the Forney and Trott encoder (\ref{encfta}), for fixed $k$ the inner product 
(the term in parentheses of (\ref{encfta})) 
is a product of terms in a single row of the \gm, 
and the outer product can be considered to be a column product.
Thus we refer to the Forney and Trott encoder as a 
{\it column-row} encoder, and the product in (\ref{encfta}) as a
column-row product.
In (\ref{enctda}), for fixed $j$ the inner product 
(the term in parentheses of (\ref{enctda})) 
is a product of terms in a single column of the \gm, 
and the outer product can be considered to be a row product.
Then we refer to encoder (\ref{enctda}) as a {\it row-column} encoder,
and its product in (\ref{enctda}) as a row-column product.
This terminology points out a distinct difference 
between the two encoders.  However note that it is easy to transform
(\ref{enctda}) to (\ref{encfta}) by merely interchanging the
inner and outer product and then reverse ordering terms in each row.

We can observe an important feature of the encoder (\ref{enctdb})
or (\ref{enctdc}).  The term in the parentheses of (\ref{enctdb}) or
(\ref{enctdc}) is a column which is some function of time $t-j$, say $h_j^{t-j}$.  
Then $b^t=\prod_{j=0}^\ell h_j^{t-j}$.  Thus the encoder has the form
of a time convolution, reminiscent of a linear system.
The Forney-Trott encoder \cite{FT} and Loeliger-Mittelholzer encoder 
\cite{LM} do not have the form of a convolution.
The term in the parentheses of (\ref{encftb}) or
(\ref{encftc}) is some function of time $t-j$ but this term is a row.
Therefore the overall encoder, the column-row product, is not a time convolution. 
This is the reason we think of the encoder (\ref{enctdb}) or(\ref{enctdc}) 
given here as a {\it time domain encoder}, while the encoders in \cite{FT,LM} are
thought of as {\it spectral domain encoders}.

We have discussed four different encoders, the forward time domain
encoder giving $b^t$ in (\ref{enctd}), the backward time domain
encoder giving $b_Y^t$ in (\ref{enctdrv}), the forward spectral domain
encoder giving $b_s^t$ in (\ref{encfta}), and the backward spectral domain
encoder giving $b_{s,Y}^t$ in (\ref{encftrv}).  Each encoder encodes an
$\bmr\in\calr$ or $\bmr_Y\in\calr_Y$ into a path $\bmb\in C$, 
and each encoder gives a 1-1 correspondence $\calr\lra C$ by $\bmr\lra\bmb$, 
or a 1-1 correspondence $\calr_Y\lra C$ by $\bmr_Y\lra\bmb$.
We show how the four encoders are related in Subsection 6.3.

\newpage

\vspace{3mm}
{\bf 6.  THE NATURAL SHIFT STRUCTURE AND CANONIC STRUCTURE}
\vspace{3mm}

\vspace{3mm}
{\bf 6.1  The tensor set $\calr$}
\vspace{3mm}

We use the time domain encoder for forward time to show the group
trellis $C$ can be reduced to tensor set $\calr$.  We think of $\calr$ 
as a second canonic form, the {\it forward time canonic form} 
of a group system $\msfc$.  We show the tensor set $\calr$ has
a natural shift structure and is a natural shift register graph $\drbinf$
which is graph isomorphic to $C$.  The paths in $\drbinf$
are tensors in $\calr$.  Then we give a dual result using the 
time domain encoder for backward time and define the 
{\it backward time canonic form}.

Note that \stm\ $R^{[t,t]}$ has the triangular form (\ref{rttf}).
We now introduce a triangle notation to describe certain subsets
of entries in $\bmr^t=R^{[t,t]}$.
For $\bmr^t\in\bmcpr^t$, we let $\tridnrt{j}{k}{t}$ be the
entries in $\bmr^t$ specified by the triangle with
lower vertex $r_{j,k}^t$ and upper vertices $r_{j,\ell}^t$ 
and $r_{j+\ell-k,\ell}^t$.  These are the entries
$r_{m,n}^t$ where $m,n$ satisfy $k\le n\le\ell$ and $j\le m\le (j+n-k)$.  
Let $\tridncrt{j}{k}{t}$ be the set of all possible triangles
$\tridnrt{j}{k}{t}$, 
$\tridncrt{j}{k}{t}\rmdef\{\tridnrt{j}{k}{t} : \bmr^t\in\bmcpr^t\}$.

A path $\bmb$ in $C$ is
\be
\label{eqncpb}
,\ldots,b^{t-1},b^t,b^{t+1},\ldots,
\ee
where $b^{t-1}=(s^{t-1},a^{t-1},s^t)$, $b^t=(s^t,a^t,s^{t+1})$, 
and $b^{t+1}=(s^{t+1},a^{t+1},s^{t+2})$.
We know $B^t/X_0^t\simeq\Sigma^t$.  We rewrite path (\ref{eqncpb}) in $C$ as
\be
\label{eqncpa}
,\ldots,(b^tX_0^t,b^t,b^{t+1}X_0^{t+1}),(b^{t+1}X_0^{t+1},b^{t+1},b^{t+2}X_0^{t+2}),\ldots.
\ee
We let the rewritten paths in (\ref{eqncpa}) give trellis $C'$.  
Clearly $C'$ is graph isomorphic to $C$, written as $C'\simeq C$.

Now replace $b^t$ in (\ref{eqncpa}) with $\bmr^t$ that encodes to it
using (\ref{enctd}).  This gives path
$$
,\ldots,(b^tX_0^t,\bmr^t,b^{t+1}X_0^{t+1}),(b^{t+1}X_0^{t+1},\bmr^{t+1},b^{t+2}X_0^{t+2}),\ldots.
$$
Call this trellis $C''$.  Then $C''\simeq C'\simeq C$.

\begin{thm}
The labels $\ldots,\bmr^t,\bmr^{t+1},\ldots$ of paths in $C''$
are the paths in $\calr$.
\end{thm}

\begin{prf}
By Corollary \ref{cor36a},
there is a 1-1 correspondence $\calr\lra C$ given by $\bmr\lra\bmb$,
where $\bmb$ is an encoding of $\bmr$ using (\ref{enctd}).  
\end{prf}

The set of transversals, $[\{r_{j,k}^t\}]$, for $0\le j\le\ell$ and
$j\le k\le\ell$, forms a \compset\ 
for the normal chain of $B^t$ given by the $X^{[t,t]}$ \stm.
We can calculate any $b^t\in B^t$ using these representatives as
in (\ref{enctd})-(\ref{enctd1}).  In terms of these representatives note that 
$b^tX_0^t=g^tX_0^t$ where 
\be
\label{rep1}
g^t=r_{\ell,\ell}^t r_{\ell-1,\ell}^t r_{\ell-1,\ell-1}^t\cdots r_{1,2}^t r_{1,1}^t.
\ee
Then all edges $\bhat^t$ out of state $b^tX_0^t$
must have $\tridnrtht{1}{1}{t}=\tridnrt{1}{1}{t}$.  Then there is a 
1-1 correspondence 
$$
B^t/X_0^t\lra\tridncrt{1}{1}{t}
$$
given by
$$
g^tX_0^t\lra\tridnrt{1}{1}{t}.
$$
So we can define $\tridnrt{1}{1}{t}$ to be the left state or left vertex
of $\bmr^t$, and $\tridnrt{1}{1}{t+1}$ to be the right state or right vertex
of $\bmr^t$.  As a result we can replace paths in $C''$ with paths
\be
\label{tnsrr}
,\ldots,(\tridnrt{1}{1}{t},\bmr^t,\tridnrt{1}{1}{t+1}),
(\tridnrt{1}{1}{t+1},\bmr^{t+1},\tridnrt{1}{1}{t+2}),\ldots.
\ee
This gives trellis $C'''$.  Then $C'''$ is graph isomorphic to $C$, since 
$C'''\simeq C''\simeq C'\simeq C$.  We rename trellis $C'''$ as $\drbinf$.  
Then we have shown $\drbinf\simeq C$.

\begin{thm}
\label{thm41}
$\drbinf$ is a graph trellis of $\calr$ and $\drbinf$ is graph isomorphic 
to group trellis $C$, $\drbinf\simeq C$.  
The isomorphism maps vertices of $\drbinf$ to vertices
of $C$.
\end{thm}

Note that $B^t/X_0^t\simeq\Sigma^t$ is a group theoretic description
of the states of $C$, and $\tridncrt{1}{1}{t}$ is a set theoretic 
description of the same states in $\drbinf$.  The following result uses the
set theoretic description of states to show that $\drbinf$ is a 
shift register trellis.

\begin{thm}
\label{fct42}
Let $\bmr=\ldots,\bmr^t,\bmr^{t+1},\ldots$ be a path in $\calr$.
In graph trellis $\drbinf$, edge $\bmr^t=(\bmr_0^t,\rrt{1}{t})$ 
has left vertex $\tridnrt{1}{1}{t}$ in $\tridncrt{1}{1}{t}$ 
and right vertex $\tridnrt{1}{1}{t+1}$ in $\tridncrt{1}{1}{t+1}$.  
We have $\bmr^{t+1}=(\bmr_0^{t+1},\bsigma\bmr^t)$,
where $\bmr_0^{t+1}$ is a new input at time $t+1$, and columns 
$\bsigma\bmr^t=(\bmr_1^{t+1},\ldots,\bmr_\ell^{t+1})$ of $\bmr^{t+1}$ 
are a shift of columns $(\bmr_0^t,\ldots,\bmr_{\ell-1}^t)$ of $\bmr^t$, i.e., 
$\bsigma\bmr_j^t=\bmr_{j+1}^{t+1}$ for $0\le j\le\ell-1$.
Note that $\tridnrt{1}{1}{t+1}=\bsigma\bmr^t$, a shift of $\bmr^t$.
Therefore the right vertex of $\bmr^t$ is completely specified by $\bmr^t$.
\end{thm}

Theorem \ref{fct42} shows that $\drbinf$ is a shift register trellis.
We can think of graph trellis $\drbinf$ as composed of trellis sections $\drtbt$.
At each time $t$, $\drbinf$ is a bipartite graph
$\drtbt$ having edges $\bmr^t\in\bmcpr^t$, left vertices
$\tridnrt{1}{1}{t}$ in vertex set $\tridncrt{1}{1}{t}$, 
and right vertices $\tridnrt{1}{1}{t+1}=\bsigma\bmr^t$ in vertex set 
$\tridncrt{1}{1}{t+1}$.

\begin{thm}
\label{thm399}
$\drtbt$ is graph isomorphic to $B^t$ given by trellis section $T^t$ 
in group trellis $C$.
\end{thm}

At each time $t$, the graph isomorphism is given by mapping left vertex
$\tridnrt{1}{1}{t}$ of $\drtbt$ to state $s^t$ in $B^t$ corresponding to
coset $g^tX_0^t\in B^t/X_0^t\simeq\Sigma^t$, where $g^t$ is given in 
(\ref{rep1}), and mapping right vertex $\tridnrt{1}{1}{t+1}=\bsigma\bmr^t$
of $\drtbt$ to state $s^{t+1}$ in $B^t$ corresponding to
coset $g^{t+1}X_0^{t+1}\in B^{t+1}/X_0^{t+1}\simeq\Sigma^{t+1}$, 
where $g^{t+1}$ is analogous to $g^t$ and only depends on $\bsigma\bmr^t$.  
And finally mapping
edge $\bmr^t$ in $\drtbt$ to edge $b^t$ in $B^t$, where $b^t$ is 
determined from $\bmr^t$ using encoding (\ref{enctd}).

We now describe two encoders of $\calr$, or equivalently $\drbinf$,
for forward time.  $\drbinf$ consists of sequences of the
form (\ref{tnsrr}).  We define a time domain encoder $\edrbinf$
of $\drbinf$ by replacing sequences of the form (\ref{tnsrr}) 
with sequences of the form
\be
\label{tnsrr1}
,\ldots,(\tridnrt{1}{1}{t},b^t,\tridnrt{1}{1}{t+1}),
(\tridnrt{1}{1}{t+1},b^{t+1},\tridnrt{1}{1}{t+2}),\ldots,
\ee
where $b^t$ is an encoding of $\bmr^t$ using time domain encoder (\ref{enctd}).
We define a spectral domain encoder $\edrbinfs$
of $\drbinf$ by replacing sequences of the form (\ref{tnsrr}) 
with sequences of the form
\be
\label{tnsrr2}
,\ldots,(\tridnrt{1}{1}{t},b_s^t,\tridnrt{1}{1}{t+1}),
(\tridnrt{1}{1}{t+1},b_s^{t+1},\tridnrt{1}{1}{t+2}),\ldots,
\ee
where $b_s^t$ is an encoding of $\bmr^t$ using spectral domain encoder 
(\ref{encfta}).

The Forney-Trott encoder in \cite{FT} is an encoding of 
the sequence $\ldots,\bmr^t,\bmr^{t+1},\ldots$ into 
$\ldots,b_s^t,b_s^{t+1},\ldots$, where 
$\bmr^{t+1}=(\bmr_0^{t+1},\bsigma\bmr^t)$ is composed of an
input $\bmr_0^{t+1}$ and a shift $\bsigma\bmr^t$ of $\bmr^t$.  Their encoder is
of the form state, input, shift to next state, next input, and so on.  
The states of their encoder are set theoretic constructions and appear 
to have no group theoretic interpretation in the spectral domain.
The state of their encoder at time $t$ can be regarded as
$\tridnrt{1}{1}{t}$ as in (\ref{tnsrr2}), and the state at time $t+1$
as $\tridnrt{1}{1}{t+1}$ as in (\ref{tnsrr2}).
Therefore the encoder $\edrbinfs$ is an exact replica
of the Forney-Trott encoder.  The Forney-Trott encoder is a minimal realization
of $\msfc$ and all minimal realizations are graph isomorphic to the 
canonic realization, or group trellis $C$ \cite{FT}.  Therefore $\edrbinfs$
is graph isomorphic to $C$.  This gives the following result.

\begin{thm}
\label{thm41x}
The time domain encoder $\edrbinf$ and spectral domain encoder $\edrbinfs$ 
are graph isomorphic to group trellis $C$.
The isomorphism maps vertices of $\drbinf$ to vertices
of $C$.
\end{thm}

We now give the dual result for backward time.
A path $\bmb$ in $C$ is given in (\ref{eqncpb}).
We know $B^t/Y_0^t\simeq\Sigma^{t+1}$.  We rewrite path (\ref{eqncpb}) in $C$ as
\be
\label{eqncpaa}
,\ldots,(b_Y^{t-1}Y_0^{t-1},b_Y^{t-1},b_Y^tY_0^t),(b_Y^tY_0^t,b_Y^t,b_Y^{t+1}Y_0^{t+1}),\ldots.
\ee
We let the rewritten paths in (\ref{eqncpaa}) give trellis $C_Y'$.  
Clearly $C_Y'$ is graph isomorphic to $C$, written as $C_Y'\simeq C$.

Now replace $b_Y^t$ in (\ref{eqncpaa}) with $\bmr_Y^t$ that encodes to it
using (\ref{enctdrv}).  This gives path
$$
,\ldots,(b_Y^{t-1}Y_0^{t-1},\bmr_Y^{t-1},b_Y^tY_0^t),(b_Y^tY_0^t,\bmr_Y^t,b_Y^{t+1}Y_0^{t+1}),\ldots.
$$
Call this trellis $C_Y''$.  Then $C_Y''\simeq C_Y'\simeq C$.

\begin{thm}
The labels $\ldots,\bmr_Y^{t-1},\bmr_Y^t,\ldots$ of paths in $C_Y''$
are the paths in $\calr_Y$.
\end{thm}

\begin{prf}
By Corollary \ref{cor36arv},
there is a 1-1 correspondence $\calr_Y\lra C$ given by $\bmr_Y\lra\bmb$,
where $\bmb$ is an encoding of $\bmr_Y$ using (\ref{enctdrv}).  
\end{prf}

The set of transversals, $[\{r_{Y,j,k}^t\}]$, for $0\le j\le\ell$ and
$j\le k\le\ell$, forms a \compset\ 
for the normal chain of $B^t$ given by the $Y^{[t,t]}$ \stm.
We can calculate any $b_Y^t\in B^t$ using these representatives as
in (\ref{enctdrv})-(\ref{enctdrv1}).  In terms of these representatives 
note that $b_Y^tY_0^t=hY_0^t$ where
\be
\label{rep2}
h=r_{Y,\ell,\ell}^t r_{Y,\ell-1,\ell}^t r_{Y,\ell-1,\ell-1}^t\cdots r_{Y,1,2}^t r_{Y,1,1}^t.
\ee
Then all edges $\bhat_Y^t$ into state $b_Y^tY_0^t$
must have $\tridnrthty{0}{1}{t}=\tridnrty{0}{1}{t}$.  Then there is a 
1-1 correspondence 
$$
B^t/Y_0^t\lra\tridncrty{0}{1}{t}
$$
given by
$$
hY_0^t\lra\tridnrty{0}{1}{t}.
$$
So we can define $\tridnrty{0}{1}{t}$ to be the right state or right vertex
of $\bmr_Y^t$, and $\tridnrty{0}{1}{t-1}$ to be the left state or left vertex
of $\bmr_Y^t$.  As a result we can replace paths in $C_Y''$ with paths
\be
\label{tnsrry}
,\ldots,(\tridnrty{0}{1}{t-2},\bmr_Y^{t-1},\tridnrty{0}{1}{t-1}),
(\tridnrty{0}{1}{t-1},\bmr_Y^t,\tridnrty{0}{1}{t}),\ldots.
\ee
This gives trellis $C_Y'''$.  Then $C_Y'''$ is graph isomorphic to $C$, since 
$C_Y'''\simeq C_Y''\simeq C_Y'\simeq C$.  We rename trellis $C_Y'''$ as $\drbinfy$.  
Then we have shown $\drbinfy\simeq C$.

\begin{thm}
\label{thm41rv}
$\drbinfy$ is a graph trellis of $\calr_Y$ and $\drbinfy$ is graph isomorphic 
to group trellis $C$, $\drbinfy\simeq C$.
The isomorphism maps vertices of $\drbinfy$ to vertices
of $C$.
\end{thm}

There are analogies of Theorems \ref{fct42} and \ref{thm399}
which show $\drbinfy$ is a shift register trellis with trellis section
$\drtbty$ graph isomorphic to $B^t$ at each time $t$.

We now describe two encoders of $\calr_Y$, or equivalently $\drbinfy$,
for backward time.  $\drbinfy$ consists of sequences of the
form (\ref{tnsrry}).  We define a time domain encoder $\edrbinfy$
of $\drbinfy$ by replacing sequences of the form (\ref{tnsrry}) 
with sequences of the form
\be
\label{tnsrry1}
,\ldots,(\tridnrty{0}{1}{t-2},b_Y^{t-1},\tridnrty{0}{1}{t-1}),
(\tridnrty{0}{1}{t-1},b_Y^t,\tridnrty{0}{1}{t}),\ldots.
\ee
where $b_Y^t$ is an encoding of $\bmr_Y^t$ using time domain encoder 
(\ref{enctdrv}).  We define a spectral domain encoder $\edrbinfsy$
of $\drbinfy$ by replacing sequences of the form (\ref{tnsrry}) 
with sequences of the form
\be
\label{tnsrry2}
,\ldots,(\tridnrty{0}{1}{t-2},b_{s,Y}^{t-1},\tridnrty{0}{1}{t-1}),
(\tridnrty{0}{1}{t-1},b_{s,Y}^t,\tridnrty{0}{1}{t}),\ldots.
\ee
where $b_{s,Y}^t$ is an encoding of $\bmr_Y^t$ using spectral domain encoder 
(\ref{encftrv}).

\begin{thm}
\label{thm41y}
The time domain encoder $\edrbinfy$ and spectral domain encoder $\edrbinfsy$ 
are graph isomorphic to group trellis $C$.
The isomorphism maps vertices of $\drbinfy$ to vertices
of $C$.
\end{thm}

\vspace{3mm}
{\bf 6.2  The tensor set $\calu$}  
\vspace{3mm}

We now describe a tensor set $\calu$ that is closely related to $\calr$.
The advantage of $\calu$ is that it is independent of basis $\bmcpb$.
There is a 1-1 correspondence $\calu\lra\calr$ for any basis $\bmcpb$.

Prevously we defined a vector basis $\calb^t$ using representatives
$\bmg^{[t,t+k]}$ of quotient group $\gamttpk$ for $0\le k\le\ell$.
We now number the cosets of $\gamttpk$ and assign an integer sequence
to \gvec\ $\bmr^{[t,t+k]}$ of $\bmg^{[t,t+k]}$.  Let integer $Q_k^t$
be the number of cosets in $\gamttpk$.  We number the cosets of
$\gamttpk$ with integers $q_k^t$ in the set $\{0,1,\ldots,|Q_k^t|-1\}$.
Define the map $\tau_k^t:  \gamttpk\ra\{0,1,\ldots,|Q_k^t|-1\}$ such that
if coset $\gamma_k^t\in\gamttpk$, then $\gamma_k^t$ is assigned an
integer $q_k^t$ in the set $\{0,1,\ldots,|Q_k^t|-1\}$; this gives assignment 
$\tau_k^t:  \gamma_k^t\mapsto q_k^t$.  The numbering is arbitrary
except we number the identity coset with integer 0.

Fix basis $\bmcpb$.  Let $\bmg^{[t,t+k]}$ be the representative of a 
coset $\gamma_k^t$ in $\gamttpk$ numbered with $q_k^t$.  We assign a 
constant integer sequence $\bmu^{[t,t+k]}$, 
\be
\bmu^{[t,t+k]}\rmdef (u_{0,k}^t, u_{1,k}^{t+1},\ldots,u_{j,k}^{t+j},\ldots,u_{k,k}^{t+k}),
\ee
to \gvec\ $\bmr^{[t,t+k]}$ of $\bmg^{[t,t+k]}$, where $u_{j,k}^t=q_k^t$ 
for $0\le j\le k$.  For $0\le k\le\ell$ and $0\le j\le k$,
$u_{j,k}^{t+j}$ is an integer in the set of integers 
$U_{j,k}^{t+j}\rmdef\{0,1,\ldots,|Q_k^t|-1\}$.  Then we define the map
$$
\lambda_{\bmcpb,k}^t:  [\gamttpk]\ra U_{0,k}^t\times U_{1,k}^{t+1}\times\cdots\times U_{j,k}^{t+j}\times\cdots\times U_{k,k}^{t+k}
$$
with assignment $\lambda_{\bmcpb,k}^t:  \bmr^{[t,t+k]}\mapsto\bmu^{[t,t+k]}$.
Then in place of \gm\ $R^{[t,t+\ell]}$ in (\ref{gmr}),
we can define a {\it shift matrix} $U^{[t,t+\ell]}$ shown in (\ref{smu}).  
The \sm\ $U^{[t,t+\ell]}$ is the same as \gm\ $R^{[t,t+\ell]}$
in (\ref{gmr}) with $r$ replaced by $u$.  The $k$-th row of matrix 
$U^{[t,t+\ell]}$, $0\le k\le\ell$ is a {\it shift vector} $\bmu^{[t,t+k]}$
which is the constant integer sequence assigned to row $\bmr^{[t,t+k]}$
of $R^{[t,t+\ell]}$.
\be
\label{smu}
\begin{array}{llllllllll}
  u_{0,\ell}^t   & u_{1,\ell}^{t+1}   & \cdots & \cdots & u_{j,\ell}^{t+j}   & \cdots & \cdots & \cdots & u_{\ell-1,\ell}^{t+\ell-1}   & u_{\ell,\ell}^{t+\ell} \\
  u_{0,\ell-1}^t & u_{1,\ell-1}^{t+1} & \cdots & \cdots & u_{j,\ell-1}^{t+j} & \cdots & \cdots & \cdots & u_{\ell-1,\ell-1}^{t+\ell-1} & \\
  \vdots & \vdots & \vdots & \vdots & \vdots & \vdots & \vdots & \vdots && \\
  u_{0,k}^t & u_{1,k}^{t+1} & \cdots & \cdots & u_{j,k}^{t+j} & \cdots & u_{k,k}^{t+k} &&& \\
  \vdots & \vdots & \vdots & \vdots & \vdots & \vdots &&&& \\
  \cdots & \cdots & \cdots & \cdots & u_{j,j}^{t+j} &&&&& \\
  \vdots & \vdots & \vdots &&&&&&& \\
  u_{0,2}^t   & u_{1,2}^{t+1} & u_{2,2}^{t+2} &&&&&&& \\
  u_{0,1}^t   & u_{1,1}^{t+1} &&&&&&&& \\
  u_{0,0}^t   &&&&&&&&&
\end{array}
\ee
We define $\bmu_j^{t+j}$ to be a column vector in (\ref{smu}), 
for $0\le j\le\ell$, where
$$
\bmu_j^{t+j}\rmdef\left(
\begin{array}{lllll}
u_{j,\ell}^{t+j} & \!\!\cdots\!\! & u_{j,k}^{t+j} & \!\!\cdots\!\! & u_{j,j}^{t+j} 
\end{array}
\right)^T.
$$
Then we can rewrite (\ref{smu}) as
\be
\label{smu1}
U^{[t,t+\ell]}=(\bmu_0^t,\bmu_1^{t+1},\ldots,\bmu_j^{t+j},\ldots,\bmu_\ell^{t+\ell}).
\ee

Fix tensor $\bmr\in\calr$.  We know $\bmr$ is defined by the collection
of \gvecs\ $\{\bmr^{[t,t+k]}:  0\le k\le\ell, t\in\bmcpz\}$.  Using
the 1-1 correspondence given by 
$\lambda_{\bmcpb,k}^t:  \bmr^{[t,t+k]}\mapsto\bmu^{[t,t+k]}$,
for $0\le k\le\ell$, for each $t\in\bmcpz$, gives a collection
of \svecs\ $\{\bmu^{[t,t+k]}:  0\le k\le\ell, t\in\bmcpz\}$.
This collection defines a {\it coset tensor} $\bmu$, shown in (\ref{utnsr}), 
which corresponds to tensor $\bmr$ in (\ref{rtnsr}).  
Let map $\blambda_\bmcpb$ give the
assignment $\blambda_\bmcpb:  \bmr\mapsto\bmu$.  Let $\calu$ be the tensor
set of all tensors $\bmu$ that can be constructed from $\bmr\in\calr$
in this way.
\be
\label{utnsr}
\setcounter{MaxMatrixCols}{7}
\begin{pmatrix}   
               &              &             & \vdots          & &          &             \\
               &              &             &                 & &          & \bmu_\ell^{(t)+\ell}  \\
               &              &             &                 & &          & \vdots      \\
               &              & \cdots      & \bmu_j^{(t)+j}  & \cdots &   &             \\
               &              &             & \vdots          & &          & \bmu_\ell^{(t-j)+\ell}  \\
\cdots         & \bmu_1^{(t)+1} & \cdots    &                 & &          &  \vdots     \\
\bmu_0^{(t)}   & \bmu_1^{(t-1)+1} & \cdots  & \bmu_j^{(t-j)+j}  & \cdots & & \bmu_\ell^{(t-\ell)+\ell} \\
\vdots         & \vdots       &             & \vdots          & &          & \vdots      \\
\cdots         & \bmu_1^{(t-j)+1} & \cdots  &                 & &          &             \\
\bmu_0^{(t-j)} & \cdots       &             &                 & &          &             \\
               &              &             & \vdots          & &          &
\end{pmatrix}
\ee

\begin{thm}
\label{thm33}
For a given basis $\bmcpb$, there is a 1-1 correspondence 
$\lambda_{\bmcpb,k}^t$ between \svecs\ $\bmu^{[t,t+k]}$ in $\bmu\in\calu$
and \gvecs\ $\bmr^{[t,t+k]}$ in $\bmr\in\calr$,
and therefore between \sms\ $U^{[t,t+\ell]}$ in $\bmu\in\calu$
and \gms\ $R^{[t,t+\ell]}$ in $\bmr\in\calr$.  This gives a 
1-1 correspondence $\blambda_\bmcpb$ between tensors $\bmu\in\calu$ and
tensors $\bmr\in\calr$, $\blambda_\bmcpb:  \bmr\mapsto\bmu$, and
therefore between tensors $\bmu\in\calu$ and paths $\bmb\in C$.
\end{thm}

Each basis $\bmcpb$ determines a tensor set $\calr$ and a map
$\blambda_\bmcpb:  \calr\ra\calu$.  As $\bmcpb$ changes, $\calr$ changes and
$\blambda_\bmcpb$ changes, but $\calu$ does not change.
Consider the 4-tuple $(\calu,\calr,C;\bmcpb)$ that includes the
triple $(\calr,C;\bmcpb)$ previously discussed in Section 5.  $\calr$
depends on choice of basis $\bmcpb$ but $\calu$ does not.  For any
basis $\bmcpb$, the map $\blambda_\bmcpb$ gives a 1-1 correspondence 
$\calu\lra\calr$.  If $\bmcpb$ is fixed, then $\calr$ is fixed, and there is a 
1-1 correspondence $\calu\lra\calr\lra C$.

The superscript parentheses terms in (\ref{utnsr}), like $(t-j)$,
indicate terms that all belong to the same \sm.  For example,
the diagonal terms 
$\bmu_0^{(t-j)},\bmu_1^{(t-j)+1},\ldots,\bmu_j^{(t-j)+j},\ldots,\bmu_\ell^{(t-j)+\ell}$
all belong to the \sm\ starting at time $t-j$, $U^{[(t-j),(t-j)+\ell]}$.  
The center row in (\ref{utnsr}) is
\be
\label{ctrrow11}
(\bmu_0^{(t)},\bmu_1^{(t-1)+1},\ldots,\bmu_j^{(t-j)+j},\ldots,\bmu_\ell^{(t-\ell)+\ell}),
\ee
where each entry is itself a column; this reduces to 
\be
\label{ctrrow22}
(\bmu_0^t,\bmu_1^t,\ldots,\bmu_j^t,\ldots,\bmu_\ell^t),
\ee
which is just the {\it static matrix} $U^{[t,t]}$.  Notice 
that each term in (\ref{ctrrow11}) and (\ref{ctrrow22}) is from one of 
$\ell+1$ different \sms.

\begin{thm}
\label{thm31u}
Fix time $t$.  A finite sequence of $\ell+1$ \sms\ $U^{[(t-j),(t-j)+\ell]}$ at
times $t-j$, for $j=0,\ldots,\ell$, uniquely determines a \stm\ 
$U^{[t,t]}$, where column $j$ of \sm\
$U^{[(t-j),(t-j)+\ell]}$, denoted $\bmu_j^{(t-j)+j}$, is column $j$ of 
\stm\ $U^{[t,t]}$, denoted $\bmu_j^t$.
\end{thm}

The \stm\ $U^{[t,t]}$ is the same as \stm\ $R^{[t,t]}$
in (\ref{rttf}) with $r$ replaced by $u$.   We define $\bmu^t$ to be a \stm\
$U^{[t,t]}$, or $\bmu^t\rmdef U^{[t,t]}$.
The set of all \stms\ $\bmu^t=U^{[t,t]}$ is the set $\bmcpu^t$ of all
triangular matrices of $\ell+1$ rows and $\ell+1$ columns
over the sets $U_{j,k}^t$, $0\le j\le k$, $0\le k\le\ell$.
Let $\bmcpu_j^t$ be the set of all $j$-th columns
of $\bmcpu^t$, $0\le j\le\ell$.  Then for $\bmu^t\in\bmcpu^t$,
we have $\bmu^t=(\bmu_0^t,\bmu_1^t,\ldots,\bmu_j^t,\ldots,\bmu_\ell^t)$, 
where $\bmu_j^t$ is a column in $\bmcpu_j^t$.
We denote any column $\bmu_j^t$ with all entries $0$ by $\bzero_j^t$.

\begin{thm}
\label{thm42}
For a given basis $\bmcpb$, there is
a 1-1 correspondence between \stms\ $\bmu^t=U^{[t,t]}\in\bmcpu^t$ 
and \stms\ $\bmr^t=R^{[t,t]}\in \bmcpr^t$, induced by the 1-1 correspondence 
between \sms\ $U^{[(t-j),(t-j)+\ell]}$ and \gms\ $R^{[(t-j),(t-j)+\ell]}$ 
at times $t-j$, for $j=0,\ldots,\ell$.
\end{thm}

We define a shift property of tensor $\bmu$
that mimics the shift property of tensor $\bmr$.
For $0\le j\le\ell$, let $\bmcpu_j^{t+j},\bmcpu_{j+1}^{t+j+1}$ be the set of all 
columns $\bmu_j^{t+j},\bmu_{j+1}^{t+j+1}$ in all possible \sms\ $\bmcpu^{[t,t+\ell]}$.
For $0\le j<\ell$, define a column shift map 
$\bsigma: \bmcpu_j^{t+j} \ra \bmcpu_{j+1}^{t+j+1}$ by the assignment
$\bsigma: \bmu_j^{t+j}\ra\bmu_{j+1}^{t+j+1}$, where this 
assignment is given by 
$\sigma:  u_{j,k}^{t+j}\mapsto u_{j+1,k}^{t+j+1}$ for $j<k\le\ell$.
Note that $\sigma u_{j,j}^{t+j}$ is not defined since $u_{j,j}^{t+j}$ ``shifts out''.
We abbreviate $\bsigma(\bmu_j^{t+j})$ as $\bsigma \bmu_j^{t+j}$
and $\sigma(u_{j,k}^{t+j})$ as $\sigma u_{j,k}^{t+j}$.  
Define
$$
\bsigma\bmu^t\rmdef 
(\bsigma\bmu_0^t,\bsigma\bmu_1^t,\ldots,\bsigma\bmu_j^t,\ldots,\bsigma\bmu_{\ell-1}^t,\bsigma\bmu_\ell^t).
$$
We have used the same shift notation in $\bsigma\bmr^t$ and $\bsigma\bmu^t$
but the meaning is clear by context.  We obtain the following result for
$\calu$ in the same way as Theorem \ref{thm29a} is obtained for $\calr$.

\begin{thm}
\label{thm29aa}
Let $\bmw=\ldots,\bmu^t,\bmu^{t+1},\ldots$ be an arbitrary sequence, not
necessarily a tensor in $\calu$, where $\bmu^t\in\bmcpu^t$
for each time $t\in\bmcpz$.  Then $\bmw$ is a tensor in $\calu$
\ifof\ for each time $t$, $\bmu^{t+1}=(\bmu_0^{t+1},\bsigma\bmu^t)$ 
where input $\bmu_0^{t+1}$ is any element of $\bmcpu_0^{t+1}$.
\end{thm}

Theorem \ref{thm29aa} shows that $\calu$ has a natural shift structure
in the same way that $\calr$ does.  In Subsection 6.1, we interpreted a 
tensor $\bmr\in\calr$ as a path in graph trellis $\drbinf$, given by 
(\ref{tnsrr}).  We define $\tridnut{j}{k}{t}$ and $\tridncut{j}{k}{t}$ 
in analogous way to $\tridnrt{j}{k}{t}$ and $\tridncrt{j}{k}{t}$.
In the same way as for $\calr$, we can interpret a
tensor $\bmu\in\calu$ as a path in a graph trellis $\duinf$, given by 
\be
\label{tnsru}
,\ldots,(\tridnut{1}{1}{t},\bmu^t,\tridnut{1}{1}{t+1}),
(\tridnut{1}{1}{t+1},\bmu^{t+1},\tridnut{1}{1}{t+2}),\ldots.
\ee
We have the following analogy to Theorem \ref{fct42}.

\begin{thm}
\label{fct42u}
Let $\bmu=\ldots,\bmu^t,\bmu^{t+1},\ldots$ be a path in $\calu$.
In graph trellis $\duinf$, edge $\bmu^t=(\bmu_0^t,\uut{1}{t})$ 
has left vertex $\tridnut{1}{1}{t}$ in $\tridncut{1}{1}{t}$ 
and right vertex $\tridnut{1}{1}{t+1}$ in $\tridncut{1}{1}{t+1}$.  
We have $\bmu^{t+1}=(\bmu_0^{t+1},\bsigma\bmu^t)$,
where $\bmu_0^{t+1}$ is a new input at time $t+1$, and columns 
$\bsigma\bmu^t=(\bmu_1^{t+1},\ldots,\bmu_\ell^{t+1})$ of $\bmu^{t+1}$ 
are a shift of columns $(\bmu_0^t,\ldots,\bmu_{\ell-1}^t)$ of $\bmu^t$, i.e., 
$\bsigma\bmu_j^t=\bmu_{j+1}^{t+1}$ for $0\le j\le\ell-1$.
Note that $\tridnut{1}{1}{t+1}=\bsigma\bmu^t$, a shift of $\bmu^t$.
Therefore the right vertex of $\bmu^t$ is completely specified by $\bmu^t$.
\end{thm}

Theorem \ref{fct42u} shows that $\duinf$ is a shift register trellis.
We can think of graph trellis $\duinf$ as composed of trellis sections $\dut$.
At each time $t$, $\duinf$ is a bipartite graph
$\dut$ having edges $\bmu^t\in\bmcpu^t$, left vertices
$\tridnut{1}{1}{t}$ in vertex set $\tridncut{1}{1}{t}$, 
and right vertices $\tridnut{1}{1}{t+1}=\bsigma\bmu^t$ in vertex set 
$\tridncut{1}{1}{t+1}$.

For each path (\ref{tnsrr}) in $\drbinf$, there is a path (\ref{tnsru}) 
in $\duinf$ induced by the 1-1 correspondence $\blambda_\bmcpb$.

\begin{thm}
\label{thm400}
$\duinf$ is graph isomorphic to $\drbinf$.
The graph isomorphism is given by the 1-1 correspondence
$\bmu\lra\bmr$ induced by $\blambda_\bmcpb$,
$$
\ldots,(\tridnut{1}{1}{t},\bmu^t,\tridnut{1}{1}{t+1}),\ldots
\lra \ldots,(\tridnrt{1}{1}{t},\bmr^t,\tridnrt{1}{1}{t+1}),\ldots.
$$  
Then we write $\duinf\simeq\drbinf$.  
The graph isomorphism maps vertices of $\duinf$ to vertices
of $\drbinf$.  For each time $t$, 
the graph isomorphism $\duinf\simeq\drbinf$ is given
by the graph isomorphism $\dut\simeq\drtbt$, where the 1-1 correspondence of 
branches and states is induced by $\blambda_\bmcpb^t$,
the time $t$ component of $\blambda_\bmcpb$.
\end{thm}

We can reverse time in the preceding results and obtain dual results
for $\bmcpb_Y$, $\calr_Y$, $\calu_Y$, and $\blambda_{\bmcpb_Y}$.  The dual
of Theorem \ref{thm400} is the following.

\begin{thm}
\label{thm400y}
$\duinfy$ is graph isomorphic to $\drbinfy$.
The graph isomorphism is given by the 1-1 correspondence
$\bmu_Y\lra\bmr_Y$ induced by $\blambda_{\bmcpb_Y}$,
$$
\ldots,(\tridnuty{0}{1}{t-1},\bmu_Y^t,\tridnuty{0}{1}{t}),\ldots
\lra \ldots,(\tridnrty{0}{1}{t-1},\bmr_Y^t,\tridnrty{0}{1}{t}),\ldots.
$$
Then we write $\duinfy\simeq\drbinfy$.  
The graph isomorphism maps vertices of $\duinfy$ to vertices
of $\drbinfy$.  For each time $t$, the graph isomorphism 
$\duinfy\simeq\drbinfy$ is given
by the graph isomorphism $\duty\simeq\drtbty$, where the 1-1 correspondence of 
branches and states is induced by $\blambda_{\bmcpb_Y}^t$,
the time $t$ component of $\blambda_{\bmcpb_Y}$.
\end{thm}

\vspace{3mm}
{\bf 6.3  Change of basis, time equivalence, and harmonic equivalence}
\vspace{3mm}

Given basis $\bmcpb=\{\calb^t:  t\in\bmcpz\}$ and encoder $E\rmdef\edrbinf$,
there is a 1-1 correspondence $\calu\lra\calr\lra C$ given by
$$
\bmu\stackrel{\blambda_\bmcpb}{\lra}\bmr\stackrel{E}{\lra}\bmb,
$$
where correspondence $\bmu\lra\bmr$ is induced by $\blambda_\bmcpb$, and 
correspondence $\bmr\lra\bmb$ is induced by encoder $E$.  Now
consider two different bases $\bmcpb_1=\{\calb_1^t:  t\in\bmcpz\}$ and 
$\bmcpb_2=\{\calb_2^t:  t\in\bmcpz\}$, and two different encoders
$E_1\rmdef\edrbinfone$ and $E_2\rmdef\edrbinftwo$.  
We say there is a {\it change of basis}.
For encoder $E_1$, there is a 1-1 correspondence $\calu\lra\calr_1\lra C$ given by
$$
\bmu_1\stackrel{\blambda_{\bmcpb_1}}{\lra}\bmr_1\stackrel{E_1}{\lra}\bmb_1,
$$
and for encoder $E_2$, there is a 1-1 correspondence $\calu\lra\calr_2\lra C$ given by
$$
\bmu_2\stackrel{\blambda_{\bmcpb_2}}{\lra}\bmr_2\stackrel{E_2}{\lra}\bmb_2.
$$
In general, if $\bmu_1=\bmu_2$, then $\bmb_1\ne\bmb_2$, and conversely,
if $\bmb_1=\bmb_2$ then $\bmu_1\ne\bmu_2$.

\begin{thm}
\label{thm42cb}
Consider two time domain encoders $E_1=\edrbinfone$ and $E_2=\edrbinftwo$
for bases $\bmcpb_1$ and $\bmcpb_2$, respectively; both encoders
go forward in time.  There is a 
graph automorphism of $\duinf$ which makes $E_1$ and $E_2$ graph isomorphic.
\end{thm}

\begin{prf}
From Theorem \ref{thm41x},
we know encoder $E_1$ is graph isomorphic to group trellis $C$ and so is
encoder $E_2$.  For encoder $E_1$, the graph isomorphism is given by a mapping
of vertices of $\drbinfone$ to states of $C$, and the same holds for $E_2$.  
Therefore there must be a mapping of vertices of $\drbinfone$ to vertices of
$\drbinftwo$ which makes $E_1$ and $E_2$ graph isomorphic. 
But from Theorem \ref{thm400}, there is a mapping of vertices of $\drbinfone$ 
to vertices of $\duinf$ which makes $\drbinfone$ and $\duinf$ graph isomorphic.  
The same holds for $\drbinftwo$ and $\duinf$.  Therefore there is a 
mapping of vertices of $\duinf$ to vertices of $\duinf$ 
which makes $E_1$ and $E_2$ graph isomorphic, or a graph automorphism of $\duinf$.
\end{prf}


Since $C$ is time invariant, we know that we can replace any basis
$\bmcpb_1=\{\calb_1^t:  t\in\bmcpz\}$ with a constant basis
$\bmcpb_{c,1}=\{\ldots,\calb_1,\calb_1,\ldots\}$.  
Similarly we can replace any basis
$\bmcpb_2=\{\calb_2^t:  t\in\bmcpz\}$ with a constant basis
$\bmcpb_{c,2}=\{\ldots,\calb_2,\calb_2,\ldots\}$.  
In general we assume $\calb_1\ne\calb_2$.  Then $C$ can
be constructed from a time domain encoder $E_{c,1}\rmdef\edrbinfcone$ and 
a time domain encoder $E_{c,2}\rmdef\edrbinfctwo$ where $\calb_1$ and
$\calb_2$ are constant vector bases.  We say a graph automorphism of 
$\duinf$ is {\it constant} if the mapping of states and
edges is constant for each time $t$.

\begin{thm}
\label{thm42cbc}
Consider two time domain encoders $E_{c,1}=\edrbinfcone$ and $E_{c,2}=\edrbinfctwo$.
There is a constant graph automorphism of $\duinf$ which makes 
$E_{c,1}$ and $E_{c,2}$ graph isomorphic.
\end{thm}

\begin{prf}
From Theorem \ref{thm41x},
we know encoder $E_{c,1}$ is graph isomorphic to group trellis $C$ and so is
encoder $E_{c,2}$.  For encoder $E_{c,1}$, the graph isomorphism is given by a 
mapping of vertices of $\drbinfcone$ to states of $C$.  Since $C$ is time 
invariant, and basis $\bmcpb_1$ is time invariant, the mapping of vertices 
of $\drbinfcone$ to states of $C$ must be time invariant.  The same holds 
for $E_{c,2}$.  Therefore there must be a time invariant mapping of vertices 
of $\drbinfcone$ to vertices of
$\drbinfctwo$ which makes $E_{c,1}$ and $E_{c,2}$ graph isomorphic. 
But from Theorem \ref{thm400}, there is a mapping of vertices of $\drbinfcone$ 
to vertices of $\duinf$ which makes $\drbinfcone$ and $\duinf$ graph isomorphic.  
This mapping is time invariant by construction of $\calu$.
The same holds for $\drbinfctwo$ and $\duinf$.  Therefore there is a 
time invariant mapping of vertices of $\duinf$ to vertices of $\duinf$ 
which makes $E_{c,1}$ and $E_{c,2}$ graph isomorphic, or a constant 
graph automorphism of $\duinf$.
\end{prf}


We now compare time domain encoders for forward time and backward time.
We consider two different bases, a basis
$\bmcpb=\{\calb^t:  t\in\bmcpz\}$ in the forward time direction and a basis
$\bmcpb_Y=\{\calb_Y^t:  t\in\bmcpz\}$ in the backward time direction.
At each time $t$, we select $\calb^t$ and $\calb_Y^t$ arbitrarily and 
independently of one another.

\begin{thm}
\label{thm42y}
Consider two time domain encoders, a forward time encoder $E=\edrbinf$
and a backward time encoder $E_Y\rmdef\edrbinfy$.
There is a graph isomorphism of $\duinf$ to $\duinfy$ which makes $E$ 
and $E_Y$ graph isomorphic.
\end{thm}

\begin{prf}
From Theorem \ref{thm41x},
we know encoder $E$ is graph isomorphic to group trellis $C$, and from
Theorem \ref{thm41y}, we know encoder $E_Y$ is graph isomorphic to group 
trellis $C$.  For encoder $E$, the graph isomorphism is given by a mapping
of vertices of $\drbinf$ to states of $C$, and for encoder $E_Y$, 
the graph isomorphism is given by a mapping
of vertices of $\drbinfy$ to states of $C$.
Therefore there must be a mapping of vertices of $\drbinf$ to vertices of
$\drbinfy$ which makes $E$ and $E_Y$ graph isomorphic.  
But from Theorem \ref{thm400}, there is a mapping of vertices of $\drbinf$ 
to vertices of $\duinf$ which makes $\drbinf$ and $\duinf$ graph isomorphic.  
And from Theorem \ref{thm400y}, the same holds for $\drbinfy$ and $\duinfy$.
Therefore there is a mapping of vertices of $\duinf$ to vertices of $\duinfy$
which makes $E$ and $E_Y$ graph isomorphic.
\end{prf}

There is a natural isomorphism of $\duinf$ to $\duinfy$.  We can look at a
generator $\bmg^{[t,t+k]}$ as beginning at time $t$ or ending at time $t+k$.
In constructing tensor $\bmu\in\calu$, at each time $t$, we have 
collected the generators $\bmg^{[t,t+k]}$, $0\le k\le\ell$, that begin at 
time $t$ to form a \sm\ $R^{[t,t+\ell]}$.  
In constructing tensor $\bmu_Y\in\calu_Y$, at each time $t$, 
we have collected the generators $\bmg^{[t-k,t]}$, $0\le k\le\ell$, 
that end at time $t$ to form a \sm\ $R_Y^{[t-\ell,t]}$.
Thus for each tensor $\bmu\in\calu$, there is a {\it natural correspondence}
$\bmu\equiv\bmu_Y$ with a tensor $\bmu_Y\in\calu_Y$ that uses the same \svecs.
The state of $\bmu$ at time $t$ is $\tridnut{1}{1}{t}$ and the state at 
time $t+1$ is $\tridnut{1}{1}{t+1}$.  
The state of $\bmu_Y$ at time $t+1$ is $\tridnuty{0}{1}{t+1}=\tridnut{0}{1}{t+1}$ 
and the state at time $t$ is $\tridnuty{0}{1}{t}=\tridnut{0}{1}{t}$.
Any graph isomorphism of $\duinf$ to $\duinfy$ is a graph automorphism of $\duinf$
composed with the {\it natural (graph) isomorphism} of $\duinf$ to $\duinfy$
given by the natural correspondence.  This gives the following result.

\begin{cor}
\label{cor42y}
Consider two time domain encoders, a forward time encoder $E=\edrbinf$ and 
a backward time encoder $E_Y=\edrbinfy$.
There is a graph automorphism of $\duinf$ composed with the natural isomorphism
to $\duinfy$ which makes $E$ and $E_Y$ graph isomorphic.
\end{cor}

We say a group system has {\it time equivalence} if,
when time domain encoder $\edrbinf$ (forward time) is loaded 
with $\bmr\in\calr$, and time domain encoder $\edrbinfy$ (backward time) 
is loaded with $\bmr_Y\in\calr_Y$, where $\bmcpb\equiv\bmcpb_Y$, 
$\calr\equiv\calr_Y$, and $\bmr\equiv\bmr_Y$,
the outputs of both encoders are the same.  In other words, if both
encoders are loaded with the same sequence of generators, 
then both encoders give the same output $\bmc\in C$.

\begin{thm}
\label{thm42yeq}
Any abelian group system has time equivalence, but this is not necessarily 
true for a nonabelian group system.  For the abelian group system, 
there is a trivial graph automorphism of $\duinf$ composed with the 
natural isomorphism to $\duinfy$ which makes $E$ and $E_Y$ graph isomorphic.
\end{thm}

\begin{prf}
We have seen in Section 4 that if $\bmr\equiv\bmr_Y$, then (\ref{enctd})
and (\ref{enctdrv}) give the same result.  But if $\bmr\equiv\bmr_Y$
then $\bmu\equiv\bmu_Y$.
\end{prf}
The standardized V.32 code is shown to be nonabelian in \cite{TR1}.
It can be shown time equivalence does not hold for this code.


Since $C$ is time invariant, we know that we can replace any basis
$\bmcpb=\{\calb^t:  t\in\bmcpz\}$ with a constant basis
$\bmcpb_c=\{\ldots,\calb,\calb,\ldots\}$.  Similarly we can replace any basis
$\bmcpb_Y=\{\calb_Y^t:  t\in\bmcpz\}$ with a constant basis
$\bmcpb_{Y,c}=\{\ldots,\calb_Y,\calb_Y,\ldots\}$.  Then $C$ can
be constructed from a time domain encoder $E_c\rmdef\edrbinfc$ (forward time) and 
a time domain encoder $E_{Y,c}\rmdef\edrbinfcy$ (backward time) where
$\bmcpb_c$ and $\bmcpb_{Y,c}$ are constant vector bases.  We say a graph 
isomorphism of $\duinf$ to $\duinfy$ is {\it constant} if the mapping of states and
edges is constant for each time $t$.

\begin{thm}
\label{thm42yc}
Consider two time domain encoders, a forward time encoder $E_c=\edrbinfc$ and 
a backward time encoder $E_{Y,c}=\edrbinfcy$.
There is a constant graph isomorphism of $\duinf$ to $\duinfy$ which makes $E_c$ 
and $E_{Y,c}$ graph isomorphic.
\end{thm}

\begin{prf}
The proof is a mix of the proof of Theorem \ref{thm42cbc}
combined with the proofs of Theorem \ref{thm42y} and Corollary \ref{cor42y}.
\end{prf}


We now compare spectral domain encoders for forward time and backward time.
The following result and proof is an analog of Theorem \ref{thm42y} and
proof for the spectral domain.

\begin{thm}
\label{thm42sy}
Consider two spectral domain encoders, a forward time encoder $E_s\rmdef\edrbinfs$ 
and a backward time encoder $E_{s,Y}\rmdef\edrbinfsy$.
There is a graph isomorphism of $\duinf$ to $\duinfy$ which makes $E_s$ 
and $E_{s,Y}$ graph isomorphic.
\end{thm}

An analog of Corollary \ref{cor42y} holds as well.

\begin{cor}
\label{cor42sy}
Consider two spectral domain encoders, a forward time encoder $E_s=\edrbinfs$ 
and a backward time encoder $E_{s,Y}=\edrbinfsy$.
There is a graph automorphism of $\duinf$ composed with the natural isomorphism
to $\duinfy$ which makes $E_s$ and $E_{s,Y}$ graph isomorphic.
\end{cor}
There are also analogs of Theorems \ref{thm42yeq} and \ref{thm42yc}.

We now compare the time and spectral domain encoders for forward time.
We consider the two different bases $\bmcpb_1$ and $\bmcpb_2$
used previously.  The following result and 
proof is similar to Theorem \ref{thm42cb}.

\begin{thm}
\label{thm42ts}
Consider a time domain encoder $E_1=\edrbinfone$ and a spectral domain
encoder $E_{s,2}\rmdef\edrbinftwos$; both encoders go forward in time.  There is a 
graph automorphism of $\duinf$ which makes $E_1$ and $E_{s,2}$ graph isomorphic.
\end{thm}

As before, we replace basis $\bmcpb_1$ with a constant basis
$\bmcpb_{c,1}$ and replace basis $\bmcpb_2$ with a constant basis
$\bmcpb_{c,2}$.  Then we have the following analog of Theorem \ref{thm42cbc}.

\begin{thm}
\label{thm42tsc}
Consider a time domain encoder $E_{c,1}=\edrbinfcone$ and a spectral domain
encoder $E_{s,c,2}\rmdef\edrbinfctwos$; both encoders go forward in time.  
There is a constant graph automorphism of $\duinf$ which makes $E_{c,1}$ 
and $E_{s,c,2}$ graph isomorphic.
\end{thm}

We say a group system has {\it harmonic equivalence} if,
when time domain encoder $E=\edrbinf$  (forward time) and 
spectral domain encoder $E_s=\edrbinfs$ (forward time) are loaded
with the same sequence of generators, i.e., the same $\bmr\in\calr$, 
then both encoders give the same output $\bmc\in C$.  In other words, 
if the group system is harmonically equivalent, then any path $\bmc$ has
a decomposition in the time domain and spectral domain into the
same $\bmr\in\calr$.

\begin{thm}
Any abelian group system has harmonic equivalence, but 
this is not necessarily true for a nonabelian group system.
For the abelian group system, there is a trivial graph automorphism 
of $\duinf$ which makes $E$ and $E_s$ graph isomorphic.
\end{thm}

\begin{prf}
For an abelian group system, we see that the rearrangement of 
$b_s^t$ in (\ref{encfta}) gives $b^t$ in (\ref{enctd}).
\end{prf}

For each of the four comparisons of encoders, $E_1$ and $E_2$, $E$ and $E_Y$,
$E_s$ and $E_{s,Y}$, and $E_1$ and $E_{s,2}$, we see there is a graph
automorphism of $\duinf$ which makes the two encoders graph isomorphic,
composed with the natural isomorphism to $\duinfy$ in the 
second and third comparisons.  If the bases are constant, the graph
automorphism of $\duinf$ is constant.
In the next section, we analyze the structure of any graph automorphism
of $\duinf$.

\newpage
\vspace{3mm}
{\bf 7.  THE FULL SYMMETRY SYSTEM OF THE COSET TENSOR SET $\calu$}
\vspace{3mm}

\vspace{3mm}
{\bf 7.1  Analysis of a symmetry permutation}
\vspace{3mm}

As defined in \cite{SMT}, the {\it full symmetry system} of $\calu$ 
is the set of all permutations or bijections of $\calu$.  This is a group
under composition operation.  Note that $\calu$ and $\duinf$ are 
equivalent:  the paths of $\calu$ are the paths of $\duinf$ and vice versa.  
Therefore the \fss\ of $\calu$ is the set of all graph automorphisms
of $\duinf$.  A {\it symmetry} $\Phi$ of $\duinf$ is a graph automorphism
of $\duinf$.  If $\bmu$ is a path in $\duinf$, then $\Phi(\bmu)$ is a path 
in $\duinf$, and we say $\Phi$ {\it preserves paths} in $\duinf$.
So a symmetry $\Phi$ of $\duinf$ is a 1-1 and onto map of the states 
and edges of $\dut$ at each time $t$ that preserves paths in $\duinf$.
In this subsection we analyze the structure of any symmetry $\Phi$, 
and then in Subsection 7.2 we show how to construct any symmetry. 
In Subsection 7.3 we study the \fss.

Let a symmetry $\Phi$ of $\duinf$ be
denoted as $\Phi=\ldots,\varphi^t,\varphi^{t+1},\ldots$, where
$\varphi^t:  \bmcpu^t\mapsto\bmcpu^t$.
Define a component form of $\varphi^t$ by 
$\varphi^t=(\varphi_{0}^t,\ldots,\varphi_{\ell}^t)$, where function 
$\varphi_{j}^t: \bmcpu^t \ra \bmcpu_j^t$ gives the $j$-th component of 
$\varphi^t$, $\je{0}$.
We say $\varphi_{j}^t$ is {\it independent of component} $\bmu_m^t$ if
$$
\varphi_{j}^t(\bmu_0^t,\ldots,\bmu_{m-1}^t,\bmu_m^t,\bmu_{m+1}^t,\ldots,\bmu_\ell^t)=
\varphi_{j}^t(\bmu_0^t,\ldots,\bmu_{m-1}^t,\bzero_m^t,\bmu_{m+1}^t,\ldots,\bmu_\ell^t)
$$
for all 
$\bmu^t=(\bmu_0^t,\ldots,\bmu_{m-1}^t,\bmu_m^t,\bmu_{m+1}^t,\ldots,\bmu_\ell^t)\in\bmcpu^t$.
We denote this property as 
$\varphi_{j}^t(\bmu_0^t,\ldots,\bmu_{m-1}^t,\bllt_m^t,\bmu_{m+1}^t,\ldots,\bmu_\ell^t)$,
where the bullet ``$\bllt_m^t$'' means $\varphi_j^t$ is independent of that component.
For $0\le j\le\ell$, 
define function $\varphi_{[j,\ell]}^t:  \bmcpu^t\ra\cuut{j}{t}$ to be the
components $\varphi_m^t$ of $\varphi^t$ for $m\in [j,\ell]$, i.e.,
$\varphi_{[j,\ell]}^t\rmdef (\varphi_{j}^t,\varphi_{j+1}^t,\ldots,\varphi_{\ell}^t)$.

First we review this important result about paths in $\duinf$, which is
a corollary of Theorem \ref{thm29aa} for tensors in $\calu$.

\begin{cor}
\label{111}
Let $\bmw=\ldots,\bmu^t,\bmu^{t+1},\ldots$ be an arbitrary sequence, not
necessarily a path in $\duinf$, where $\bmu^t\in\bmcpu^t$
for each time $t\in\bmcpz$.  Then $\bmw$ is a path in $\duinf$
\ifof\ for each time $t$, $\bmu^{t+1}=(\bmu_0^{t+1},\bsigma\bmu^t)$ 
where input $\bmu_0^{t+1}$ is any element of $\bmcpu_0^{t+1}$.
\end{cor}

We know if $\bmw$ is a path, then the symmetry $\Phi(\bmw)$
is also a path, and Corollary \ref{111} applies to both $\bmw$ and
$\Phi(\bmw)$.  Therefore the commutative diagram 
Figure \ref{commutpath} holds.

\begin{figure}[h]
\centering
\vspace{3ex}

\begin{picture}(100,100)
\put(0,20){\vector(0,1){60}}
\put(20,100){\vector(1,0){60}}
\put(100,20){\vector(0,1){60}}
\put(20,0){\vector(1,0){60}}
\put(0,100){\makebox(0,0){$\bmuht^t$}}
\put(0,0){\makebox(0,0){$\bmu^t$}}
\put(100,0){\makebox(0,0){$\bmu^{t+1}$}}
\put(100,100){\makebox(0,0){$\bmuht^{t+1}$}}

\put(50,105){\makebox(0,0)[b]{$\bsigma$}}
\put(50,5){\makebox(0,0)[b]{$\bsigma$}}
\put(-5,50){\makebox(0,0)[r]{$\varphi^t$}}
\put(105,50){\makebox(0,0)[l]{$\varphi^{t+1}$}}
\end{picture}

\caption{Commutative diagram for $\varphi^t$ and $\varphi^{t+1}$, 
where $\bmu^{t+1}=(\bmu_0^{t+1},\bsigma\bmu^t)$ and 
$\bmuht^{t+1}=(\bmuht_0^{t+1},\bsigma\bmuht^t)$, and
$\bmu_0^{t+1}$ and $\bmuht_0^{t+1}$ are inputs at time $t+1$.}
\label{commutpath}

\end{figure}

For each $t$, a component $\varphi^t$ of $\Phi$ must be a 1-1 and onto map
$\varphi^t:  \bmcpu^t\mapsto\bmcpu^t$.  Therefore the maps 
$\varphi^t:  \bmcpu^t\mapsto\bmcpu^t$ and
$\varphi^{t+1}:  \bmcpu^{t+1}\mapsto\bmcpu^{t+1}$ in Figure \ref{commutpath}
must be 1-1 and onto.  The map $\varphi^{t+1}$ must be 1-1 and onto,
but we know as well that all branches $\bmu^{t+1}$ and $\bmuht^{t+1}$
which split from states $\bsigma\bmu^t$ and $\bsigma\bmuht^t$ must map
to each other.  Or, in other words, state $\bsigma\bmu^t$ must
map to state $\bsigma\bmuht^t$.  This gives
commutative diagram Figure \ref{commutpath1} and Theorem \ref{thm51}.

\begin{figure}[h]
\centering

\begin{picture}(100,100)
\put(0,20){\vector(0,1){60}}
\put(20,100){\vector(1,0){60}}
\put(20,0){\vector(1,0){60}}
\put(100,20){\vector(0,1){60}}
\put(0,0){\makebox(0,0){$\bmu^t$}}
\put(0,100){\makebox(0,0){$\bmuht^t$}}
\put(100,100){\makebox(0,0){$\bsigma\bmuht^t$}}
\put(100,0){\makebox(0,0){$\bsigma\bmu^t$}}
\put(-5,50){\makebox(0,0)[r]{$\varphi^t$}}
\put(50,105){\makebox(0,0)[b]{$\bsigma$}}
\put(50,5){\makebox(0,0)[b]{$\bsigma$}}
\put(105,50){\makebox(0,0)[l]{$\varphi_{[1,\ell]}^{t+1} (\bllt_0^{t+1},\bsigma\bmu^t)$}}
\end{picture}

\caption{Commutative diagram for $\varphi^t$ and $\varphi_{[1,\ell]}^{t+1}$.}
\label{commutpath1}

\end{figure}

\begin{thm}
\label{thm51}
$\Phi=\ldots,\varphi^t,\varphi^{t+1},\ldots$ is a symmetry of $\duinf$ 
\ifof\ the following two conditions hold for each $t\in\bmcpz$:

(i) $\varphi^t:  \bmcpu^t\ra\bmcpu^t$ is 1-1 and onto,

(ii) 
\be
\label{kshft1}
\bsigma\varphi^t(\bmu^t)
=\varphi_{[1,\ell]}^{t+1} (\bllt_0^{t+1},\bsigma\bmu^t),
\ee
for each $\bmu^t\in\bmcpu^t$.
\end{thm}

\begin{cor}
\label{cor51x}
$\Phi=\ldots,\varphi^t,\varphi^{t+1},\ldots$ is a symmetry of $\duinf$ 
\ifof\ the following three conditions hold for each $t\in\bmcpz$:

(i) $\varphi^t:  \bmcpu^t\ra\bmcpu^t$ is 1-1 and onto,

(ii) for $1\le j\le\ell$, $\varphi_j^t$ is independent of $\bmu_0^t$, e.g.,  
$\varphi_j^t(\bllt_0^{t+1},\uut{1}{t})$,

(iii) for $0\le j\le\ell-1$, for each $\bmu^t\in\bmcpu^t$,
$$
\sigma\varphi_j^t(\bmu^t)=\varphi_{j+1}^{t+1}(\bllt_0^{t+1},\bsigma\bmu^t).
$$
\end{cor}

\begin{prf}
We can write (\ref{kshft1}) in component form as
\be
\label{kshft2}
\sigma\varphi_j^t(\bmu^t)=\varphi_{j+1}^{t+1}(\bllt_0^{t+1},\bsigma\bmu^t),
\ee
for $j=0,\ldots,\ell-1$, for each $\bmu^t\in\bmcpu^t$.
\end{prf}

We can use Corollary \ref{cor51x} to further characterize a symmetry
$\Phi$ of $\duinf$ as follows.  Using (ii), we can rewrite (iii) as
\begin{equation}
\label{fv4}
\pvphit{0}{t}(\bmu_0^t,\uut{1}{t})=\varphi_{1}^{t+1}(\bllt_0^{t+1},\pvpvt{0}{t}),
\end{equation}
for $j=0$.  Since (ii) and (iii) hold for each $t\in\bmcpz$, we have
\begin{equation}
\label{fv5}
\pvphit{j}{t+j}(\bllt_0^{t+j},\uut{1}{t+j})=\varphi_{j+1}^{t+j+1}(\bllt_0^{t+j+1},\pvpvt{0}{t+j}),
\end{equation}
for $j=1,\ldots,\ell-1$.  We can reduce the set of equations (\ref{fv5}) 
further.  Start with $j=1$,
$$
\pvphit{1}{t+1}(\bllt_0^{t+1},\uut{1}{t+1})=\varphi_{2}^{t+2}(\bllt_0^{t+2},\pvpvt{0}{t+1}).
$$
Fix $\uut{1}{t+1}$ on the left hand side; then the value of the left hand 
side is fixed.  Since $\uut{1}{t+1}$ are fixed on the left hand side, 
$\pvpvt{1}{t+1}$ are fixed on the right hand side.  Then to have equality, 
$\varphi_{2}^{t+2}$ must be independent of $\pvt{0}{t+1}$, or
\be
\label{fv6}
\pvphit{1}{t+1}(\bllt_0^{t+1},\uut{1}{t+1})=\varphi_{2}^{t+2}(\bllt_0^{t+2},\bllt_1^{t+2},\pvpvt{1}{t+1}).  
\ee
Now look at the case $j=2$,
$$
\pvphit{2}{t+2}(\bllt_0^{t+2},\uut{1}{t+2})=\varphi_{3}^{t+3}(\bllt_0^{t+3},\pvpvt{0}{t+2}).
$$
Using the result (\ref{fv6}), we obtain
$$
\pvphit{2}{t+2}(\bllt_0^{t+2},\bllt_1^{t+2},\uut{2}{t+2})=\varphi_{3}^{t+3}(\bllt_0^{t+3},\pvpvt{0}{t+2}).
$$
Now fix $\uut{2}{t+2}$ on the left hand side.  Then to have equality, 
$\varphi_{3}^{t+3}$ must be independent of $\pvt{0}{t+2}$ and $\pvt{1}{t+2}$, 
so we have
$$
\pvphit{2}{t+2}(\bllt_0^{t+2},\bllt_1^{t+2},\uut{2}{t+2})=\varphi_{3}^{t+3}(\bllt_0^{t+3},\bllt_1^{t+3},\bllt_2^{t+3},\pvpvt{2}{t+2}).
$$
Continuing this process in the same manner, we finally reduce the last 
equation, $j=\ell$, to
$$
\pvphit{\ell-1}{t+\ell-1}(\bllt_0^{t+\ell-1},\bllt_1^{t+\ell-1},\ldots,\bllt_{\ell-2}^{t+\ell-1},\bmu_{\ell-1}^{t+\ell-1},\bmu_\ell^{t+\ell-1})
=\varphi_{\ell}^{t+\ell}(\bllt_0^{t+\ell},\bllt_1^{t+\ell},\ldots,\bllt_{\ell-1}^{t+\ell},\pvt{\ell-1}{t+\ell-1}).
$$
Summarizing our results, we can rewrite (\ref{fv5}) as
\be
\label{fv8}
\pvphit{j}{t+j}(\bllt_0^{t+j},\bllt_1^{t+j},\ldots,\bllt_{j-1}^{t+j},\uut{j}{t+j})
=\varphi_{j+1}^{t+j+1}(\bllt_0^{t+j+1},\bllt_1^{t+j+1},\ldots,\bllt_j^{t+j+1},\pvpvt{j}{t+j})
\ee
for $j=1,\ldots,\ell-1$.  With the understanding that the left hand side
of (\ref{fv8}) is $\pvphit{0}{t}(\bmu_0^t,\uut{1}{t})$ when $j=0$, then 
(\ref{fv8}) also includes (\ref{fv4}), and we can assume (\ref{fv8}) holds
for $j=0,\ldots,\ell-1$.  Note that (\ref{fv8}) 
can be explained using a commutative diagram.

Equation (\ref{fv8}) shows that $\varphi_{j}^{t+j}$ is
independent of components $\bmu_0^{t+j},\bmu_1^{t+j},\ldots,\bmu_{j-1}^{t+j}$, 
for $\je{1}$.  This means that
\be
\label{form0}
\varphi_j^{t+j}(\bmu_0^{t+j},\ldots,\bmu_{j-1}^{t+j},\uut{j}{t+j})
=\varphi_j^{t+j}(\bzero_0^{t+j},\ldots,\bzero_{j-1}^{t+j},\uut{j}{t+j})
\ee
for all $\bmu^{t+j}\in\bmcpu^{t+j}$.  We refer to this property by saying
$\varphi_j^{t+j}$ is a function of the form 
\be
\label{form}
\varphi_j^{t+j}:  (\bllt_0^{t+j},\ldots,\bllt_{j-1}^{t+j})\times\cuut{j}{t+j}\ra\bmcpu_j^{t+j},
\ee
where $(\bllt_0^{t+j},\ldots,\bllt_{j-1}^{t+j})$ means the function is independent
of these components.  

For $\je{0}$, let $_r\varphi_j^{t+j}$ be the restriction
of $\varphi_j^{t+j}$ to $\cuut{j}{t+j}$.  Then $_r\varphi_j^{t+j}$ is a function
\be
\label{form1}
_r\varphi_j^{t+j}:  \cuut{j}{t+j}\ra\bmcpu_j^{t+j}.
\ee
For $j=0$, $\varphi_j^{t+j}$ and $_r\varphi_j^{t+j}$ are the same.  With $_r\varphi_j^{t+j}$
the restriction of $\varphi_j^{t+j}$, we have that  (\ref{fv8}) holds
for $j=0,\ldots,\ell-1$ \ifof\ 
\be
\label{fv8r}
\sigma\,_r\varphi_j^{t+j}(\uut{j}{t+j})= {_r\varphi_{j+1}^{t+j+1}(\pvpvt{j}{t+j})}
\ee
holds for $j=0,\ldots,\ell-1$.

If $\varphi_j^{t+j}$ is a function of the form (\ref{form}), then $_r\varphi_j^{t+j}$
is uniquely defined.  Conversely if $_r\varphi_j^{t+j}$ is a function defined as
in (\ref{form1}), then there is a unique function $\varphi_j^{t+j}$ of the
form (\ref{form}) whose restriction is $_r\varphi_j^{t+j}$.

\begin{lem}
\label{lem22}
Fix time $t\in\bmcpz$.  For $\je{1}$, suppose that $\varphi_{j}^t$ 
has the property in (\ref{form0})-(\ref{form}) that $\varphi_{j}^t$ is
independent of components $\bmu_0^t,\bmu_1^t,\ldots,\bmu_{j-1}^t$.  
With this property of $\varphi_{j}^t$, we have that
$\varphi^t=(\varphi_0^t,\ldots,\varphi_j^t,\ldots,\varphi_\ell^t)$ is 
1-1 and onto \ifof\ the restriction $_r\varphi_j^t$ of $\varphi_j^t$ has
the property that $_r\varphi_j^t:  \bmcpu_j^t\times (\uut{j+1}{t}) \ra \bmcpu_j^t$ 
is 1-1 and onto for each fixed $(\uut{j+1}{t}) \in \cuut{j+1}{t}$,
for $j$ such that $0\le j\le\ell$.  For $j=\ell$, this is understood to mean
$_r\varphi_\ell^t: \bmcpu_\ell^t\ra \bmcpu_\ell^t$ is 1-1 and onto.
\end{lem}

\begin{prf}
Assume $\varphi^t$ is 1-1 and onto.  Fix $(\uut{j+1}{t})\in\cuut{j+1}{t}$.  
It is clear $\varphi^t$ cannot be onto unless 
$_r\varphi_j^t: \bmcpu_j^t\times (\uut{j+1}{t})\ra\bmcpu_j^t$
is onto.  But if $\varphi_j^t$ is onto, it must be 1-1.

Conversely, suppose
$_r\varphi_j^t: \bmcpu_j^t\times (\uut{j+1}{t})\ra\bmcpu_j^t$
is 1-1 and onto for each fixed $(\uut{j+1}{t})\in\cuut{j+1}{t}$,
for $j$ such that $0\le j\le\ell$.  We show $\varphi^t$ is 1-1 and onto.
We use proof by induction.  Consider the function
$({_r\varphi_{j+1}^t},\ldots,{_r\varphi_\ell^t}):  \cuut{j+1}{t}\ra\cuut{j+1}{t}$.
Assume this function is 1-1 and onto.  We show the function
$({_r\varphi_j^t},{_r\varphi_{j+1}^t},\ldots,{_r\varphi_\ell^t}):  \cuut{j}{t}\ra\cuut{j}{t}$
is 1-1 and onto.  By hypothesis, the restriction 
$_r\varphi_j^t: \bmcpu_j^t\times (\uut{j+1}{t})\ra\bmcpu_j^t$,
is 1-1 and onto for each fixed $(\uut{j+1}{t})\in\cuut{j+1}{t}$.  Then it follows that
$({_r\varphi_j^t},{_r\varphi_{j+1}^t},\ldots,{_r\varphi_\ell^t})$ is 1-1 and onto.
But by hypothesis the restriction $_r\varphi_\ell^t:  \bmcpu_\ell^t\ra\bmcpu_\ell^t$
is 1-1 and onto.  Then by induction the function
$({_r\varphi_0^t},{_r\varphi_1^t},\ldots,{_r\varphi_\ell^t}):  \cuut{0}{t}\ra\cuut{0}{t}$
is 1-1 and onto.  But the function 
$({_r\varphi_0^t},{_r\varphi_1^t},\ldots,{_r\varphi_\ell^t})$ has the same values
as $(\varphi_0^t,\varphi_1^t,\ldots,\varphi_\ell^t)$.  This proves that 
$\varphi^t: \bmcpu^t\ra\bmcpu^t$ is 1-1 and onto.
\end{prf}

We now formalize the properties of $\varphi_j^t$ and $_r\varphi_j^t$.

\begin{defn}
\label{defn1}
{\bf (Definition of $\bomga_j^{t+j}:  0\le j\le\ell$)}

Fix $j$ such that $0\le j\le\ell$.  We define a function 
$\bomga_j^{t+j}:  \bmcpu^{t+j}\ra\bmcpu_j^{t+j}$ with the
following two properties: \\

(i) The function $\bomga_j^{t+j}$ is a function of the form
$$
\bomga_j^{t+j}:  (\bllt_0^{t+j},\ldots,\bllt_{j-1}^{t+j})\times\cuut{j}{t+j}\ra\bmcpu_j^{t+j}.
$$

(ii) The restriction of $\bomga_j^{t+j}$ to $\cuut{j}{t+j}$ is a 
function $\beta_{j}^{t+j}:  \cuut{j}{t+j}\ra\bmcpu_j^{t+j}$ which is a 
1-1 and onto function 
$$
\beta_j^{t+j}:  \bmcpu_j^{t+j}\times (\uut{j+1}{t+j})\ra\bmcpu_j^{t+j}
$$
from $\bmcpu_j^{t+j}$ to $\bmcpu_j^{t+j}$ for each fixed 
$(\uut{j+1}{t+j})\in\cuut{j+1}{t+j}$.
\end{defn}

We call a function $\bomga_j^{t+j}$ with the properties in Definition
\ref{defn1} a {\it separating function}, and function $\beta_j^{t+j}$
a {\it restricted separating function}.

\begin{lem}
\label{lemxyz}
The function $\varphi_j^{t+j}$ is a separating function, and $_r\varphi_j^t$
is a restriced separating function.
\end{lem}

Using Lemma \ref{lem22} and \ref{lemxyz}, we are able to characterize 
a symmetry $\Phi$ of $\duinf$ as follows.

\begin{thm}[Analysis]
\label{thm23} 
$\Phi=\ldots,\varphi^t,\varphi^{t+1},\ldots$ is a symmetry of $\duinf$ 
\ifof, for each $t\in\bmcpz$, 
$\varphi^t=(\varphi_0^t,\ldots,\varphi_j^t,\ldots,\varphi_\ell^t)$ 
where the following two equivalent conditions hold:

(i) function $\varphi_j^t$ is a separating function,
for $j$ such that $0\le j\le\ell$, and 
(\ref{fv8}) is satisfied for $j=0,\ldots,\ell-1$;

(ii) function $_r\varphi_j^t$ is a restricted separating function,
for $j$ such that $0\le j\le\ell$, and 
(\ref{fv8r}) is satisfied for $j=0,\ldots,\ell-1$.
\end{thm}

For each $t$, $t\in\bmcpz$, a sequence of functions
\be
\label{twr1}
\Psi^t\rmdef (\varphi_0^t,\varphi_1^{t+1},\ldots,\varphi_j^{t+j},
\varphi_{j+1}^{t+j+1},\ldots,\varphi_{\ell}^{t+\ell})
\ee
such that for $j=0,\ldots,\ell-1$, each pair
$(\varphi_{j}^{t+j},\varphi_{j+1}^{t+j+1})$ satisfies (\ref{fv8}),
and such that for $j=0,\ldots,\ell$, $\varphi_j^t$ is a separating function,
is called a $t$-{\it tower} $\Psi^t$ or just tower.  Then an essential
conclusion of Theorem \ref{thm23} is that any symmetry of
$\duinf$ gives rise to a sequence of $t$-towers
$\ldots,\Psi^t,\Psi^{t+1},\ldots$, that is, a $t$-tower
for each $t\in\bmcpz$.  We utilize $t$-towers in the construction 
algorithm below.

\be
\label{twrphi}
\setcounter{MaxMatrixCols}{7}
\begin{pmatrix}   
                  &              &               & \vdots          & &          &             \\
                  &              &               &                 & &          & \varphi_\ell^{(t)+\ell}  \\
                  &              &               &                 & &          & \vdots      \\
                  &              & \cdots        & \varphi_j^{(t)+j} & \cdots & &             \\
                  &              &               & \vdots          & &          & \varphi_\ell^{(t-j)+\ell}  \\
\cdots            & \varphi_1^{(t)+1} & \cdots   &                 & &          &  \vdots     \\
\varphi_0^{(t)}   & \varphi_1^{(t-1)+1} & \cdots & \varphi_j^{(t-j)+j} & \cdots & & \varphi_\ell^{(t-\ell)+\ell} \\
\vdots            & \vdots       &               & \vdots          & &          & \vdots      \\
\cdots            & \varphi_1^{(t-j)+1} & \cdots &                 & &          &             \\
\varphi_0^{(t-j)} & \cdots       &               &                 & &          &             \\
                  &              &               & \vdots          & &          &
\end{pmatrix}
\ee

Equation (\ref{twrphi}) shows a $t$-tower $\Psi^t$ and $(t-j)$-tower
$\Psi^{t-j}$ as diagonals in an infinite matrix of towers.  The
theorem shows that the sequence of towers 
$\ldots,\Psi^{t-j},\ldots,\Psi^t,\ldots$ in (\ref{twrphi})
defines a symmetry and any such matrix (\ref{twrphi}) of towers
defines a symmetry.  A component $\varphi^t$ of the symmetry
is defined by going ``across the row'' in (\ref{twrphi}),
\begin{align}
\label{eq61}
\varphi^t 
&=(\varphi_0^{(t)},\varphi_1^{(t-1)+1},\ldots,\varphi_j^{(t-j)+j},\ldots,\varphi_{\ell}^{(t-\ell)+\ell}) \\
\label{eq62}
&=(\varphi_0^t,\varphi_1^t,\ldots,\varphi_j^t,\ldots,\varphi_{\ell}^t),
\end{align}
and more explicitly, $\varphi^t:  \bmcpu^t\ra\bmcpu^t$ is defined by
$$
\varphi^t(\bmu^t)=(\varphi_0^t(\bmu^t),\varphi_1^t(\bmu^t),\ldots,
\varphi_j^t(\bmu^t),\ldots,\varphi_\ell^t(\bmu^t)).
$$
Note that each component $\varphi_j^t$ in (\ref{eq62}) is selected from
a different tower.  For example, $\varphi_j^t$ in (\ref{eq62}) is component 
$\varphi_j^{(t-j)+j}$ in $(t-j)$-tower $\Psi^{t-j}$:
$$
\varphi_j^t(\bmu^t)=\varphi_j^{(t-j)+j}(\bmu^{(t-j)+j}).
$$

\vspace{3mm}
{\bf 7.2  Construction of a symmetry permutation}
\vspace{3mm}

We now use Theorem \ref{thm23} to construct any symmetry
$\bomga=\ldots,\bomga^t,\bomga^{t+1},\ldots$ of $\duinf$ .
We can solve the set of equations (\ref{fv8}) or (\ref{fv8r}) by starting 
with $j=\ell$ and working backwards, for $j=\ell,\ell-1,\ldots,1,0$.  
At each step $j$, we want to find a separating function
$\bomga_j^{t+j}:  \bmcpu^{t+j}\ra\bmcpu_j^{t+j}$ that satisfies
(\ref{fv8}), or a restricted separating function 
$\beta_{j}^{t+j}:  \cuut{j}{t+j}\ra\bmcpu_j^{t+j}$ that satisfies (\ref{fv8r}).

\begin{alg}[Construction] 
\label{alg54} 
Any solution of the set of equations (\ref{fv8}) or (\ref{fv8r}) which is a symmetry 
$\bomga=\ldots,\bomga^t,\bomga^{t+1},\ldots$ of $\duinf$ can be found 
as follows.  

\vspace{2mm}
\noindent {\bf DO} \\
1.  Fix time $t$.

\vspace{2mm}
\noindent 2.  Let $\beta_\ell^{t+\ell}:  \bmcpu_\ell^{t+\ell}\ra\bmcpu_\ell^{t+\ell}$ 
be any restricted separating function.  Define $\bomga_\ell^{t+\ell}$ to be the 
unique separating function whose restriction is $\beta_{\ell}^{t+\ell}$.

\vspace{2mm}
\noindent 3. \\
{\bf FOR} $j=\ell-1,\ldots,0$ (counting down in order), \\
find a restricted separating function 
$\beta_{j}^{t+j}:  \cuut{j}{t+j}\ra\bmcpu_j^{t+j}$ such that 
\be
\label{fv14b}
\sigma\beta_j^{t+j}(\uut{j}{t+j})=\beta_{j+1}^{t+j+1}(\pvpvt{j}{t+j}).
\ee
\noindent Define $\bomga_j^{t+j}$ to be the unique separating function
whose restriction is $\beta_{j}^{t+j}$. \\
{\bf ENDFOR} \\
{\bf ENDDO}

\vspace{2mm}
\noindent 4.  For each time $t$, steps 1-3 produce a $t$-tower $\Upsilon^t$,
where
\be
\Upsilon^t\rmdef (\bomga_0^t,\bomga_1^{t+1},\ldots,\bomga_j^{t+j},\ldots,\bomga_{\ell}^{t+\ell}).
\ee
A sequence of any $t$-towers $\ldots,\Upsilon^t,\Upsilon^{t+1},\ldots$ defines 
a symmetry $\bomga=\ldots,\bomga^t,\bomga^{t+1},\ldots$ of $\duinf$ in the 
following manner.  For each $t$, define a function 
$\bomga^t:  \bmcpu^t\ra\bmcpu^t$ by
\be
\bomga^t=(\bomga_0^t,\bomga_1^t,\ldots,\bomga_j^t,\ldots,\bomga_{\ell}^t),
\ee
where $\bomga_j^t$ is component function $\bomga_j^{(t-j)+j}$ in 
$(t-j)$-tower $\Upsilon^{t-j}$:
$$
\bomga_j^t(\bmu^t)=\bomga_j^{(t-j)+j}(\bmu^{(t-j)+j}).
$$
Then $\bomga=\ldots,\bomga^t,\bomga^{t+1},\ldots$ is a symmetry
of $\duinf$, and the set of all possible symmetries $\bomga$ obtained this
way is the \fss\ of $\duinf$.
\end{alg}

\begin{prf}
Note that if $\beta_{j+1}^{t+j+1}$ on the right hand side of
(\ref{fv14b}) is a restricted separating function, then
$$
\beta_{j+1}^{t+j+1}:  \bmcpu_{j+1}^{t+j+1}\times (\pvpvt{j+1}{t+j})\ra\bmcpu_{j+1}^{t+j+1}
$$
is a 1-1 and onto function for each fixed
$(\pvpvt{j+1}{t+j})\in\cuut{j+2}{t+j+1}$.
This means we can find a function $\beta_j^{t+j}$ on the left hand side of
(\ref{fv14b}) such that 
$$
\beta_j^{t+j}: \bmcpu_j^{t+j}\times (\uut{j+1}{t+j})\ra\bmcpu_j^{t+j}
$$
is a 1-1 and onto function for each fixed
$(\uut{j+1}{t+j})\in\cuut{j+1}{t+j}$.  Then $\beta_{j}^{t+j}$ is a
restricted separating function.
\end{prf}

A sequence of any $t$-towers $\ldots,\Upsilon^t,\Upsilon^{t+1},\ldots$ defines 
a symmetry.  Note that in general, $\Upsilon^t$ can be different for each
$t$, i.e., we need not have $\Upsilon^t=\Upsilon^{t+1}$.

Define $\btridnut{j}{k}{t}$ to be
the same as $\tridnut{j}{k}{t}$ except missing
entry $u_{j,k}^t$, and likewise define $\btridncut{j}{k}{t}$ to be
the same as $\tridncut{j}{k}{t}$ except missing
entry $u_{j,k}^t$.  Define $\btriuput{j}{k}{t}$ to be all
the entries $u_{m,n}^t$ in $\bmu^t$ except those in $\tridnut{j}{k}{t}$,
and define $\btriupcut{j}{k}{t}$ to be 
$\btriupcut{j}{k}{t}\rmdef\{\btriuput{j}{k}{t} : \bmu^t\in\bmcpu^t\}$.
If a function is independent of entries in $\btriupcut{j}{k}{t}$,
we denote this by $\btriuptblt{j}{k}{t}$.

We now define functions $\omega_{j,k}^{t+j}$ and $\beta_{j,k}^{t+j}$
and then show these functions can be used to construct a separating 
function $\bomga_j^{t+j}$.

\begin{defn}
\label{defn2}
{\bf (Definition of $\omega_{j,k}^{t+j}:  0\le j\le k,0\le k\le\ell$)}

Fix $k$ such that $0\le k\le\ell$.  Fix $j$ such that $0\le j\le k$.
We define a function 
$\omega_{j,k}^{t+j}:  \bmcpu^{t+j}\ra U_{j,k}^{t+j}$ with the
following two properties: \\

(i) The function $\omega_{j,k}^{t+j}$ is a function of the form
$$
\omega_{j,k}^{t+j}:  \btriuptblt{j}{k}{t+j}\times U_{j,k}^{t+j}\times\btridncut{j}{k}{t+j}\ra U_{j,k}^{t+j}.
$$

(ii) The restriction of $\omega_{j,k}^{t+j}$ to $\tridncut{j}{k}{t+j}$ is a function 
$\beta_{j,k}^{t+j}:  U_{j,k}^{t+j}\times\btridncut{j}{k}{t+j}\ra U_{j,k}^{t+j}$
which is a 1-1 and onto function
$$
\beta_{j,k}^{t+j}:  U_{j,k}^{t+j}\times\btridnut{j}{k}{t+j}\ra U_{j,k}^{t+j}
$$
from $U_{j,k}^{t+j}$ to $U_{j,k}^{t+j}$ for each fixed 
$\btridnut{j}{k}{t+j}\in\btridncut{j}{k}{t+j}$.

For $j=k=\ell$, (i) is understood to mean $\omega_{j,k}^{t+j}$ 
is a function of the form
$$
\omega_{\ell,\ell}^{t+\ell}:
\btriuptblt{\ell}{\ell}{t+\ell}\times U_{\ell,\ell}^{t+\ell}\ra U_{\ell,\ell}^{t+\ell}.
$$
For $j=k=0$, (i) is understood to mean $\omega_{j,k}^{t+j}$ 
is a function of the form
$$
\omega_{0,0}^{t}:
U_{0,0}^{t}\times\btridncut{0}{0}{t}\ra U_{0,0}^{t}.
$$
\end{defn}

Again we call a function $\omega_{j,k}^{t+j}$ with the properties in 
Definition \ref{defn2} a {\it separating function} and $\beta_{j,k}^{t+j}$
a {\it restricted separating function}.

We now use these definitions and results to simplify Algorithm \ref{alg54}
by solving (\ref{fv14b}) of Step 3 in Algorithm \ref{alg54}.  
Given a separating function $\beta_{j+1}^{t+j+1}$, we want to find a
separating function $\beta_j^{t+j}$ that satisfies (\ref{fv14b}).  
We first find properties of any function $\beta_j^{t+j}$ that 
satisfies (\ref{fv14b}) and then give a necessary and sufficient
condition that it be a separating function.

It is sufficient to construct $\beta_j^{t+j}$ 
for arbitrary fixed $(\uut{j+1}{t+j})\in\cuut{j+1}{t+j}$.  For fixed 
$(\uut{j+1}{t+j})$, $\beta_j^{t+j}$ is a function with domain 
$\bmcpu_j^{t+j}$ and range $\bmcpu_j^{t+j}$:
\be
\label{beta1}
\beta_j^{t+j}: \bmcpu_j^{t+j}\times (\uut{j+1}{t+j})\ra\bmcpu_j^{t+j}.
\ee
We decompose (\ref{beta1}) into two functions by dividing range
$\bmcpu_j^{t+j}$ into two pieces:  $\bmcpu_{j,>j}^{t+j}$ and $U_{j,j}^{t+j}$,
where $\bmcpu_{j,>j}^{t+j}$ are all the vectors $\bmu_j^{t+j}\in\bmcpu_j^{t+j}$ 
except component $u_{j,j}^{t+j}$ is deleted; denote these vectors by 
$\bmu_{j,>j}^{t+j}$.
The first function, defined to be $\beta_{j,>j}^{t+j}$, has domain 
$\bmcpu_j^{t+j}$ and range $\bmcpu_{j,>j}^{t+j}$ for fixed 
$(\uut{j+1}{t+j})\in\times\cuut{j+1}{t+j}$:
\be
\label{beta2}
\beta_{j,>j}^{t+j}: \bmcpu_j^{t+j}\times (\uut{j+1}{t+j})\ra\bmcpu_{j,>j}^{t+j}.
\ee
The second function,
defined to be $\beta_{j,j}^{t+j}$, has domain $\bmcpu_j^{t+j}$ and
range $U_{j,j}^{t+j}$ for fixed 
$(\uut{j+1}{t+j})\in\cuut{j+1}{t+j}$:
\be
\label{beta3}
\beta_{j,j}^{t+j}: \bmcpu_j^{t+j}\times (\uut{j+1}{t+j})\ra U_{j,j}^{t+j}.
\ee
(At this point we do not assume that $\beta_{j,j}^{t+j}$ is a restricted
separating function.)
Since any $\bmu_j^{t+j}\in\bmcpu_j^{t+j}$ can be uniquely expressed as
$\bmu_j^{t+j}=(\bmu_{j,>j}^{t+j},u_{j,j}^{t+j})$, it is clear that given
$\beta_j^{t+j}$ in (\ref{beta1}), then $\beta_{j,>j}^{t+j}$ in (\ref{beta2})
and $\beta_{j,j}^{t+j}$ in (\ref{beta3}) are completely specified, and
the reverse is also true.  Thus specifying $\beta_{j,>j}^{t+j}$ 
and $\beta_{j,j}^{t+j}$ will completely specify $\beta_j^{t+j}$.
We will see that $\beta_{j,>j}^{t+j}$ is completely specified by 
(\ref{fv14b}), but (\ref{fv14b}) has nothing to say about 
$\beta_{j,j}^{t+j}$.

We first determine $\beta_{j,>j}^{t+j}$.  Since the range of
$\beta_{j,>j}^{t+j}$ is $\bmcpu_{j,>j}^{t+j}$, we can rewrite
(\ref{fv14b}) as
\be
\label{fv14b2}
\beta_{j,>j}^{t+j}(\uut{j}{t+j})=\beta_{j+1}^{t+j+1}(\pvpvt{j}{t+j}).
\ee
The function $\beta_{j,>j}^{t+j}$ on the left hand side of (\ref{fv14b2})
is a function of $\tridnut{j}{j}{t+j}$, and
$\beta_{j+1}^{t+j+1}$ on the right hand side is a function of 
$\tridnutbs{j+1}{j+1}{t+j}$.  Clearly $\tridnutbs{j+1}{j+1}{t+j}$
is a shift of $\tridnut{j}{j}{t+j}$.  We can divide 
$\tridnut{j}{j}{t+j}$ into two pieces.  One piece is $\tridnut{j}{j+1}{t+j}$,
and the other piece is the remaining diagonal terms in 
$\tridnut{j}{j}{t+j}-\tridnut{j}{j+1}{t+j}$; denote the later piece
by $\nearrow_{j,j}(\bmu^{t+j})$.  Note that the
first piece shifts to $\tridnutbs{j+1}{j+1}{t+j}$, and the second
piece, the diagonal terms, shifts out.  In fact, aside from time
index, the first piece is identical in integer values to 
$\tridnutbs{j+1}{j+1}{t+j}$.  We refer to this by saying $\tridnut{j}{j}{t+j}$
is {\it shift equivalent} to $\tridnutbs{j+1}{j+1}{t+j}$ on $\tridnut{j}{j+1}{t+j}$,
written as $\tridnut{j}{j}{t+j}\cong\tridnutbs{j+1}{j+1}{t+j}$.

For fixed $\tridnutbs{j+1}{j+1}{t+j}$, the elements 
$\tridnwt{j}{j}{t+j}\in\tridncut{j}{j}{t+j}$
which shift to $\tridnutbs{j+1}{j+1}{t+j}$ are all the elements in
which $\tridnwt{j}{j+1}{t+j}$ is the same as 
$\tridnut{j}{j+1}{t+j}$, but the remaining diagonal of terms
$\nearrow_{j,j}(\bmw^{t+j})$ can be anything.  We refer to
such elements by saying $\tridnwt{j}{j}{t+j}$
is {\it shift equivalent} to $\tridnutbs{j+1}{j+1}{t+j}$ on $\tridnwt{j}{j+1}{t+j}$,
written as $\tridnwt{j}{j}{t+j}\cong\tridnutbs{j+1}{j+1}{t+j}$.
Therefore any element $\tridnwt{j}{j}{t+j}$ which is shift equivalent to
$\tridnutbs{j+1}{j+1}{t+j}$ on $\tridnwt{j}{j+1}{t+j}$
will shift to $\tridnutbs{j+1}{j+1}{t+j}$.

Again fix $\tridnutbs{j+1}{j+1}{t+j}$.  Then the value
$\bmu_{j+1}^{t+j+1}$ of $\beta_{j+1}^{t+j+1}$ on the right hand side 
of (\ref{fv14b2}) is fixed.  Now examine $\beta_{j,>j}^{t+j}$ on 
the left hand side.  With the right hand side fixed, the value
$\bmu_{j,>j}^{t+j}$ of $\beta_{j,>j}^{t+j}(\bmw^{t+j})$ must be
the same for the set of all elements $\bmw^{t+j}$ such that 
$\tridnwt{j}{j}{t+j}\cong\tridnutbs{j+1}{j+1}{t+j}$.
We can look at the function $\beta_{j,>j}^{t+j}(\bmw^{t+j})$
of all these elements $\bmw^{t+j}$ in a slightly different way.
The only components of $\bmw^{t+j}$ which remain fixed in this set
are $\tridnwt{j}{j+1}{t+j}$.  Therefore we can regard
$\beta_{j,>j}^{t+j}$ as a map from these fixed 
values $\tridnut{j}{j+1}{t+j}$ to $\bmu_{j,>j}^{t+j}$,
\be
\label{tn7}
\beta_{j,>j}^{t+j}:  \tridnut{j}{j+1}{t+j}\mapsto\bmu_{j,>j}^{t+j},
\ee
which ignores components in the diagonal of terms $\nearrow_{j,j}(\bmw^{t+j})$.  
In other words, $\beta_{j,>j}^{t+j}$ is a function of the form
\be
\label{eqncc}
\beta_{j,>j}^{t+j}:  \nearrow_{j,j}(\bllt^{t+j})\times
\tridncut{j}{j+1}{t+j}\ra\bmcpu_{j,>j}^{t+j}.
\ee
On the right hand side of (\ref{fv14b2}) we know that $\beta_{j+1}^{t+j+1}$
is an assignment of $\tridnutbs{j+1}{j+1}{t+j}$ to $\bmu_{j+1}^{t+j+1}$,
\be
\label{tn8}
\beta_{j+1}^{t+j+1}:  \tridnutbs{j+1}{j+1}{t+j}\mapsto\bmu_{j+1}^{t+j+1}.
\ee
Therefore (\ref{fv14b2}) reduces to 
\be
\label{tn78}
\beta_{j,>j}^{t+j}(\nearrow_{j,j}(\bllt^{t+j}),\tridnut{j}{j+1}{t+j})
=\beta_{j+1}^{t+j+1}(\tridnutbs{j+1}{j+1}{t+j}).
\ee

But we know that for $\beta_{j+1}^{t+j+1}$, the assignment of
(\ref{tn8}) is an assignment
$$
\beta_{j+1}^{t+j+1}:  \bsigma\bmu_j^{t+j}\times
\tridnutbs{j+2}{j+2}{t+j}\mapsto\bmu_{j+1}^{t+j+1},
$$
which is 1-1 and onto from $\bmcpu_{j+1}^{t+j+1}$ to 
$\bmcpu_{j+1}^{t+j+1}$ for each fixed $\tridnutbs{j+2}{j+2}{t+j}$.
Moreover from (\ref{tn78}), the assignment of (\ref{tn7}) must be
identical, aside from time index, to the assignment of (\ref{tn8}).
Therefore we must have that the assignment of (\ref{tn7}) is
$$
\beta_{j,>j}^{t+j}:  \bmu_{j,>j}^{t+j}
\times\tridnut{j+1}{j+2}{t+j}\mapsto\bmu_{j,>j}^{t+j},
$$
which is 1-1 and onto from $\bmcpu_{j,>j}^{t+j}$ to 
$\bmcpu_{j,>j}^{t+j}$ for each fixed $\tridnut{j+1}{j+2}{t+j}$.
Then we can rewrite (\ref{tn78}) as
\be
\label{tn78a}
\beta_{j,>j}^{t+j}(\nearrow_{j,j}(\bllt^{t+j}),\bmu_{j,>j}^{t+j},\tridnut{j+1}{j+2}{t+j})
=\beta_{j+1}^{t+j+1}(\bsigma\bmu_{j,>j}^{t+j},\tridnutbs{j+2}{j+2}{t+j}).
\ee
Note that $\tridnut{j}{j}{t+j}$ is shift equivalent
to $(\bsigma\bmu_{j,>j}^{t+j},\tridnutbs{j+2}{j+2}{t+j})$
on $\tridnut{j}{j+1}{t+j}=(\bmu_{j,>j}^{t+j},\tridnut{j+1}{j+2}{t+j})$.  
Therefore we refer to the property of $\beta_{j,>j}^{t+j}$ given in (\ref{tn78a})
by saying $\beta_{j,>j}^{t+j}$ is shift equivalent to 
$\beta_{j+1}^{t+j+1}$ on $\tridncut{j}{j+1}{t+j}$, and write this
as $\beta_{j,>j}^{t+j}\cong\beta_{j+1}^{t+j+1}$.

We know $\beta_{j,>j}^{t+j}$ is a function of the form (\ref{eqncc})
which is 1-1 and onto from $\bmcpu_{j,>j}^{t+j}$ to 
$\bmcpu_{j,>j}^{t+j}$ for each fixed $\tridnut{j+1}{j+2}{t+j}$.
It follows that $\beta_{j,>j}^{t+j}$ is a function
which is 1-1 and onto from $\bmcpu_{j,>j}^{t+j}$ to 
$\bmcpu_{j,>j}^{t+j}$ for each fixed $(\uut{j+1}{t+j})\in\cuut{j+1}{t+j}$.
But $\beta_j^{t+j}$ in (\ref{fv14b}) must be a restricted separating
function which is 1-1 and onto from $\bmcpu_j^{t+j}$ to $\bmcpu_j^{t+j}$
for each fixed $(\uut{j+1}{t+j})\in\cuut{j+1}{t+j}$.
Therefore in order for $\beta_j^{t+j}$ to have this property, 
it is necessary and sufficient that $\beta_{j,j}^{t+j}$ be any 
function of the form
\be
\label{beta6a}
\beta_{j,j}^{t+j}:  \tridncut{j}{j}{t+j}\ra U_{j,j}^{t+j}
\ee
which is 1-1 and onto from $U_{j,j}^{t+j}$ to $U_{j,j}^{t+j}$ for each fixed 
$(\bmu_{j,>j}^{t+j},\uut{j+1}{t+j})\in\bmcpu_{j,>j}^{t+j}\times\cuut{j+1}{t+j}$,
\be
\label{beta6}
\beta_{j,j}^{t+j}:
U_{j,j}^{t+j}\times (\bmu_{j,>j}^{t+j},\uut{j+1}{t+j})\ra U_{j,j}^{t+j}.
\ee
In other words, $\beta_{j,j}^{t+j}$ must be a restricted separating function.
With $\beta_{j,>j}^{t+j}$ specified as in (\ref{eqncc}) and (\ref{tn78a}),
and $\beta_{j,j}^{t+j}$ specified as in (\ref{beta6a}) and (\ref{beta6}),
$\beta_j^{t+j}$ is completely determined, and $\beta_j^{t+j}$ is a restricted
separating function with the desired properties.

We can summarize these results as follows.

\begin{thm}
\label{thm55a}
The solution $\beta_j^{t+j}$ of (\ref{fv14b}) is a restricted separating
function composed of a function $\beta_{j,>j}^{t+j}$, 
given in (\ref{eqncc}) and (\ref{tn78a}),
and a function $\beta_{j,j}^{t+j}$, 
given in (\ref{beta6a}) and (\ref{beta6}).  The function $\beta_{j,>j}^{t+j}$ is
shift equivalent to $\beta_{j+1}^{t+j+1}$ on $\tridncut{j}{j+1}{t+j}$.
Therefore, it is 1-1 and onto from $\bmcpu_{j,>j}^{t+j}$ to 
$\bmcpu_{j,>j}^{t+j}$ for each fixed $\tridnut{j+1}{j+2}{t+j}$.
The function $\beta_{j,j}^{t+j}$ is a restricted separating function.
\end{thm}

This gives the following algorithm.

\begin{alg}
\label{alg55} 
Any solution of the set of equations (\ref{fv8}) or (\ref{fv8r}) which is a symmetry 
$\bomga=\ldots,\bomga^t,\bomga^{t+1},\ldots$ of $\duinf$ can be found 
as follows.  

\vspace{2mm}
\noindent {\bf DO} \\
1.  Fix time $t$.

\vspace{2mm}
\noindent 2.  Let $\beta_\ell^{t+\ell}:  \bmcpu_\ell^{t+\ell}\ra\bmcpu_\ell^{t+\ell}$ 
be any restricted separating function.  Define $\bomga_\ell^{t+\ell}$ to be the 
unique separating function whose restriction is $\beta_{\ell}^{t+\ell}$.

\vspace{2mm}
\noindent 3. \\
{\bf FOR} $j=\ell-1,\ldots,0$ (counting down in order),

(i) Find the unique function 
$\beta_{j,>j}^{t+j}:  \cuut{j}{t+j}\ra\bmcpu_{j,>j}^{t+j}$ of the form
(\ref{eqncc}) that satisfies (\ref{tn78a}), or in other words, $\beta_{j,>j}^{t+j}$ 
is shift equivalent to $\beta_{j+1}^{t+j+1}$ on $\tridncut{j}{j+1}{t+j}$,
$$
\beta_{j,>j}^{t+j}\cong\beta_{j+1}^{t+j+1}.
$$

(ii) Define any restricted separating function
$\beta_{j,j}^{t+j}: \cuut{j}{t+j}\ra U_{j,j}^{t+j}$.

\vspace{2mm}
\noindent Now combine the $\beta_{j,>j}^{t+j}$ and $\beta_{j,j}^{t+j}$ to 
form $\beta_j^{t+j}$. \\
{\bf FOR} each $(\uut{j}{t+j})\in\cuut{j}{t+j}$, \\

(i) Define $\bmuht_{j,>j}^{t+j}\in\bmcpu_{j,>j}^{t+j}$ by
$$
\bmuht_{j,>j}^{t+j}\rmdef\beta_{j,>j}^{t+j}(\uut{j}{t+j}),
$$ 
and define $\uht_{j,j}^{t+j}\in U_{j,j}^{t+j}$ by
$$
\uht_{j,j}^{t+j}\rmdef\beta_{j,j}^{t+j}(\uut{j}{t+j}).
$$ 

(ii) Define $\beta_j^{t+j}:  \cuut{j}{t+j}\ra\bmcpu_j^{t+j}$
by $\beta_j^{t+j}(\uut{j}{t+j})\rmdef\bmuht_j^{t+j}$, where
$$
\bmuht_j^{t+j}=((\bmuht_{j,>j}^{t+j})^T,\uht_{j,j})^T.
$$
{\bf ENDFOR}

\vspace{2mm}
\noindent Define $\bomga_j^{t+j}$ to be the unique separating function
whose restriction is $\beta_{j}^{t+j}$. \\
{\bf ENDFOR} \\
{\bf ENDDO}

\vspace{2mm}
\noindent 4.  Step 4 same as in Algorithm \ref{alg54}.
\end{alg}

We now show that Theorem \ref{thm55a} and Algorithm \ref{alg55} can be refined
to use separating functions $\omega_{j,k}^{t+j}$ and restricted
separating functions $\beta_{j,k}^{t+j}$.  We first show how to construct 
a function $f_j^{t+j}:  \cuut{j}{t+j}\ra\bmcpu_j^{t+j}$ 
(not necessarily a restricted separating function) using the set of 
restricted separating functions $\{\beta_{j,k}^{t+j}:  j\le k\le\ell\}$.

\begin{defn}
\label{defn3}
{\bf (Construction of $f_j^{t+j}$)}

Let $\{\beta_{j,k}^{t+j}:  j\le k\le\ell\}$ be a set of restricted separating
functions, as defined in Definition \ref{defn2}.
Define a function $f_j^{t+j}:  \cuut{j}{t+j}\ra\bmcpu_j^{t+j}$
as follows.

\vspace{2mm}
\noindent {\bf FOR} each fixed $\tridnut{j}{j}{t+j}\in\tridncut{j}{j}{t+j}$, \\
{\bf FOR} each $k$ such that $j\le k\le\ell$, \\
define $v_{j,k}^{t+j}\in U_{j,k}^{t+j}$ by \\
\be
\label{constr1}
v_{j,k}^{t+j}\rmdef\beta_{j,k}^{t+j}(\tridnut{j}{k}{t+j}).
\ee
{\bf ENDFOR} \\

\vspace{2mm}
\noindent Define $f_j^{t+j}(\tridnut{j}{j}{t+j})$
to be the vector $\bmv_j^{t+j}$ in $\bmcpu_j^{t+j}$ given by
\be
\label{constr2}
\bmv_j^{t+j}=\left(
\begin{array}{lllll}
v_{j,\ell}^{t+j} & \!\!\cdots\!\! & v_{j,k}^{t+j} & \!\!\cdots\!\! & v_{j,j}^{t+j} 
\end{array}
\right)^T.
\ee
{\bf ENDFOR}
\end{defn}

Consistent with (\ref{constr1}) and (\ref{constr2}), 
we can represent $f_j^{t+j}$ by the vector of functions
$$
f_j^{t+j}=\left(
\begin{array}{lllll}
\beta_{j,\ell}^{t+j} & \!\!\cdots\!\! & \beta_{j,k}^{t+j} & \!\!\cdots\!\! & \beta_{j,j}^{t+j} 
\end{array}
\right)^T.
$$
If Definition \ref{defn3} holds, we say $f_j^{t+j}$ is
{\it constructed from} the set of restricted separating
functions $\{\beta_{j,k}^{t+j}:  j\le k\le\ell\}$.
Given a set of restricted separating functions $\{\beta_{j,k}^{t+j}:  j\le k\le\ell\}$
as defined in Definition \ref{defn2}, the construction in 
Definition \ref{defn3} gives a unique function $f_j^{t+j}$.

\begin{thm}[Induction hypothesis]
\label{ind1}
Assume the function $\beta_{j+1}^{t+j+1}$ on the right hand side 
of (\ref{fv14b}) is a restricted separating function 
$\beta_{j+1}^{t+j+1}:  \cuut{j+1}{t+j+1}\ra\bmcpu_{j+1}^{t+j+1}$ such that
$\beta_{j+1}^{t+j+1}$ is constructed from a set of restricted separating
functions $\{\beta_{j+1,k}^{t+j+1}\}$, where $\beta_{j+1,k}^{t+j+1}$
is shift equivalent to $\beta_{k,k}^{t+k}$ on $\tridncut{j+1}{k}{t+j+1}$,
\be
\label{confrm1}
\beta_{j+1,k}^{t+j+1}\cong\beta_{k,k}^{t+k},
\ee
for $k$ such that $j+1<k\le\ell$, and where 
\be
\label{confrm2}
\beta_{j+1,j+1}^{t+j+1}(\bmu^{t+j+1})
\ee
is any restricted separating function, for $k=j+1$.

Then there exists a solution $\beta_j^{t+j}$ on the left hand side 
of (\ref{fv14b}) which is a restricted separating function 
$\beta_j^{t+j}:  \cuut{j}{t+j}\ra\bmcpu_j^{t+j}$ such that
$\beta_j^{t+j}$ is constructed from the set of restricted separating
functions $\{\beta_{j,k}^{t+j}\}$, where $\beta_{j,k}^{t+j}$
is shift equivalent to $\beta_{k,k}^{t+k}$ on $\tridncut{j}{k}{t+j}$,
\be
\label{confrm3}
\beta_{j,k}^{t+j}\cong\beta_{k,k}^{t+k},
\ee
for $k$ such that $j<k\le\ell$, and where 
\be
\label{confrm4}
\beta_{j,j}^{t+j}(\bmu^{t+j})
\ee
is any restricted separating function, for $k=j$.
\end{thm}

\begin{prf}
We use proof by induction.  Assume we have found
$\beta_{j+1}^{t+j+1}$ on the right hand side of (\ref{fv14b}), 
and assume $\beta_{j+1}^{t+j+1}$ can be constructed from the set of 
restricted separating functions $\{\beta_{j+1,k}^{t+j+1}:  j+1\le k\le\ell\}$.  
We then show the solution $\beta_j^{t+j}$ on the left hand side of 
(\ref{fv14b}) can be constructed from a set of restricted separating functions
$\{\beta_{j,k}^{t+j}:  j\le k\le\ell\}$, which are related to the set 
$\{\beta_{j+1,k}^{t+j+1}:  j+1\le k\le\ell\}$.

From (\ref{tn78a}), we have that
\be
\label{tn10}
\beta_{j,>j}^{t+j}(\nearrow_{j,j}(\bllt^{t+j}),\bmu_{j,>j}^{t+j},\tridnut{j+1}{j+2}{t+j})
=\beta_{j+1}^{t+j+1}(\bsigma\bmu_{j,>j}^{t+j},\tridnutbs{j+2}{j+2}{t+j}),
\ee
where
$$
\bmu_{j,>j}^{t+j}=\left(
\begin{array}{lllll}
u_{j,\ell}^{t+j} & \!\!\cdots\!\! & u_{j,k}^{t+j} & \!\!\cdots\!\! & u_{j,j+1}^{t+j} 
\end{array}
\right)^T,
$$
and 
$$
\bsigma\bmu_{j,>j}^{t+j}=\left(
\begin{array}{lllll}
\sigma u_{j,\ell}^{t+j} & \!\!\cdots\!\! & \sigma u_{j,k}^{t+j} & \!\!\cdots\!\! & \sigma u_{j,j+1}^{t+j} 
\end{array}
\right)^T.
$$
On the right hand side of (\ref{tn10}), we know that
$\beta_{j+1}^{t+j+1}$ is constructed from the set of restricted 
separating functions $\{\beta_{j+1,k}^{t+j+1}:  j+1\le k\le\ell\}$.
If $\beta_{j+1}^{t+j+1}(\bsigma\bmu^{t+j})
=\bmv_{j+1}^{t+j+1}\in\bmcpu_{j+1}^{t+j+1}$, where
\be
\bmv_{j+1}^{t+j+1}=\left(
\begin{array}{lllll}
v_{j+1,\ell}^{t+j+1} & \!\!\cdots\!\! & v_{j+1,k}^{t+j+1} & \!\!\cdots\!\! & v_{j+1,j+1}^{t+j+1} 
\end{array}
\right)^T,
\ee
then $\beta_{j+1}^{t+j+1}$ can be represented by the vector of functions
$$
\beta_{j+1}^{t+j+1}=\left(
\begin{array}{lllll}
\beta_{j+1,\ell}^{t+j+1} & \!\!\cdots\!\! & \beta_{j+1,k}^{t+j+1} & \!\!\cdots\!\! & \beta_{j+1,j+1}^{t+j+1} 
\end{array}
\right)^T,
$$
where the $k$-th coordinate $\beta_{j+1,k}^{t+j+1}$ of $\beta_{j+1}^{t+j+1}$
gives the $k$-th coordinate $v_{j+1,k}^{t+j+1}$ of $\bmv_{j+1}^{t+j+1}$.
Fix $k$ such that $j+1\le k\le\ell$.  Then
\begin{align*}
v_{j+1,k}^{t+j+1} &=\beta_{j+1,k}^{t+j+1}(\bsigma\bmu^{t+j}) \\
&=\beta_{j+1,k}^{t+j+1}(\sigma u_{j,k}^{t+j},\btridnutbs{j+1}{k}{t+j}).
\end{align*}

Let $\beta_{j,>j}^{t+j}(\tridnut{j}{j}{t+j})=\bmv_{j,>j}^{t+j}\in\bmcpu_{j,>j}^{t+j}$, 
where
\be
\bmv_{j,>j}^{t+j}=\left(
\begin{array}{lllll}
v_{j,\ell}^{t+j} & \!\!\cdots\!\! & v_{j,k}^{t+j} & \!\!\cdots\!\! & v_{j,j+1}^{t+j} 
\end{array}
\right)^T.
\ee
For $k$ such that $j+1\le k\le\ell$, let 
$f_{j,k}^{t+j}:  \tridncut{j}{j}{t+j}\ra U_{j,k}^{t+j}$
be the function which gives the $k$-th coordinate $v_{j,k}^{t+j}$
of $\bmv_{j,>j}^{t+j}$.  Then we can represent $\beta_{j,>j}^{t+j}$ 
by the vector of functions
$$
\beta_{j,>j}^{t+j}=\left(
\begin{array}{lllll}
f_{j,\ell}^{t+j} & \!\!\cdots\!\! & f_{j,k}^{t+j} & \!\!\cdots\!\! & f_{j,j+1}^{t+j} 
\end{array}
\right)^T.
$$
Then from (\ref{tn10}), we must have
\begin{multline}
\label{tn11}
f_{j,k}^{t+j}(\nearrow_{j,j}(\bllt^{t+j}),\bmu_{j,>j}^{t+j},\tridnut{j+1}{j+2}{t+j}) \\
=\beta_{j+1,k}^{t+j+1}(\sigma u_{j,k}^{t+j},\btridnutbs{j+1}{k}{t+j}).
\end{multline}
Now use the same argument as given for finding $\beta_j^{t+j}$
given $\beta_{j+1}^{t+j+1}$.  The right hand side of (\ref{tn11})
is a function of $(\sigma u_{j,k}^{t+j},\btridnutbs{j+1}{k}{t+j})$.
Therefore the left hand side must be a function
of the set $(u_{j,k}^{t+j},\btridnut{j}{k}{t+j})$ which is shift 
equivalent to $(\sigma u_{j,k}^{t+j},\btridnutbs{j+1}{k}{t+j})$,
and independent of other components.  Therefore the left hand side is
some function $h_{j,k}^{t+j}:  \tridncut{j}{k}{t+j}\ra U_{j,k}^{t+j}$
such that
\be
\label{tn12}
h_{j,k}^{t+j}(u_{j,k}^{t+j},\btridnut{j}{k}{t+j})
=\beta_{j+1,k}^{t+j+1}(\sigma u_{j,k}^{t+j},\btridnutbs{j+1}{k}{t+j}).
\ee
And since $\beta_{j+1,k}^{t+j+1}$ is a function 1-1 and onto 
from $U_{j+1,k}^{t+j+1}$ to $U_{j+1,k}^{t+j+1}$ for each fixed
$\btridnutbs{j+1}{k}{t+j}$, then $h_{j,k}^{t+j}$ must be a
function 1-1 and onto from $U_{j,k}^{t+j}$ to $U_{j,k}^{t+j}$ 
for each fixed $\btridnut{j}{k}{t+j}$.  In other words,
$h_{j,k}^{t+j}$ is a restricted separating function $\beta_{j,k}^{t+j}$,
and (\ref{tn12}) gives $\beta_{j,k}^{t+j}\cong\beta_{j+1,k}^{t+j+1}$
on $\tridncut{j}{k}{t+j}$.  Since $\beta_{j+1,k}^{t+j+1}\cong\beta_{k,k}^{t+k}$
on $\tridncut{j+1}{k}{t+j+1}$ from (\ref{confrm1}), then we see that
$\beta_{j,k}^{t+j}\cong\beta_{k,k}^{t+k}$ on $\tridncut{j}{k}{t+j}$.
\end{prf}

Clearly the induction hypothesis holds for $j+1=\ell$ because
$\beta_\ell^{t+\ell}$ is a restricted separating function 
$\beta_{\ell,\ell}^{t+\ell}$.  This completes the proof by induction.
Thus we have proven the following algorithm, using results 
(\ref{confrm3}) and (\ref{confrm4}) above.

\begin{alg} 
\label{alg58}
Any solution of the set of equations (\ref{fv8}) which is a symmetry 
$\bomga=\ldots,\bomga^t,\bomga^{t+1},\ldots$ of $\duinf$ can be found 
as follows.  

\vspace{2mm}
\noindent {\bf DO} \\
1.  Fix time $t$.

\vspace{2mm}
\noindent 2. \\
{\bf FOR} $k=\ell,\ldots,0$, \\
define any separating function
$\omega_{k,k}^{t+k}: \bmcpu^{t+k}\ra U_{k,k}^{t+k}$.

\vspace{2mm}
\noindent {\bf FOR} $j$ satisfying $0\le j<k$, \\
define a separating function
$\omega_{j,k}^{t+j}: \bmcpu^{t+j}\ra U_{j,k}^{t+j}$ by
\be
\label{eq99h}
\omega_{j,k}^{t+j}\cong\omega_{k,k}^{t+k}
\ee
on $\tridncut{j}{k}{t+j}$. \\
{\bf ENDFOR}

\vspace{2mm}
\noindent {\bf ENDFOR} \\
{\bf ENDDO}

\vspace{2mm}
\noindent 3.  Now combine the $\omega_{j,k}^t$ directly 
to form $\bomga^t$.
\end{alg}

\begin{prf}
For $0\le j\le k$, $0\le k\le\ell$, let $\omega_{j,k}^{t+j}$ be the unique
separating function whose restriction is $\beta_{j,k}^{t+j}$.  Now use the
induction hypothesis Theorem \ref{ind1}.
\end{prf}

Algorithm \ref{alg58} shows that we only need a few separating
functions to determine $\bomga^t$.

\begin{thm}
\label{sep}
Any symmetry $\bomga$ of $\duinf$ is uniquely specified 
by the collection of separating functions 
$\omega_{k,k}^{t+k}(\bmu^{t+k})$, for $k$ such that $0\le k\le\ell$, 
for each $t\in\bmcpz$.
\end{thm}

In this subsection we have given three algorithms to construct all
the symmetries of $\duinf$.  Algorithm \ref{alg54} is the basic algorithm.
It can be shown that Algorithm \ref{alg55} is the best algorithm to construct
any group system $C$.  Algorithm \ref{alg58} is a very simple
algorithm and the best for finding all the symmetries of $\duinf$.

\vspace{3mm}
{\bf 7.3  The full symmetry system}
\vspace{3mm}


In the same way as (\ref{twrphi}), we can diagram Step 4 of Algorithms
\ref{alg54} and \ref{alg55} as shown in (\ref{twrbeta}).
\be
\label{twrbeta}
\setcounter{MaxMatrixCols}{7}
\begin{pmatrix}   
               &              &             & \vdots          & &          &             \\
               &              &             &                 & &          & \bomga_\ell^{(t)+\ell}  \\
               &              &             &                 & &          & \vdots      \\
               &              & \cdots      & \bomga_j^{(t)+j} & \cdots &  &             \\
               &              &             & \vdots          & &          & \bomga_\ell^{(t-j)+\ell}  \\
\cdots         & \bomga_1^{(t)+1} & \cdots  &                 & &          &  \vdots     \\
\bomga_0^{(t)} & \bomga_1^{(t-1)+1} & \cdots & \bomga_j^{(t-j)+j} & \cdots & & \bomga_\ell^{(t-\ell)+\ell} \\
\vdots         & \vdots       &             & \vdots          & &          & \vdots      \\
\cdots         & \bomga_1^{(t-j)+1} & \cdots &                & &          &             \\
\bomga_0^{(t-j)} & \cdots     &             &                 & &          &             \\
               &              &             & \vdots          & &          &
\end{pmatrix}
\ee
Equation (\ref{twrbeta}) shows a $t$-tower $\Upsilon^t$
and $(t-j)$-tower $\Upsilon^{t-j}$
as diagonals in an infinite matrix of towers.  Algorithms
\ref{alg54} and \ref{alg55} show that the sequence of towers 
$\ldots,\Upsilon^{t-j},\ldots,\Upsilon^t,\ldots$ in (\ref{twrbeta})
defines a symmetry and any such matrix (\ref{twrbeta}) of towers
defines a symmetry.  A component $\bomga^t$ of the symmetry $\bomga$
is defined by going ``across the row'' in (\ref{twrbeta}),
\begin{align}
\label{eq61a}
\bomga^t 
&=(\bomga_0^{(t)},\bomga_1^{(t-1)+1},\ldots,\bomga_j^{(t-j)+j},\ldots,\bomga_{\ell}^{(t-\ell)+\ell}) \\
\label{eq62a}
&=(\bomga_0^t,\bomga_1^t,\ldots,\bomga_j^t,\ldots,\bomga_{\ell}^t),
\end{align}
and more explicitly, $\bomga^t:  \bmcpu^t\ra\bmcpu^t$ is defined by
$$
\bomga^t(\bmu^t)=(\bomga_0^t(\bmu^t),\bomga_1^t(\bmu^t),\ldots,
\bomga_j^t(\bmu^t),\ldots,\bomga_\ell^t(\bmu^t)).
$$
Note that each component $\bomga_j^t$ in (\ref{eq62a}) is selected from
a different tower.

Let $\calm$ be the \fss\ of $\calu$ obtained using Algorithms \ref{alg54}, \ref{alg55}, 
or \ref{alg58} to find each symmetry $\bomga\in \calm$.  As just discussed,
we have the following.

\begin{thm}
The full symmetry system $\calm$ is a set of tensors.
\end{thm}

And at the beginning of Section 7, we noted this.

\begin{thm}
The full symmetry system $\calm$ is a group system.
\end{thm}

Consider the 5-tuple family $(\calm,\calu,\calr,C;\bmcpb)$, which includes the 
4-tuple family $(\calu,\calr,C;\bmcpb)$ already considered in Section 6.
$\calm$ is a tensor set like $\calu$ and $\calr$.  And $\calm$ is also a
group system like $C$.  $\calm$ only depends on $\calu$ and does not depend
on basis $\bmcpb$.  $\calm$ acts on $\calu$.  Since there is a 1-1 correspondence
$\calu\lra\calr\lra C$, $\calm$ implicitly acts on $\calr$ and $C$ also.
The induced action of $\calm$ on $C$ means symmetry $\bomga\in\calm$ gives
a permutation of the paths of $C$.

From the form of (\ref{twrbeta}) and (\ref{eq61a})-(\ref{eq62a}),
we can regard $\bomga_0^{(t)}$, or $\bomga_0^t$, 
as the {\it input} at time $t$, and
$(\bomga_1^{(t-1)+1},\ldots,\bomga_j^{(t-j)+j},\ldots,\bomga_{\ell}^{(t-\ell)+\ell})$,
or $(\bomga_1^t,\ldots,\bomga_j^t,\ldots,\bomga_{\ell}^t)$,
as the {\it state} at time $t$.  Note that the state at time $t$ is composed
of shifts of previous inputs, i.e., $\bomga_j^{(t-j)+j}$ is a shift of
the input $\bomga_j^{(t-j)}$ at time $t-j$.

Because the structure of a symmetry $\bomga$ mirrors the structure of
a tensor $\bmu\in\calu$, we see that component $\bomga^t$ has the same
form as component $\bmu^t$ in tensor $\bmu$, which is a \stm\
$U^{[t,t]}$; therefore we also call component $\bomga^t$ a
{\it static matrix} $\Omega^{[t,t]}$.  The set of all
\stms\ $\Omega^{[t,t]}$ is $\bcomega^t$.

Recall that $\bomga_j^{t+j}$ is a column vector with components
$\omega_{j,k}^{t+j}$,
$$
\bomga_j^{t+j}=\left(
\begin{array}{lllll}
\omega_{j,\ell}^{t+j} & \!\!\cdots\!\! & \omega_{j,k}^{t+j} & \!\!\cdots\!\! & \omega_{j,j}^{t+j}
\end{array}
\right)^T,
$$
where $\omega_{j,k}^{t+j}$ is defined in Definition \ref{defn2}.
Therefore in the same way as (\ref{utnsr}), we regard the diagonals of 
(\ref{twrbeta}) as columns $\bomga_j^{t+j}$ in a 
{\it shift matrix} $\Omega^{[t,t+\ell]}$,
$$
\Omega^{[t,t+\ell]}=(\bomga_0^t,\bomga_1^{t+1},\ldots,\bomga_j^{t+j},\ldots,\bomga_\ell^{t+\ell}).
$$
A diagonal of (\ref{twrbeta}) is a $t$-tower.  Therefore
a $t$-tower $\Upsilon^t$ is a \sm\ $\Omega^{[t,t+\ell]}$ at time $t$.
A {\it shift vector} $\bomga^{[t,t+k]}$ is a row in $\Omega^{[t,t+\ell]}$,
for $0\le k\le\ell$, where
$$
\bomga^{[t,t+k]}
\rmdef (\omega_{0,k}^t,\omega_{1,k}^{t+1},\ldots,\omega_{j,k}^{t+j},\ldots,\omega_{k,k}^{t+k}).
$$  
The \svec\ is determined by shifts of the separating permutation $\omega_{k,k}^{t+k}$.

Step 4 of Algorithms \ref{alg54} and \ref{alg55}
can be viewed as the construction of the \stm\ $\bomga^t=\Omega^{[t,t]}$
using a sequence of \sms, exactly analogous to the procedure in Theorem
\ref{thm31} for \gms.

A path in $C$ is denoted $\bmc$, where
$$
\bmc=\ldots,c^t,\ldots.
$$
Each component $c^t$ is a branch.
Every group system has an identity sequence.    The identity path of
$C$ is the path where each component, or branch, $c^t$, is the
identity $\bone^t$.

A path in $\calm$ is denoted $\bomga$, where
$$
\bomga=\ldots,\bomga^t,\ldots,
$$
and $\bomga^t=(\bomga_0^t,\bomga_1^t,\ldots,\bomga_{\ell}^t)$.
We can think of component $\bomga^t$ as a {\it branch}
in a bipartite or unipartite graph.  The vertices (states) of
the graph are given by $(\bomga_1^t,\ldots,\bomga_{\ell}^t)$,
and the input is $\bomga_0^t$.  The next state is
$\bsigma\bomga^t=\bsigma(\bomga_0^t,\bomga_1^t,\ldots,\bomga_{\ell}^t)$.
This mimics the description of the graph $\dut$.  The identity path
of $\calm$ is the path $\bomga$ where each component, or branch,
$\bomga^t$, is given by $(\bone_0^t,\bone_1^t,\ldots,\bone_{\ell}^t)$.
The identity sequence is obtained using inputs $\bomga_0^t$,
where $\bomga_0^t$ is the identity $\bone_0^t$ for each time $t$.

The equation (\ref{fv8}) was used in the analyis of a
symmetry permutation.  We can think of this equation in
shorthand form as $\sigma\varphi_j^{t+j}=\varphi_{j+1}^{t+j+1}$
for $j=0,\ldots,\ell-1$, and $\varphi_{j+1}^{t+j+1}$ can be
regarded as a ``shift" of $\varphi_j^{t+j}$.  
In the construction of a symmetry, we solved the same equation 
$\sigma\bomga_j^{t+j}=\bomga_{j+1}^{t+j+1}$ going backwards, from
$j=\ell-1$ to $j=0$.  However it is clear that we can also go forward,
and once $\bomga_0^t$ is found, we can find all $\bomga_j^{t+j}$,
$1\le j\le\ell$.  Thus a \sm\ $\Omega^{[t,t+\ell]}$,
$$
\Omega^{[t,t+\ell]}=(\bomga_0^t,\bomga_1^{t+1},\ldots,\bomga_j^{t+j},\ldots,\bomga_\ell^{t+\ell}),
$$
is completely determined by $\bomga_0^t$.  This situation is completely 
analogous to that for a \sm\ $U^{[t,t+\ell]}$, where (\ref{twrbeta})
is analogous to (\ref{utnsr}), \sm\ $U^{[t,t+\ell]}$ is completely
determined by $\bmu_0^t$, and an analogous equation 
$\sigma\bmu_j^{t+j}=\bmu_{j+1}^{t+j+1}$ holds.
This gives the following result.

\begin{prop}
\label{prop55}
A symmetry $\bomga$ in $\calm$ is completely determined by
a sequence of inputs $\bomga_0^t$, for $t\in\bmcpz$.
\end{prop}

Using (\ref{twrbeta}) and Proposition \ref{prop55}, it is easy
to define a sliding block encoder of the \fss.  The encoder
slides along the matrix in (\ref{twrbeta}) from left to right
as time increases.  At each time $t$, a new input $\bomga_0^t$
is selected from a set of inputs.
The encoder output, component $\bomga^t$ of symmetry $\bomga$,
is defined by going ``across the row'' in (\ref{twrbeta}),
as given in (\ref{eq61a})-(\ref{eq62a}).  Note that this
is equivalent to just forming the \stm\ $\Omega^{[t,t]}$.

\begin{thm}
The full symmetry system $\calm$ of $C$ is \ellctl, the same as $C$.
\end{thm}

\begin{prf}
$\calm$ is completely determined by a sequence of inputs, which can be
selected arbitrarily.  Therefore we can go from any state of $\calm$ to 
any other state in $\ell$ steps, by a suitable choice of inputs.
\end{prf}


$\calm^t\rmdef\chi^t(\calm)$ are the time $t$ components of the symmetries 
in $\calm$,
$$
\calm^t\rmdef\{\bomga^t : \bomga=\ldots,\bomga^t,\ldots,\bomga\in \calm\}.
$$
$\calm^t$ is called a {\it branch group}.

An element $\bmu^t\in\bmcpu^t$ is a \stm\ $U^{[t,t]}$.
The \stm\ $U^{[t,t]}$ is permuted by component $\bomga^t$ in 
symmetry $\bomga$, and $\bomga^t$ is a \stm\ $\Omega^{[t,t]}$,
with components $\omega_{j,k}^t$ for $0\le j\le k$, $0\le k\le\ell$.
Component $\omega_{j,k}^t$ in $\Omega^{[t,t]}$ permutes 
component $u_{j,k}^t$ in $U^{[t,t]}$.

We now study the action of the \fss\ on $\calu$.  Fix symmetry $\bomga\in\calm$.
Fix tensor $\bmu\in\calu$.  Fix time $t$ and fix $k$, $0\le k\le\ell$.  
Let $\bmu$ have \svec\ $\bmu^{[t,t+k]}$.  The \svec\ $\bmu^{[t,t+k]}$ 
is a finite sequence
\be
\label{gran1}
(u_{0,k}^t, u_{1,k}^{t+1},\ldots,u_{j,k}^{t+j},\ldots,u_{k,k}^{t+k}),
\ee
where $u_{j,k}^{t+j}$ is the same integer for $0\le j\le k$.
From the form of the solution of the \fss, we know symmetry $\bomga$
acts on this finite sequence with the finite sequence of permutations
\be
\label{perm1}
(\omega_{0,k}^{t},\omega_{1,k}^{t+1},\ldots,\omega_{j,k}^{t+j},\ldots,\omega_{k,k}^{t+k}),
\ee
which is a \svec\ $\bomga^{[t,t+k]}$ in $\bomga$.
Then the action of (\ref{perm1}) on (\ref{gran1}) gives
$$
(\omega_{0,k}^t(u_{0,k}^t;\btridnut{0}{k}{t}),
\omega_{1,k}^{t+1}(u_{1,k}^{t+1};\btridnut{1}{k}{t+1}),\ldots,
\omega_{j,k}^{t+j}(u_{j,k}^{t+j};\btridnut{j}{k}{t+j}),\ldots,
\omega_{k,k}^{t+k}(u_{k,k}^{t+k};\btridnut{k}{k}{t+k})).
$$
But from (\ref{eq99h}), we have
$$
\omega_{j,k}^{t+j}(u_{j,k}^{t+j};\btridnut{j}{k}{t+j})  
=\omega_{k,k}^{t+k}(u_{j,k}^{t+j};\btridnut{j}{k}{t+j})
$$
for $0\le j\le k$.
But the contents of memory $\btridnut{j}{k}{t+j}$ is the same
for $0\le j\le k$, and integer $u_{j,k}^{t+j}$ is the same for $0\le j\le k$.
Then the action of (\ref{perm1}) on (\ref{gran1}) gives 
\be
\label{gran2}
(\uht_{0,k}^t, \uht_{1,k}^{t+1},\ldots,\uht_{j,k}^{t+j},\ldots,\uht_{k,k}^{t+k}),
\ee
where $\uht_{j,k}^{t+j}$ is the same integer for $0\le j\le k$.
But then (\ref{gran2}) is a \svec\ $\bmuht^{[t,t+k]}$ in $\calu$.
Thus the \svec\ (\ref{gran1}) has been changed to \svec\ (\ref{gran2}).

It is clear that the action of $\bomga^{[t,t+k]}$ on $\bmu^{[t,t+k]}$
is completely determined by the first component 
$\omega_{0,k}^t(u_{0,k}^t;\btridnut{0}{k}{t})$.
The argument of $\omega_{0,k}^t$ is a function of $\bmu^t$.
The state of $\bmu$ at time $t$ is $\btridnut{0}{1}{t}$, and the
input at time $t$ is $\bmu_0^t$.  We see that $\omega_{0,k}^t$
is only a function of part of the state, $\btridnut{0}{k}{t}$,
and part of the input, $u_{0,m}^t$, for $k\le m\le\ell$.  Some
special cases are of interest.  For $k=\ell$, $\omega_{0,k}^t$
is only a function of $u_{0,\ell}^t$ and not a function of any
part of the state.  For $k=0$, $\omega_{0,k}^t$ is a function of
all of the state and all of the input.

\begin{thm}
\label{gransub}
Fix symmetry $\bomga\in\calm$.  Fix tensor $\bmu\in\calu$.  
Fix time $t$ and fix $k$, $0\le k\le\ell$.  
Let $\bmu$ have \svec\ $\bmu^{[t,t+k]}$.  The symmetry $\bomga$
permutes \svec\ $\bmu^{[t,t+k]}$ to another \svec\ $\bmuht^{[t,t+k]}$ in $\calu$.
The permutation is solely determined by component $\omega_{0,k}^t$
of symmetry input $\bomga_0^t$ at time $t$.  The argument of 
$\omega_{0,k}^t$ is $\btridnut{0}{k}{t}$ and $u_{0,m}^t$, for $k\le m\le\ell$,
which is part of the state $\btridnut{0}{1}{t}$ of $\bmu$,
and part of the input $\bmu_0^t$ of $\bmu$, respectively, at time $t$.

Then for each time $t\in\bmcpz$ and each $k$, $0\le k\le\ell$,
symmetry $\bomga$ permutes \svec\ $\bmu^{[t,t+k]}$ in $\bmu$ 
to another \svec\ $\bmuht^{[t,t+k]}$ in $\calu$.  The collection of \svecs\
$\{\bmuht^{[t,t+k]}:  t\in\bmcpz, 0\le k\le\ell\}$ specifies
a unique tensor $\bmuht\in\calu$.  Thus $\bomga$ permutes tensor
$\bmu$ to tensor $\bmuht$.
\end{thm}

\begin{thm}
\label{gransub1}
Fix symmetry $\bomga\in\calm$.  Fix tensor $\bmu\in\calu$.  
Fix time $t$ and fix $k$, $0\le k\le\ell$.  
Let $\bmu$ have \svec\ $\bmu^{[t,t+k]}$.  The symmetry $\bomga$
permutes \svec\ $\bmu^{[t,t+k]}$ to another \svec\ $\bmuht^{[t,t+k]}$ in $\calu$.
Fix basis $\bmcpb$.  There is a 1-1 correspondence $\calu\lra\calr$.
From this correspondence, let $\bmu\lra\bmr$ and
$\bmu^{[t,t+k]}\lra\bmr^{[t,t+k]}$.  Then through the 1-1 correspondence
$\calu\lra\calr$, symmetry $\bomga$ induces an assignment that takes
\gvec\ $\bmr^{[t,t+k]}$ to another \gvec\ $\bmrht^{[t,t+k]}$ in $\calr$,
where $\bmrht^{[t,t+k]}\lra\bmuht^{[t,t+k]}$.
The permutation is solely determined by component $\omega_{0,k}^t$
of symmetry input $\bomga_0^t$ at time $t$.  Since $\bmu^t\lra\bmr^t$,
the permutation effectively depends on
$\btridnrt{0}{k}{t}$, a part of the state $\btridnrt{0}{1}{t}$
of $\bmr$ at time $t$, and $r_{0,m}^t$, for $k\le m\le\ell$, 
a part of the input $\bmr_0^t$ of $\bmr$ at time $t$.

Then for each time $t\in\bmcpz$ and each $k$, $0\le k\le\ell$,
symmetry $\bomga$ induces a permutation of \gvec\ $\bmr^{[t,t+k]}$ in $\bmr$ 
to another \gvec\ $\bmrht^{[t,t+k]}$ in $\calr$.  The collection of \gvecs\
$\{\bmrht^{[t,t+k]}:  t\in\bmcpz, 0\le k\le\ell\}$ specifies
a unique tensor $\bmrht\in\calr$.  Thus $\bomga$ effectively permutes tensor
$\bmr$ to tensor $\bmrht$.
\end{thm}

In Subsection 6.3, 
for each of the four comparisons of encoders, $E_1$ and $E_2$, $E$ and $E_Y$,
$E_s$ and $E_{s,Y}$, and $E_1$ and $E_{s,2}$, we saw there was a graph
automorphism of $\duinf$ which made the two encoders graph isomorphic,
composed with the natural isomorphism to $\duinfy$ in the 
second and third comparisons.  If the bases are constant, the graph
automorphism of $\duinf$ is constant.
In this section, we analyzed the structure of any graph automorphism
of $\duinf$.  Theorem \ref{gransub} shows that any
graph automorphism of $\duinf$ is a symmetry $\bomga$ which permutes
\svec\ $\bmu^{[t,t+k]}$ in tensor $\bmu\in\calu$ to \svec\ $\bmuht^{[t,t+k]}$ 
in tensor $\bmuht\in\calu$.  Theorem \ref{gransub1} shows that
symmetry $\bomga$ induces a permutation of $\calr$ which permutes
\gvec\ $\bmr^{[t,t+k]}$ in tensor $\bmr\in\calr$ to \gvec\ $\bmrht^{[t,t+k]}$ 
in tensor $\bmrht\in\calr$.  For each time $t$, this means 
\gvec\ $\bmr^{[t,t+k]}$ in vector basis $\calb^t$ is taken to
\gvec\ $\bmrht^{[t,t+k]}$ in $\calb^t$.  The permutation depends on
$\btridnrt{0}{k}{t}$, a part of the state of $\bmr$ at time $t$, and
$r_{0,m}^t$, for $k\le m\le\ell$, a part of the input of $\bmr$ at time $t$.
For a constant basis, the permutation is constant.
In the case of the comparison $E$ and $E_s$, the permutation of
\gvecs\ gives a transformation between the time domain
and spectral domains.

\newpage
\vspace{3mm}
{\bf 8.  THE NATURAL SYMMETRY SYSTEM}
\vspace{3mm}

\vspace{3mm}
{\bf 8.1  The natural symmetry system}
\vspace{3mm}

Rotman \cite{ROT} gives the Cayley theorem and proof for finite groups
(Theorem 3.12).  Let $S_n$ be the symmetric group on integers $\{1,\ldots,n\}$.

\begin{thm}[Cayley theorem] 
Let $|G|=n$.  Every group $G$ can be imbedded as a subgroup of $S_n$.
\end{thm}

\begin{prf}
Note that a bijection is a permutation and a permutation is a bijection.
Left translation $L_g: G\ra G$ defined by assignment $h\mapsto gh$ 
is a bijection, so $L_g\in S_n$.  The map $L:  G\ra S_n$ 
defined by the assignment $g\mapsto L_g$ is an injection 
and homomorphism.  Then $G\simeq {\rm im}(L)$.
\end{prf}

We have just seen the set $\{L_g:  g\in G\}$ is a group 
${\rm im}(L)$ and $G\simeq {\rm im}(L)$ under the 1-1 correspondence
$g\mapsto L_g$.  The operation in ${\rm im}(L)$ is composition
defined as follows.  If $L_{g_1}\in {\rm im}(L)$ and 
$L_{g_2}\in {\rm im}(L)$, then $L_{g_1}\circ L_{g_2}\in {\rm im}(L)$,
and in fact $L_{g_1}\circ L_{g_2}=L_{g_1g_2}$.

We now want to extend the Cayley theorm for finite groups to group system $C$.
The following result is just the Cayley theorem and proof restated for group 
system $C$.  Let $S_C$ be the symmetric group on group system $C$.
This is the group of all permutations of paths in $C$ with composition operation.

\begin{thm}
Every group system $C$ can be imbedded as a subgroup of $S_C$.
\end{thm}

\begin{prf}
Note that a bijection is a permutation and a permutation is a bijection.
Left translation $L_\bmb: C\ra C$ defined by assignment $\bmc\mapsto \bmb\bmc$ 
is a bijection, so $L_\bmb\in S_C$.  
The map $L:  C\ra S_C$ defined by the assignment $\bmb\mapsto L_\bmb$
is an injection and homomorphism.  Then $C\simeq {\rm im}(L)$.
\end{prf}

Note that left translation $L_\bmb$ is essentially just $\bmb C$, and
the map $L:  C\ra S_C$ defined by the assignment $\bmb\mapsto L_\bmb$
is essentially just $\bmb\mapsto \bmb C$.  Then
we have just seen the set $\{\bmb C:  \bmb\in C\}$ is a group 
${\rm im}(L)$ and $C\simeq {\rm im}(L)$ under the 1-1 correspondence
$\bmb\mapsto \bmb C$.  The operation in ${\rm im}(L)$ is composition
defined as follows.  If $\bmb_1 C\in {\rm im}(L)$ and 
$\bmb_2 C\in {\rm im}(L)$, then $\bmb_1 C\circ\bmb_2 C\in {\rm im}(L)$,
and in fact $\bmb_1 C\circ\bmb_2 C=(\bmb_1\bmb_2)C$.
We now show $\bmb C$ is essentially a symmetry.

\begin{lem}
\label{lem747}
Left translation $L_b$, a bijection on $C$, induces a symmetry $\bomga_\bmb$ 
of $\duinf$, a bijection on $\calu$.
\end{lem}

\begin{prf}
The paths of $C$ are described by sequences of the encoder $\edrbinf$.
Then multiplication by $\bmb$ in product $\bmb C$ permutes the
sequences of $\edrbinf$, and therefore the vertices of $\drbinf$
so the sequences are preserved.  But $\drbinf$ is graph isomorphic
to $\duinf$.  Therefore the product $\bmb C$ must induce a permutation of 
vertices of $\duinf$ that preserves paths.
\end{prf}

Lemma \ref{lem747} shows we can define an isomorphism from 
${\rm im}(L)$ into $\calm$.  Let $\bmb C\stackrel{\alpha}{\mapsto}\bomga_\bmb$.
Then $\bmb_1 C\circ\bmb_2 C\stackrel{\alpha}{\mapsto}\bomga_{\bmb_1}\circ\bomga_{\bmb_2}$.
Therefore ${\rm im}(\alpha)$ is a subgroup of $\calm$ with composition
operation.  We let ${\rm im}(\alpha)$ be $\caln$, the 
{\it natural symmetry system} of $C$.  A symmetry in $\caln$ is denoted
by $\bomga_\bmb$, where $\bomga_\bmb$ is the symmetry induced by $\bmb C$.

\begin{thm}
There is an isomorphism $C\simeq {\rm im}(L)\stackrel{\alpha}{\simeq}\caln$,
where $\caln$ is the group of symmetries induced by the iterated mapping
$\bmb\mapsto\bmb C\stackrel{\alpha}{\mapsto}\bomga_\bmb$, where $\bmb\in C$ and
$\bmb C\in {\rm im}(L)$.  Thus
every group system $C$ can be imbedded as a subgroup $\caln$ of $\calm$.
\end{thm}

The isomorphism $C\simeq {\rm im}(L)\stackrel{\alpha}{\simeq}\caln$
gives the assignments $\bmb\mapsto\bmb C\stackrel{\alpha}{\mapsto}\bomga_\bmb$.  
If $\bmb$ is 
a generator $\bmg^{[t,t+k]}$ in $C$, then we have the assignments
$$
\bmg^{[t,t+k]}\mapsto\bmg^{[t,t+k]} C\stackrel{\alpha}{\mapsto}\bomga_{\bmg^{[t,t+k]}}.
$$
For a generator $\bmg^{[t,t+k]}$ in $C$, $\bmg^{[t,t+k]} C$ is a generator 
in ${\rm im}(L)$ and the corresponding symmetry $\bomga_{\bmg^{[t,t+k]}}$ 
is a generator in $\caln$.  
We see that a generator in $\caln$ can be more complicated than a 
generator $\bmg^{[t,t+k]}$ in $C$ because it involves multiplication 
$\bmg^{[t,t+k]} C$.

Based on \cite{FT}, (\ref{encft}) gives a decomposition of any
path $\bmb\in C$ as a product of generators $\bmg^{[t,t+k]}$, $t\in\bmcpz$,
$0\le k\le\ell$.  We now consider the equality (\ref{encft}) in ${\rm im}(L)$.  
$L$ gives the assignments
$$
L:  \bmb\mapsto\bmb C,
$$
and
$$
L:  \prod_{0\le k\le\ell} \prod_t \bmg^{[t,t+k]}\mapsto
\bigotimes_{0\le k\le\ell} \bigotimes_t \bmg^{[t,t+k]}C,
$$
where $\bigotimes$ indicates an iterated series of compositions in ${\rm im}(L)$.
Then we can rewrite (\ref{encft}) in ${\rm im}(L)$ as
$$
\bmb C=\bigotimes_{0\le k\le\ell} \bigotimes_t \bmg^{[t,t+k]}C.
$$
Using the isomorphism ${\rm im}(L)\stackrel{\alpha}{\simeq}\caln$,
this gives
$$
\bomga_\bmb=\bigotimes_{0\le k\le\ell} \bigotimes_t \bomga_{\bmg^{[t,t+k]}}.
$$
Thus a symmetry $\bomga_\bmb$ in $\caln$ is a composition of
generators $\bomga_{\bmg^{[t,t+k]}}$ in $\caln$.  Then to study any
symmetry $\bomga_\bmb$, it is sufficient to study the generator 
$\bomga_{\bmg^{[t,t+k]}}$.  Alternatively we may study the product
$\bmg^{[t,t+k]}C$ or the product $\bmg^{[t,t+k]}\bmc$ for any $\bmc\in C$.

\begin{thm}
\label{thm83}
Let $\bmb\in C$ be composed of generators $\bmg^{[t,t+k]}\in C$.
A symmetry $\bomga_\bmb$ in $\caln$ is a composition of
generators $\bomga_{\bmg^{[t,t+k]}}$ in $\caln$, where $\bomga_\bmb$
and $\bomga_{\bmg^{[t,t+k]}}$ satisfy the 1-1 correspondence
in the isomorphism $C\simeq {\rm im}(L)\stackrel{\alpha}{\simeq}\caln$.
\end{thm}

Consider the 5-tuple family $(\caln,\calu,\calr,C;\bmcpb)$.  Like $\calm$
in 5-tuple family $(\calm,\calu,\calr,C;\bmcpb)$,
$\caln$ is a tensor set and group system.  
And like $\calm$, $\caln$ acts on $\calu$ and through the 1-1 correspondence
$\calu\lra\calr\lra C$, $\caln$ implicitly acts on $\calr$ and $C$ also.
Unlike $\calm$, $\caln\simeq C$ and $\caln$ depends on basis $\bmcpb$.

In Section 8 we use a second notation to denote \svecs\ $\bmr^{[t,t+k]}$,
$\bmu^{[t,t+k]}$, and $\bomga^{[t,t+k]}$.  If \svec\ $\bmr^{[t,t+k]}$
is in a tensor $\bmr\in\calr$, we let $\bmr^{[t,t+k]}$ be denoted by
$$
\bmv^{[t,t+k]}(\bmr)\rmdef\bmr^{[t,t+k]}.
$$
Similarly, if \svec\ $\bmu^{[t,t+k]}$ is in tensor $\bmu\in\calu$, let 
$$
\bmv^{[t,t+k]}(\bmu)\rmdef\bmu^{[t,t+k]},
$$
and if \svec\ $\bomga^{[t,t+k]}$ is in tensor $\bomga\in\calm$, let 
$$
\bmv^{[t,t+k]}(\bomga)\rmdef\bomga^{[t,t+k]}.
$$

Left translation $L_\bmb:  C\ra C$ gives the assignment 
$L_\bmb:  \bmc\mapsto\bmb\bmc$.  Let $\bmb\bmc=\bmcbr$.
Consider the 1-1 correspondences $\bmu\lra\bmr\lra\bmc$ 
and $\bmubr\lra\bmrbr\lra\bmcbr$.  $L_b$ gives the 
assignment $L_\bmb:  \bmc\mapsto\bmcbr$.  
The symmetry $\bomga_\bmb$ corresponding to $L_\bmb$
gives the corresponding assignment 
$\bomga_\bmb:  \bmu\mapsto\bmubr$.  The commmutative diagram 
Figure \ref{commutrel} relates $L_\bmb$ and $\bomga_\bmb$.

Through the 1-1 correspondence $\calu\lra\calr$, a symmetry
$\bomga_\bmb:  \calu\ra\calu$ induces a function
$\bvarpi_\bmb:  \calr\ra\calr$ such that if $\bomga_\bmb:  \bmu\mapsto\bmubr$,
then $\bvarpi_\bmb:  \bmr\mapsto\bmrbr$, as shown by the commutative
diagram Figure \ref{commutrel}.  Tensor $\bmr$
is composed of \svecs\ $\bmv^{[t,t+k]}(\bmr)$ for $t\in\bmcpz$
and $0\le k\le\ell$.  If
\be
\label{map55}
\bvarpi_\bmb:  \bmr\mapsto\bmrbr,
\ee
then \svec\ $\bmv^{[t,t+k]}(\bmr)$ in $\bmr$ is changed to
\svec\ $\bmv^{[t,t+k]}(\bmrbr)$ in $\bmrbr$ for $t\in\bmcpz$
and $0\le k\le\ell$.  We abuse notation (\ref{map55}) slightly and indicate this as
\be
\label{map66}
\bvarpi_\bmb:  \bmv^{[t,t+k]}(\bmr)\mapsto\bmv^{[t,t+k]}(\bmrbr).
\ee
Let $L_\bmb^t$, $\bvarpi_\bmb^t$, and $\bomga_\bmb^t$ be the time $t$
components of $L_\bmb$, $\bvarpi_\bmb$, and $\bomga_\bmb$ respectively.

\begin{figure}[h]
\centering

\begin{picture}(100,200)
\put(0,20){\vector(0,1){60}}
\put(0,80){\vector(0,-1){60}}
\put(0,120){\vector(0,1){60}}
\put(0,180){\vector(0,-1){60}}
\put(20,200){\vector(1,0){60}}
\put(20,100){\vector(1,0){60}}
\put(20,0){\vector(1,0){60}}
\put(100,20){\vector(0,1){60}}
\put(100,80){\vector(0,-1){60}}
\put(100,120){\vector(0,1){60}}
\put(100,180){\vector(0,-1){60}}
\put(0,0){\makebox(0,0){$\bmr$}}
\put(0,100){\makebox(0,0){$\bmu$}}
\put(0,200){\makebox(0,0){$\bmc$}}
\put(100,200){\makebox(0,0){$\bmcbr=\bmb\bmc$}}
\put(100,100){\makebox(0,0){$\bmubr$}}
\put(100,0){\makebox(0,0){$\bmrbr$}}
\put(50,195){\makebox(0,0)[t]{$L_\bmb(\bmc)$}}
\put(50,95){\makebox(0,0)[t]{$\bomga_\bmb(\bmu)$}}
\put(50,5){\makebox(0,0)[b]{$\bvarpi_\bmb(\bmr)$}}
\end{picture}

\caption{Commutative diagram relating $L_\bmb$, $\bomga_\bmb$, and $\bvarpi_\bmb$.}
\label{commutrel}

\end{figure}


From Theorems \ref{gransub} and \ref{gransub1}, we know that
each symmetry in $\calm$ takes each \svec\ in $\bmu\in\calu$ to
another \svec\ in $\bmuht\in\calu$ and each \gvec\ in $\bmr\in\calr$
to another \gvec\ in $\bmrht\in\calr$, where $\bmu\lra\bmr$ and
$\bmuht\lra\bmrht$ are in 1-1 correspondence.  We can now give a 
result on multiplication in $C$.

\begin{thm}
\label{thm80}
The multiplication by $\bmb$ in $\bmb C$ corresponds to changing 
each \gvec\ in $\calr$ to another, at each time $t$ and length $k$, 
$0\le k\le\ell$.
\end{thm}

\begin{prf}
From Lemma \ref{lem747}, left translation induces a symmetry
of $\duinf$.
\end{prf}
We consider the effect of multiplication in $C$ on tensor set $\calr$ 
further in the next section.

\begin{prop}
The \nss\ $\caln$ of $C$ is \ellctl, the same as $C$.
\end{prop}

\begin{prf}
Fix any $\bmb_1,\bmb_2\in C$.  Consider any $\bmb_1^{(-\infty,t]}$ and 
any $\bmb_2^{[t+\ell,\infty)}$.  Since $C$ is \ellctl, there is always a
path $\bmb$ such that $\chi^{(-\infty,t]}(\bmb)=\bmb_1^{(-\infty,t]}$
and $\chi^{[t+\ell,\infty)}(\bmb)=\bmb_2^{[t+\ell,\infty)}$.  Now fix any 
$\bmb_1^{(-\infty,t]}C$ and any $\bmb_2^{[t+\ell,\infty)}C$.
Since $C$ is \ellctl, there is always a
path $\bmb$ such that $\chi^{(-\infty,t]}(\bmb C)=\bmb_1^{(-\infty,t]} C$
and $\chi^{[t+\ell,\infty)}(\bmb C)=\bmb_2^{[t+\ell,\infty)} C$.
Thus $\caln$ is \ellctl.
\end{prf}

\vspace{3mm}
{\bf 8.2  Multiplication in $\calr$}
\vspace{3mm}

We study multiplication in $\calr$ and show this is related to the abelian
and nonabelian structure of a group system and the structure of the 
\nss\ $\caln$.

Multiplication in $C$ is easy.  For $\bmb,\bmc\in C$, product $\bmb\bmc$ is given
by $b^tc^t$ for each $t\in\bmcpz$.  We now want to consider multiplication in
$\calr$.  Multiplication in $\calr$ gives more insight into the structure
of a group system than multiplication in $C$.  In Theorem \ref{thm83},
we showed that products in $C$ can be decomposed into terms of the form
$\bmg^{[t,t+k]}C$ or $\bmg^{[t,t+k]}\bmc$ for any $\bmc\in C$.  In this 
subsection we study the term $\bmg^{[t,t+k]}\bmc$ using the time domain
and \nss\ $\caln$.  First we give some useful definitions.

For any time $t\in\bmcpz$, we have given an expansion, or \crepc, of
branch $b^t$ in terms of \creps\ in (\ref{enctd1}):
$$
b^t=r_{\ell,\ell}^t r_{\ell-1,\ell}^t r_{\ell-1,\ell-1}^t\cdots 
r_{j,\ell}^t\cdots r_{j,k}^t\cdots r_{j,j}^t\cdots
r_{2,2}^tr_{1,\ell}^t\cdots r_{1,1}^tr_{0,\ell}^t\cdots r_{0,2}^tr_{0,1}^tr_{0,0}^t.  
$$
Fix representative $r_{j,k}^t$.  The components $r_{m,n}^t$ to the left of 
$r_{j,k}^t$ in (\ref{enctd1}) are called {\it ascendants} of $r_{j,k}^t$.  These
are ``above'' $r_{j,k}^t$ in the \crepc.
The components $r_{m,n}^t$ to the right of $r_{j,k}^t$
in (\ref{enctd1}) are called {\it descendants} of $r_{j,k}^t$.  These
are ``below'' $r_{j,k}^t$ in the \crepc.

We say two time intervals $[t,t+k]$ and $[t',t'+n]$ {\it overlap}
if $[t,t+k]\cap [t',t'+n]$ is not empty.  We say two generators
$\bmg^{[t,t+k]}$ and $\bmg^{[t',t'+n]}$ {\it overlap} if the
time intervals $[t,t+k]$ and $[t',t'+n]$ overlap.  We now give
conditions under which a component $r_{m,n}^{t'+m}$ of
\gvec\ $\bmg^{[t',t'+n]}$ is an ascendant and a descendant of 
component $r_{j,k}^{t+j}$ in \gvec\ $\bmg^{[t,t+k]}$.

\begin{lem}
\label{lem113}
Fix $\bmr$.  Fix time $t+j$.  Fix $r_{j,k}^{t+j}\in\bmg^{[t,t+k]}$.
Fix $r_{m,n}^{t'+m}\in\bmg^{[t',t'+n]}$.  Then
$r_{m,n}^{t'+m}$ is an ascendant of $r_{j,k}^{t+j}$ \ifof\
these 3 conditions hold:  $[t,t+k]$ and $[t',t'+n]$ overlap with
$t'\le t$, if $t=t'$ then $n>k$, and $t'+m=t+j$.  And
$r_{m,n}^{t'+m}$ is a descendant of $r_{j,k}^{t+j}$ \ifof\
these 3 conditions hold:  $[t,t+k]$ and $[t',t'+n]$ overlap with
$t'\ge t$, if $t=t'$ then $n<k$, and $t'+m=t+j$.
\end{lem}

We say a \svec\ $\bmv^{[t',t'+n]}(\bmrht)$ in $\calr$ is {\it subordinate}
to $[t,t+k]$ if $[t',t'+n]\subset [t,t+k]$ and $n<k$.
We say a \svec\ $\bmv^{[t',t'+n]}(\bmrht)$ in $\calr$ is {\it superordinate}
to $[t,t+k]$ if $[t,t+k]\subset [t',t'+n]$ and $n>k$.

We say representative $r_{m,n}^{t'+m}$ is a {\it direct ascendant} of $r_{j,k}^{t+j}$,
where $t'+m=t+j$, if it is a component of a \svec\ $\bmv^{[t',t'+n]}(\bmrht)$
superordinate to $[t,t+k]$.  
We say representative $r_{m,n}^{t'+m}$ is a {\it direct descendant} of $r_{j,k}^{t+j}$,
where $t'+m=t+j$, if it is a component of a \svec\ $\bmv^{[t',t'+n]}(\bmrht)$
subordinate to $[t,t+k]$.
By Lemma \ref{lem113}, it is easy to see both definitions are well defined.
Ascendants that are not direct ascendants are
called {\it indirect ascendants}.  Descendants that are not direct descendants 
are called {\it indirect descendants}.

\begin{prop}
\label{prop67}
The direct ascendants of $r_{j,k}^{t+j}$ are all the components in 
$\btridnrt{j}{k}{t+j}$.
\end{prop}
It can be seen the components that are direct descendants give a
parallelogram shape in $\tridnrt{0}{0}{t+j}$ with upper right corner $r_{j,k}^{t+j}$.

\begin{lem}
Fix time $t$.  Consider components $c^t$ and $\cbr^t$ from two paths $\bmc$ and $\bmcbr$.
Let $\bmc\lra\bmr$ and $\bmcbr\lra\bmrbr$.  We have $c^t=\cbr^t$ \ifof\ 
$r_{j,k}^t=\rbr_{j,k}^t$ for $0\le j\le k$ and $0\le k\le\ell$.
\end{lem}

\begin{lem}
\label{lem71}
Fix time $t$.  Consider components $c^t$ and $\cbr^t$ from two paths $\bmc$ and $\bmcbr$.
Let $\bmc\lra\bmr$ and $\bmcbr\lra\bmrbr$.  We have $c^t=\cbr^t$ \ifof\ 
\svec\ $\bmv^{[t',t'+\alpha]}(\bmr)$ in $\bmr$ and 
\svec\ $\bmv^{[t',t'+\alpha]}(\bmrbr)$
in $\bmrbr$ satisfy $\bmv^{[t',t'+\alpha]}(\bmr)=\bmv^{[t',t'+\alpha]}(\bmrbr)$ for 
any $[t',t'+\alpha]$ such that $t\in [t',t'+\alpha]$.
\end{lem}

\begin{prf}
Shift vector $\bmv^{[t',t'+\alpha]}(\bmrbr)$ is uniquely determined by any
of its components $\rbr_{j,k}^t$, $t'\le t\le t'+k$, for fixed basis $\bmcpb$.
\end{prf}

\begin{thm}
\label{thm104a}
Consider the product $\bmg^{[t,t+k]}\bmc=\bmcbr$ 
or $L_{\bmg^{[t,t+k]}}(\bmc)$.  Let $\bmc\lra\bmr$ and $\bmcbr\lra\bmrbr$.  
After multiplication, the decomposition $\bmr$ of $\bmc$ changes to
that of $\bmrbr$.  The only \svecs\ in $\bmr$ which can change 
from $\bmr$ to $\bmrbr$ are those subordinate to $[t,t+k]$.
\end{thm}

\begin{prf}
From Theorem \ref{thm80}, we know the product $\bmg^{[t,t+k]}\bmc$ 
changes \svecs\ in $\bmr$ to \svecs\ in $\bmrbr$, 
for each time $t$ and length $k$, $0\le k\le\ell$.
Since $c^{t'}=\cbr^{t'}$ for $t'$ outside time interval $[t,t+k]$, then we can
apply Lemma \ref{lem71}.  This means the only \svecs\ in $\bmr$ 
which can change from $\bmr$ to $\bmrbr$ are those subordinate to $[t,t+k]$.
\end{prf}

\begin{cor}
Consider the product $\bmg^{[t,t+k]}\bmc=\bmcbr$ 
or $L_{\bmg^{[t,t+k]}}(\bmc)$.  Let $\bmc\lra\bmr$ and $\bmcbr\lra\bmrbr$.  
After multiplication, the decomposition $\bmr$ of $\bmc$ changes to
that of $\bmrbr$.  The only representatives in $\bmr$ which can change 
from $\bmr$ to $\bmrbr$ are $r_{j,k}^{t+j}$ and direct descendants of
$r_{j,k}^{t+j}$, for $0\le j\le k$.
\end{cor}

\begin{prf}
The representative $r_{j,k}^{t+j}$ and direct descendants of
$r_{j,k}^{t+j}$, for $0\le j\le k$, are the representatives of \svecs\
in $\bmr$ which are subordinate to $[t,t+k]$.
\end{prf}

We can use these results to find the form of the symmetry $\bomga_{\bmg^{[t,t+k]}}$ 
corresponding to the product $\bmg^{[t,t+k]} C$.  In particular,
we want to find $\bmg^{[t,t+k]} \bmc$, or $L_{\bmg^{[t,t+k]}}(\bmc)$,
for each $\bmc\in C$.  Let $\bmu\lra\bmr\lra\bmc$ and $\bmubr\lra\bmrbr\lra\bmcbr$.
As a consequence of Theorem \ref{thm104a},
we know the form of function $\bomga_{\bmg^{[t,t+k]}}(\bmu)$ corresponding to
$L_{\bmg^{[t,t+k]}}(\bmc)$.  All functions $\omega_{m,n}^{t'+m}$ 
in $\bomga_{\bmg^{[t,t+k]}}$ are trivial 
except possibly those belonging to any \svec\
\be
\label{svec1}
\bmv^{[t',t'+n]}(\bomga_{\bmg^{[t,t+k]}})=
(\omega_{0,n}^{t'},\omega_{1,n}^{t'+1},\ldots,\omega_{m,n}^{t'+m},\ldots,\omega_{n,n}^{t'+n}),
\ee
where $[t',t'+n]\subset [t,t+k]$.  Then the multiplication 
$\bmg^{[t,t+k]}\bmc=\bmcbr$ induces a change from \svec\
\be
\label{xyzb}
\bmv^{[t',t'+n]}(\bmu)=
(u_{0,n}^{t'},u_{1,n}^{t'+1},\ldots,u_{m,n}^{t'+m},\ldots,u_{n,n}^{t'+n})
\ee
in $\bmu$ to \svec\ $\bmv^{[t',t'+n]}(\bmubr)$ in $\bmubr$,
\be
\label{map1}
\bomga_{\bmg^{[t,t+k]}}:  \bmv^{[t',t'+n]}(\bmu)\mapsto\bmv^{[t',t'+n]}(\bmubr),
\ee
if $[t',t'+n]\subset [t,t+k]$, but all other \svecs\ are unchanged.
Consequently this multiplication also induces a change from \svec\
$\bmv^{[t',t'+n]}(\bmr)$ in $\bmr$ to \svec\ $\bmv^{[t',t'+n]}(\bmrbr)$ in $\bmrbr$,
\be
\label{map1a}
\bvarpi_{\bmg^{[t,t+k]}}:  \bmv^{[t',t'+n]}(\bmr)\mapsto\bmv^{[t',t'+n]}(\bmrbr),
\ee
if $[t',t'+n]\subset [t,t+k]$, but all other \svecs\ are unchanged.

In any component function 
$\omega_{m,n}^{t'+m}(u_{m,n}^{t'+m},\btridnut{m}{n}{t'+m})$ of a
\svec\ (\ref{svec1}), partial argument $\btridnut{m}{n}{t'+m}$ has a
triangle shape.  As $\bmc$ varies among elements of $C$ in product 
$\bmg^{[t,t+k]}\bmc$, $\bmu$ changes and therefore entries in 
$\btridnut{m}{n}{t'+m}$ change.  We know that entries in $\btridnut{m}{n}{t'+m}$
correspond 1-1 with entries in $\btridnrt{m}{n}{t'+m}$. 
From Proposition \ref{prop67}, an entry 
in the triangle $\btridnrt{m}{n}{t'+m}$ corresponds to 
a \svec\ $\bmv^{[t'',t''+i]}(\bmr)$ in $\bmr$ superordinate to $[t',t'+n]$, 
and each \svec\ $\bmv^{[t'',t''+i]}(\bmr)$ in $\bmr$ superordinate to $[t',t'+n]$
corresponds to an entry in the triangle.  This means that in the product 
$\bmg^{[t,t+k]}\bmc=\bmcbr$, the change from $\bmv^{[t',t'+n]}(\bmr)$
in $\bmr$ to $\bmv^{[t',t'+n]}(\bmrbr)$ in $\bmrbr$ is only affected 
by \svecs\ in $\bmr$ which are superordinate to $[t',t'+n]$.

\begin{thm}
\label{thm699}
Let $[t',t'+n]\subset [t,t+k]$.  In the product 
$\bmg^{[t,t+k]}\bmc=\bmcbr$, the change in (\ref{map1a}) from 
$\bmv^{[t',t'+n]}(\bmr)$ in $\bmr$ to $\bmv^{[t',t'+n]}(\bmrbr)$ 
in $\bmrbr$ is only affected by \svecs\ in $\bmr$ superordinate to $[t',t'+n]$.
\end{thm}

\begin{cor}
Let $[t',t'+n]\subset [t,t+k]$.  Fix $m$ such that $t'\le t'+m\le t'+n$.
Let $r_{m,n}^{t'+m}$ be a component in \svec\ $\bmv^{[t',t'+n]}(\bmr)$.
In the product $\bmg^{[t,t+k]}\bmc=\bmcbr$, 
the change from representative $r_{m,n}^{t'+m}$
in $\bmr$ to representative $\rbr_{m,n}^{t'+m}$ in $\bmrbr$ is only affected 
by representatives in $\bmr$ which are direct ascendants of $r_{m,n}^{t'+m}$.
\end{cor}

\begin{prf}
Consider $r_{m,n}^{t'+m}(c^{t'+m})$.  Since $\bmg^{[t,t+k]}\bmc=\bmcbr$
is only affected by \svecs\ in $\bmr$ superordinate to $[t',t'+n]$,
then for $0\le m\le n$, $r_{m,n}^{t'+m}(c^{t'+m})$ is only affected 
by representatives in $\bmr$ which are direct ascendants of $r_{m,n}^{t'+m}$.
\end{prf}

In particular we now want to study the effect of multiplication 
$\bmg^{[t,t+k]} \bmc$ on \svec\ $\bmv^{[t',t'+n]}(\bmu)$ in $\bmu$
when $[t',t'+n]=[t,t+k]$.  Then (\ref{svec1})-(\ref{map1a}) become
\be
\label{svec11}
\bmv^{[t,t+k]}(\bomga_{\bmg^{[t,t+k]}})=
(\omega_{0,k}^t,\omega_{1,k}^{t+1},\ldots,\omega_{j,k}^{t+j},\ldots,\omega_{k,k}^{t+k}),
\ee
\be
\label{xyzb1}
\bmv^{[t,t+k]}(\bmu)=
(u_{0,k}^t,u_{1,k}^{t+1},\ldots,u_{j,k}^{t+j},\ldots,u_{k,k}^{t+k}),
\ee
\be
\label{map11}
\bomga_{\bmg^{[t,t+k]}}:  \bmv^{[t,t+k]}(\bmu)\mapsto\bmv^{[t,t+k]}(\bmubr),
\ee
\be
\label{map1a1}
\bvarpi_{\bmg^{[t,t+k]}}:  \bmv^{[t,t+k]}(\bmr)\mapsto\bmv^{[t,t+k]}(\bmrbr),
\ee
where $0\le j\le k$.  We have just shown the assignment in (\ref{map11})
only depends on direct ascendants of $u_{j,k}^{t+j}$, for $0\le j\le k$,
and the assignment in (\ref{map1a1})
only depends on direct ascendants of $r_{j,k}^{t+j}$, for $0\le j\le k$.

We now consider two different choices for $\bmc$, $\bmcdt$ and $\bmcddt$.
Let $\bmudt\lra\bmrdt\lra\bmcdt$ and $\bmuddt\lra\bmrddt\lra\bmcddt$.
We select $\bmcdt$ and $\bmcddt$ so that
$$
\bmv^{[t,t+k]}(\bmudt)=\bmv^{[t,t+k]}(\bmuddt).
$$
In $\bmuddt$, all the ascendants of $\uddt_{j,k}^{t+j}$, for $0\le j\le k$,
are trivial.
In $\bmudt$, all the direct ascendants of $\udt_{j,k}^{t+j}$, for $0\le j\le k$,
are trivial, but the indirect ascendants can be arbitrary.
Let $\bmg^{[t,t+k]}\bmcddt=\bmcgr$ and
$\bmg^{[t,t+k]}\bmcdt=\bmcac$.  
Let $\bmugr\lra\bmrgr\lra\bmcgr$ and $\bmuac\lra\bmrac\lra\bmcac$.
Then for multiplication $\bmg^{[t,t+k]}\bmcddt=\bmcgr$, we have
\be
\label{map3}
\bomga_{\bmg^{[t,t+k]}}:  \bmv^{[t,t+k]}(\bmuddt)\mapsto\bmv^{[t,t+k]}(\bmugr),
\ee
\be
\label{map3a}
\bvarpi_{\bmg^{[t,t+k]}}:  \bmv^{[t,t+k]}(\bmrddt)\mapsto\bmv^{[t,t+k]}(\bmrgr),
\ee
and for multiplication $\bmg^{[t,t+k]}\bmcdt=\bmcac$, we have
\be
\label{map2}
\bomga_{\bmg^{[t,t+k]}}:  \bmv^{[t,t+k]}(\bmudt)\mapsto\bmv^{[t,t+k]}(\bmuac),
\ee
\be
\label{map2a}
\bvarpi_{\bmg^{[t,t+k]}}:  \bmv^{[t,t+k]}(\bmrdt)\mapsto\bmv^{[t,t+k]}(\bmrac).
\ee
But since $\bmv^{[t,t+k]}(\bmudt)$ in $\bmudt$ is the same as 
$\bmv^{[t,t+k]}(\bmuddt)$ in $\bmuddt$, and since the direct ascendants 
of $\udt_{j,k}^{t+j}$ in $\bmudt$ 
are the same as the direct ascendants of $\uddt_{j,k}^{t+j}$
in $\bmuddt$, for $0\le j\le k$, we must have 
$\bmv^{[t,t+k]}(\bmugr)=\bmv^{[t,t+k]}(\bmuac)$.  
This gives the commutative diagram Figure \ref{commut4}.
But since $\bmv^{[t,t+k]}(\bmugr)=\bmv^{[t,t+k]}(\bmuac)$, we  must have
$\bmv^{[t,t+k]}(\bmrgr)=\bmv^{[t,t+k]}(\bmrac)$, and consequently 
commutative diagram Figure \ref{commut5} also holds.
Thus we have shown the following.

\begin{lem}
Consider the products $\bmg^{[t,t+k]}\bmcddt=\bmcgr$ and
$\bmg^{[t,t+k]}\bmcdt=\bmcac$.  
Let $\bmugr\lra\bmrgr\lra\bmcgr$ and $\bmuac\lra\bmrac\lra\bmcac$.
Then $\bmv^{[t,t+k]}(\bmrgr)=\bmv^{[t,t+k]}(\bmrac)$.  In other words,
we have $\rgr_{j,k}^{t+j}=\rac_{j,k}^{t+j}$ for $0\le j\le k$.
\end{lem}

We know that $\rgr_{j,k}^{t+j}$ is the $(j,k)$ component of $\bmrgr^{t+j}$
and $\rac_{j,k}^{t+j}$ is the $(j,k)$ component of $\bmrac^{t+j}$.
And using Figure \ref{commutrel}, we know that $\bmrgr^{t+j}$ is the 
decomposition of $\bmcgr^{t+j}$, where
\begin{align*}
\bmcgr^{t+j} &=\chi^{t+j}(\bmg^{[t,t+k]})\cddt^{t+j} \\
&=r_{j,k}^{t+j}\cddt^{t+j},
\end{align*}
as shown in Figure \ref{commut6}.  Similarly, $\bmrac^{t+j}$ is the 
decomposition of $\bmcac^{t+j}$, where
$$
\bmcac^{t+j}=r_{j,k}^{t+j}\cdt^{t+j}.
$$

\begin{thm}
\label{thm69}
Consider the products $\bmg^{[t,t+k]}\bmcddt=\bmcgr$ and
$\bmg^{[t,t+k]}\bmcdt=\bmcac$.  
Let $\bmugr\lra\bmrgr\lra\bmcgr$ and $\bmuac\lra\bmrac\lra\bmcac$.
Fix $j$ such that $0\le j\le k$.
The $(j,k)$ component $\rgr_{j,k}^{t+j}$ of the decomposition $\bmrgr^{t+j}$
of product $r_{j,k}^{t+j}(\cddt^{t+j})$
is the same as the $(j,k)$ component $\rac_{j,k}^{t+j}$ 
of the decomposition $\bmrac^{t+j}$ of product 
$r_{j,k}^{t+j}(\cdt^{t+j})$, that is, $\rgr_{j,k}^{t+j}=\rac_{j,k}^{t+j}$.
\end{thm}

\begin{figure}[h]
\centering

\begin{picture}(100,100)
\put(0,20){\vector(0,1){60}}
\put(0,80){\vector(0,-1){60}}
\put(20,100){\vector(1,0){60}}
\put(20,0){\vector(1,0){60}}
\put(100,20){\vector(0,1){60}}
\put(100,80){\vector(0,-1){60}}
\put(10,0){\makebox(0,0)[r]{$\bmv^{[t,t+k]}(\bmuddt)$}}
\put(10,100){\makebox(0,0)[r]{$\bmv^{[t,t+k]}(\bmudt)$}}
\put(90,100){\makebox(0,0)[l]{$\bmv^{[t,t+k]}(\bmuac)$}}
\put(90,0){\makebox(0,0)[l]{$\bmv^{[t,t+k]}(\bmugr)$}}
\put(-5,50){\makebox(0,0)[r]{$=$}}
\put(105,50){\makebox(0,0)[l]{$=$}}
\put(50,95){\makebox(0,0)[t]{$\bomga_{\bmg^{[t,t+k]}}$}}
\put(50,5){\makebox(0,0)[b]{$\bomga_{\bmg^{[t,t+k]}}$}}
\end{picture}

\caption{Commutative diagram for $\bomga_{\bmg^{[t,t+k]}}$.}
\label{commut4}

\end{figure}

\begin{figure}[h]
\centering

\begin{picture}(100,100)
\put(0,20){\vector(0,1){60}}
\put(0,80){\vector(0,-1){60}}
\put(20,100){\vector(1,0){60}}
\put(20,0){\vector(1,0){60}}
\put(100,20){\vector(0,1){60}}
\put(100,80){\vector(0,-1){60}}
\put(10,0){\makebox(0,0)[r]{$\bmv^{[t,t+k]}(\bmrddt)$}}
\put(10,100){\makebox(0,0)[r]{$\bmv^{[t,t+k]}(\bmrdt)$}}
\put(90,100){\makebox(0,0)[l]{$\bmv^{[t,t+k]}(\bmrac)$}}
\put(90,0){\makebox(0,0)[l]{$\bmv^{[t,t+k]}(\bmrgr)$}}
\put(-5,50){\makebox(0,0)[r]{$=$}}
\put(105,50){\makebox(0,0)[l]{$=$}}
\put(50,95){\makebox(0,0)[t]{$\bvarpi_{\bmg^{[t,t+k]}}$}}
\put(50,5){\makebox(0,0)[b]{$\bvarpi_{\bmg^{[t,t+k]}}$}}
\end{picture}

\caption{Commutative diagram for $\bvarpi_{\bmg^{[t,t+k]}}$.}
\label{commut5}

\end{figure}

\begin{figure}[h]
\centering

\begin{picture}(100,100)
\put(0,20){\vector(0,1){60}}
\put(0,80){\vector(0,-1){60}}
\put(20,100){\vector(1,0){60}}
\put(20,0){\vector(1,0){60}}
\put(100,20){\vector(0,1){60}}
\put(100,80){\vector(0,-1){60}}
\put(0,0){\makebox(0,0){$\bmrddt$}}
\put(0,100){\makebox(0,0){$\bmcddt$}}
\put(90,100){\makebox(0,0)[l]{$\cgr^{t+j}=r_{j,k}^{t+j}\cddt^{t+j}$}}
\put(90,0){\makebox(0,0)[l]{$\bmrgr^{t+j}$}}
\put(50,95){\makebox(0,0)[t]{$L_{\bmg^{[t,t+k]}}^{t+j}$}}
\put(50,5){\makebox(0,0)[b]{$\bvarpi_{\bmg^{[t,t+k]}}^{t+j}$}}
\end{picture}

\caption{Commutative diagram used to calculate $\bmrgr^{t+j}$.}
\label{commut6}

\end{figure}

We now evaluate $r_{j,k}^{t+j}\cddt^{t+j}$ and $r_{j,k}^{t+j}\cdt^{t+j}$
and use these results to show that Theorem \ref{thm69} explains a 
commutative property of any group system $C$.
We can calculate $r_{j,k}^{t+j}\cddt^{t+j}=\cgr^{t+j}$ as 
\be
\label{map5}
r_{j,k}^{t+j}(\cddt^{t+j})
=r_{j,k}^{t+j}(\prod_{m=0}^\ell\prod_{n=m}^\ell \rddt_{m,n}^{t+j}),
\ee
for $0\le j\le k$.  The representatives $\rddt_{m,n}^{t+j}$ are the identity
except for $\rddt_{j,k}^{t+j}$.  Then
\be
\label{aeqn}
r_{j,k}^{t+j}(\cddt^{t+j})=r_{j,k}^{t+j}(\rddt_{j,k}^{t+j}).
\ee
We can calculate
$r_{j,k}^{t+j}\cdt^{t+j}=\cac^{t+j}$ as
\be
\label{map6}
r_{j,k}^{t+j}(\cdt^{t+j})
=r_{j,k}^{t+j}(\prod_{m=0}^\ell\prod_{n=m}^\ell \rdt_{m,n}^{t+j}),
\ee
for $0\le j\le k$.

Fix $j$ such that $0\le j\le k$.   Consider an $\bmrdt$ such that 
$$
\cdt^{t+j}=\rdt_{p,q}^{t+j} \rdt_{j,k}^{t+j} \rdt_{j,k-1}^{t+j} \cdots \rdt_{0,0}^{t+j},
$$
where for some $p,q$, $\rdt_{p,q}^{t+j}$ is a nontrivial ascendant of 
$\rdt_{j,k}^{t+j}$ but not a direct ascendant.  The remaining ascendants 
are trivial.  Since $\rddt_{j,k}^{t+j}=\rdt_{j,k}^{t+j}$, we can rewrite
$\cdt^{t+j}$ as 
$\cdt^{t+j}
=\rdt_{p,q}^{t+j} \rddt_{j,k}^{t+j} \rdt_{j,k-1}^{t+j} \cdots \rdt_{0,0}^{t+j}$.
Then 
\be
\label{bbeqn}
r_{j,k}^{t+j}(\cdt^{t+j})
=r_{j,k}^{t+j}(\rdt_{p,q}^{t+j} \rddt_{j,k}^{t+j} \rdt_{j,k-1}^{t+j} \cdots \rdt_{0,0}^{t+j}).
\ee

From (\ref{qg2}) we know $r_{j,k}^{t+j}$ is a representative of quotient group
\be
\label{qg118m}
\frac{\calf^j(\Delta_k^t)}{\calf^j(\Delta_{k-1}^t)}=
\frac{X_{j-1}^{t+j}(X_j^{t+j}\cap Y_{k-j}^{t+j})}{X_{j-1}^{t+j}(X_j^{t+j}\cap Y_{k-j-1}^{t+j})},
\ee
for $j=0,1,\ldots,k$.  Consider the quotient group (\ref{qg118}) 
determined by representative $r_{j,k}^{t+j}$ and (\ref{qg118m}),  
\be
\label{qg118}
\frac{B^{t+j}}{X_{j-1}^{t+j}(X_j^{t+j}\cap Y_{k-j-1}^{t+j})}.
\ee
This quotient group contains the cosets of (\ref{qg118m}).
Consider representatives $r_{m,n}^{t+j}$ for $(m,n)$ satisfying
$m=j$, $n\ge k$, and $j<m\le\ell$, $m\le n\le\ell$.  Then $r_{m,n}^{t+j}$
is a representative of some coset (\ref{qg118}).
If $\call(r_{m,n}^{t+j})$ is a coset in (\ref{qg118}) such that representative 
$r_{m,n}^{t+j}\in \call(r_{m,n}^{t+j})$, then $r_{m,n}^{t+j}$ is a {\it lifting} 
of $\call(r_{m,n}^{t+j})$.  Going the other way, given a representative $r_{m,n}^{t+j}$
of a coset $\call(r_{m,n}^{t+j})$ in (\ref{qg118}), we say $\call(r_{m,n}^{t+j})$ 
is a {\it reverse lifting} of $r_{m,n}^{t+j}$.
Let $\call(r_{j,k}^{t+j})$ be the reverse lifting of $r_{j,k}^{t+j}$
to the quotient group (\ref{qg118}).  Similarly let $\call(\rdt_{p,q}^{t+j})$ be the 
reverse lifting of $\rdt_{p,q}^{t+j}$ to the quotient group (\ref{qg118}).

The $(j,k)$ component $\rgr_{j,k}^{t+j}$ of the decomposition $\bmrgr^{t+j}$
of product (\ref{aeqn}) is the same as the $(j,k)$ component $\rac_{j,k}^{t+j}$ 
of the decomposition $\bmrac^{t+j}$ of product (\ref{bbeqn}).  This means
that $r_{j,k}^{t+j}(\rddt_{j,k}^{t+j})$ must be in the same coset of 
(\ref{qg118m}) in (\ref{qg118}) as $r_{j,k}^{t+j}(\rdt_{p,q}^{t+j} \rddt_{j,k}^{t+j})$.
Let $\rht_{j,k}^{t+j}$ be the representative that satisfies
\be
\label{ceqn}
r_{j,k}^{t+j} \rdt_{p,q}^{t+j}=\rdt_{p,q}^{t+j} \rht_{j,k}^{t+j}.
\ee
Then we must have $\rht_{j,k}^{t+j} \rddt_{j,k}^{t+j}$ is in the same coset of 
(\ref{qg118m}) in (\ref{qg118}) as $r_{j,k}^{t+j} \rddt_{j,k}^{t+j}$.  This is true
\ifof\ $\rht_{j,k}^{t+j}$ is in the same coset of 
(\ref{qg118m}) in (\ref{qg118}) as $r_{j,k}^{t+j}$.  This is true \ifof\ 
$\call(\rht_{j,k}^{t+j})$, the reverse lifting of $\rht_{j,k}^{t+j}$ 
to the quotient group (\ref{qg118}), is the same coset of 
(\ref{qg118m}) in (\ref{qg118}) as $\call(r_{j,k}^{t+j})$.  Then from (\ref{ceqn}), this
is true \ifof\ coset $\call(r_{j,k}^{t+j})$ commutes with
coset $\call(\rdt_{p,q}^{t+j})$ in (\ref{qg118}).  This gives a
commutative property that holds for any strongly controllable
group system.

\begin{thm}
Fix any representative $r_{j,k}^{t+j}$.  Fix any $j,k$ such that 
$0\le j\le\ell$, $j\le k\le\ell$.  Let $\bmrdt$ be any tensor in
$\calr$ which has $r_{j,k}^{t+j}$ as a component.  Let $\rdt_{p,q}^{t+j}$ 
be any representative in $\bmrdt\in\calr$ such that $\rdt_{p,q}^{t+j}$
is an ascendant but not a direct ascendant of $r_{j,k}^{t+j}$.
Then the coset $\call(r_{j,k}^{t+j})$ in quotient group (\ref{qg118})
determined by $r_{j,k}^{t+j}$ commutes with coset $\call(\rdt_{p,q}^{t+j})$
in (\ref{qg118}).
\end{thm}

There are 3 extreme cases of this result.  Element $r_{\ell,\ell}^{t+j}$
has no ascendants so this result does not apply.  However 
$\rdt_{\ell,\ell}^{t+j}$ is an indirect ascendant of any representative
in $\tridnrt{0}{1}{t+j}$, and so there is a commutative property
with all these representatives.  Element $r_{0,\ell}^{t+j}$
has no direct ascendants so there is a commutative property with
all representatives in $\tridnrtdt{1}{1}{t+j}$.  Element $r_{0,0}^{t+j}$ 
has no indirect ascendants so this result does not apply.
In general $\rdt_{p,q}^{t+j}$ is an indirect ascendant of any
representative $r_{j,k}^{t+j}$ that is not a direct descendant of
$\rdt_{p,q}^{t+j}$, and so a commutative property holds.

\end{document}